\documentclass[useAMS,usenatbib]{mn2e}

\def\setsymbol#1#2{\expandafter\def\csname #1\endcsname{#2}}
\def\getsymbol#1{\csname #1\endcsname}

\def\Planck{{\it Planck\/}}

\newbox\tablebox    \newdimen\tablewidth
\def\leaderfil{\leaders\hbox to 5pt{\hss.\hss}\hfil}
\def\tablenote#1 #2\par{\begingroup \parindent=0.8em
    \abovedisplayshortskip=0pt\belowdisplayshortskip=0pt
    \noindent
    $$\hss\vbox{\hsize\tablewidth \hangindent=\parindent \hangafter=1 \noindent
    \hbox to \parindent{$^#1$\hss}\strut#2\strut\par}\hss$$
    \endgroup}

\def\L2{\ifmmode L_2\else $L_2$\fi}

\def\DeltaT{\ifmmode \Delta T\else $\Delta T$\fi}
\def\deltat{\ifmmode \Delta t\else $\Delta t$\fi}
\def\fknee{\ifmmode f_{\rm knee}\else $f_{\rm knee}$\fi}
\def\Fmax{\ifmmode F_{\rm max}\else $F_{\rm max}$\fi}
\def\solar{\ifmmode{\rm M}_{\mathord\odot}\else${\rm M}_{\mathord\odot}$\fi}
\def\Msolar{\ifmmode{\rm M}_{\mathord\odot}\else${\rm M}_{\mathord\odot}$\fi}
\def\Lsolar{\ifmmode{\rm L}_{\mathord\odot}\else${\rm L}_{\mathord\odot}$\fi}

\def\inv{\ifmmode^{-1}\else$^{-1}$\fi}
\def\mo{\ifmmode^{-1}\else$^{-1}$\fi}
\def\sup#1{\ifmmode ^{\rm #1}\else $^{\rm #1}$\fi}
\def\expo#1{\ifmmode \times 10^{#1}\else $\times 10^{#1}$\fi}
\def\,{\thinspace}
\def\lsim{\mathrel{\raise .4ex\hbox{\rlap{$<$}\lower 1.2ex\hbox{$\sim$}}}}
\def\gsim{\mathrel{\raise .4ex\hbox{\rlap{$>$}\lower 1.2ex\hbox{$\sim$}}}}

\def\simprop{\mathrel{\raise .4ex\hbox{\rlap{$\propto$}\lower 1.2ex\hbox{$\sim$}}}}
\def\deg{\ifmmode^\circ\else$^\circ$\fi}
\def\pdeg{\ifmmode $\setbox0=\hbox{$^{\circ}$}\rlap{\hskip.11\wd0 .}$^{\circ}
          \else \setbox0=\hbox{$^{\circ}$}\rlap{\hskip.11\wd0 .}$^{\circ}$\fi}
\def\arcs{\ifmmode {^{\scriptstyle\prime\prime}}
          \else $^{\scriptstyle\prime\prime}$\fi}
\def\arcm{\ifmmode {^{\scriptstyle\prime}}
          \else $^{\scriptstyle\prime}$\fi}
\newdimen\sa  \newdimen\sb
\def\parcs{\sa=.07em \sb=.03em
     \ifmmode \hbox{\rlap{.}}^{\scriptstyle\prime\kern -\sb\prime}\hbox{\kern -\sa}
     \else \rlap{.}$^{\scriptstyle\prime\kern -\sb\prime}$\kern -\sa\fi}
\def\parcm{\sa=.08em \sb=.03em
     \ifmmode \hbox{\rlap{.}\kern\sa}^{\scriptstyle\prime}\hbox{\kern-\sb}
     \else \rlap{.}\kern\sa$^{\scriptstyle\prime}$\kern-\sb\fi}
\def\ra[#1 #2 #3.#4]{#1\sup{h}#2\sup{m}#3\sup{s}\llap.#4}
\def\dec[#1 #2 #3.#4]{#1\deg#2\arcm#3\arcs\llap.#4}
\def\deco[#1 #2 #3]{#1\deg#2\arcm#3\arcs}
\def\rra[#1 #2]{#1\sup{h}#2\sup{m}}

\def\dots{\relax\ifmmode \ldots\else $\ldots$\fi}
\def\WHzsr{\ifmmode $W\,Hz\mo\,sr\mo$\else W\,Hz\mo\,sr\mo\fi}
\def\mHz{\ifmmode $\,mHz$\else \,mHz\fi}
\def\GHz{\ifmmode $\,GHz$\else \,GHz\fi}
\def\mKs{\ifmmode $\,mK\,s$^{1/2}\else \,mK\,s$^{1/2}$\fi}
\def\muKs{\ifmmode \,\mu$K\,s$^{1/2}\else \,$\mu$K\,s$^{1/2}$\fi}
\def\muKRJs{\ifmmode \,\mu$K$_{\rm RJ}$\,s$^{1/2}\else \,$\mu$K$_{\rm RJ}$\,s$^{1/2}$\fi}
\def\muKHz{\ifmmode \,\mu$K\,Hz$^{-1/2}\else \,$\mu$K\,Hz$^{-1/2}$\fi}
\def\MJysr{\ifmmode \,$MJy\,sr\mo$\else \,MJy\,sr\mo\fi}
\def\MJysrmK{\ifmmode \,$MJy\,sr\mo$\,mK$_{\rm CMB}\mo\else \,MJy\,sr\mo\,mK$_{\rm CMB}\mo$\fi}
\def\microns{\ifmmode \,\mu$m$\else \,$\mu$m\fi}
\def\micron{\microns}
\def\muK{\ifmmode \,\mu$K$\else \,$\mu$\hbox{K}\fi}
\def\microK{\ifmmode \,\mu$K$\else \,$\mu$\hbox{K}\fi}
\def\muW{\ifmmode \,\mu$W$\else \,$\mu$\hbox{W}\fi}
\def\kms{\ifmmode $\,km\,s$^{-1}\else \,km\,s$^{-1}$\fi}
\def\kmsMpc{\ifmmode $\,\kms\,Mpc\mo$\else \,\kms\,Mpc\mo\fi}
\setsymbol{LFI:center:frequency:70GHz:units}{70.3\,GHz}
\setsymbol{LFI:center:frequency:44GHz:units}{44.1\,GHz}
\setsymbol{LFI:center:frequency:30GHz:units}{28.5\,GHz}

\setsymbol{LFI:center:frequency:70GHz}{70.3}
\setsymbol{LFI:center:frequency:44GHz}{44.1}
\setsymbol{LFI:center:frequency:30GHz}{28.5}

\setsymbol{LFI:center:frequency:LFI18:Rad:M:units}{71.7\GHz}
\setsymbol{LFI:center:frequency:LFI19:Rad:M:units}{67.5\GHz}
\setsymbol{LFI:center:frequency:LFI20:Rad:M:units}{69.2\GHz}
\setsymbol{LFI:center:frequency:LFI21:Rad:M:units}{70.4\GHz}
\setsymbol{LFI:center:frequency:LFI22:Rad:M:units}{71.5\GHz}
\setsymbol{LFI:center:frequency:LFI23:Rad:M:units}{70.8\GHz}
\setsymbol{LFI:center:frequency:LFI24:Rad:M:units}{44.4\GHz}
\setsymbol{LFI:center:frequency:LFI25:Rad:M:units}{44.0\GHz}
\setsymbol{LFI:center:frequency:LFI26:Rad:M:units}{43.9\GHz}
\setsymbol{LFI:center:frequency:LFI27:Rad:M:units}{28.3\GHz}
\setsymbol{LFI:center:frequency:LFI28:Rad:M:units}{28.8\GHz}
\setsymbol{LFI:center:frequency:LFI18:Rad:S:units}{70.1\GHz}
\setsymbol{LFI:center:frequency:LFI19:Rad:S:units}{69.6\GHz}
\setsymbol{LFI:center:frequency:LFI20:Rad:S:units}{69.5\GHz}
\setsymbol{LFI:center:frequency:LFI21:Rad:S:units}{69.5\GHz}
\setsymbol{LFI:center:frequency:LFI22:Rad:S:units}{72.8\GHz}
\setsymbol{LFI:center:frequency:LFI23:Rad:S:units}{71.3\GHz}
\setsymbol{LFI:center:frequency:LFI24:Rad:S:units}{44.1\GHz}
\setsymbol{LFI:center:frequency:LFI25:Rad:S:units}{44.1\GHz}
\setsymbol{LFI:center:frequency:LFI26:Rad:S:units}{44.1\GHz}
\setsymbol{LFI:center:frequency:LFI27:Rad:S:units}{28.5\GHz}
\setsymbol{LFI:center:frequency:LFI28:Rad:S:units}{28.2\GHz}

\setsymbol{LFI:center:frequency:LFI18:Rad:M}{71.7}
\setsymbol{LFI:center:frequency:LFI19:Rad:M}{67.5}
\setsymbol{LFI:center:frequency:LFI20:Rad:M}{69.2}
\setsymbol{LFI:center:frequency:LFI21:Rad:M}{70.4}
\setsymbol{LFI:center:frequency:LFI22:Rad:M}{71.5}
\setsymbol{LFI:center:frequency:LFI23:Rad:M}{70.8}
\setsymbol{LFI:center:frequency:LFI24:Rad:M}{44.4}
\setsymbol{LFI:center:frequency:LFI25:Rad:M}{44.0}
\setsymbol{LFI:center:frequency:LFI26:Rad:M}{43.9}
\setsymbol{LFI:center:frequency:LFI27:Rad:M}{28.3}
\setsymbol{LFI:center:frequency:LFI28:Rad:M}{28.8}
\setsymbol{LFI:center:frequency:LFI18:Rad:S}{70.1}
\setsymbol{LFI:center:frequency:LFI19:Rad:S}{69.6}
\setsymbol{LFI:center:frequency:LFI20:Rad:S}{69.5}
\setsymbol{LFI:center:frequency:LFI21:Rad:S}{69.5}
\setsymbol{LFI:center:frequency:LFI22:Rad:S}{72.8}
\setsymbol{LFI:center:frequency:LFI23:Rad:S}{71.3}
\setsymbol{LFI:center:frequency:LFI24:Rad:S}{44.1}
\setsymbol{LFI:center:frequency:LFI25:Rad:S}{44.1}
\setsymbol{LFI:center:frequency:LFI26:Rad:S}{44.1}
\setsymbol{LFI:center:frequency:LFI27:Rad:S}{28.5}
\setsymbol{LFI:center:frequency:LFI28:Rad:S}{28.2}

\setsymbol{LFI:white:noise:sensitivity:70GHz:units}{134.7\muKs}
\setsymbol{LFI:white:noise:sensitivity:44GHz:units}{164.7\muKs}
\setsymbol{LFI:white:noise:sensitivity:30GHz:units}{143.4\muKs}

\setsymbol{LFI:white:noise:sensitivity:70GHz}{134.7}
\setsymbol{LFI:white:noise:sensitivity:44GHz}{164.7}
\setsymbol{LFI:white:noise:sensitivity:30GHz}{143.4}

\setsymbol{LFI:white:noise:sensitivity:LFI18:Rad:M:units}{512.0\muKs}
\setsymbol{LFI:white:noise:sensitivity:LFI19:Rad:M:units}{581.4\muKs}
\setsymbol{LFI:white:noise:sensitivity:LFI20:Rad:M:units}{590.8\muKs}
\setsymbol{LFI:white:noise:sensitivity:LFI21:Rad:M:units}{455.2\muKs}
\setsymbol{LFI:white:noise:sensitivity:LFI22:Rad:M:units}{492.0\muKs}
\setsymbol{LFI:white:noise:sensitivity:LFI23:Rad:M:units}{507.7\muKs}
\setsymbol{LFI:white:noise:sensitivity:LFI24:Rad:M:units}{462.2\muKs}
\setsymbol{LFI:white:noise:sensitivity:LFI25:Rad:M:units}{413.6\muKs}
\setsymbol{LFI:white:noise:sensitivity:LFI26:Rad:M:units}{478.6\muKs}
\setsymbol{LFI:white:noise:sensitivity:LFI27:Rad:M:units}{277.7\muKs}
\setsymbol{LFI:white:noise:sensitivity:LFI28:Rad:M:units}{312.3\muKs}
\setsymbol{LFI:white:noise:sensitivity:LFI18:Rad:S:units}{465.7\muKs}
\setsymbol{LFI:white:noise:sensitivity:LFI19:Rad:S:units}{555.6\muKs}
\setsymbol{LFI:white:noise:sensitivity:LFI20:Rad:S:units}{623.2\muKs}
\setsymbol{LFI:white:noise:sensitivity:LFI21:Rad:S:units}{564.1\muKs}
\setsymbol{LFI:white:noise:sensitivity:LFI22:Rad:S:units}{534.4\muKs}
\setsymbol{LFI:white:noise:sensitivity:LFI23:Rad:S:units}{542.4\muKs}
\setsymbol{LFI:white:noise:sensitivity:LFI24:Rad:S:units}{399.2\muKs}
\setsymbol{LFI:white:noise:sensitivity:LFI25:Rad:S:units}{392.6\muKs}
\setsymbol{LFI:white:noise:sensitivity:LFI26:Rad:S:units}{418.6\muKs}
\setsymbol{LFI:white:noise:sensitivity:LFI27:Rad:S:units}{302.9\muKs}
\setsymbol{LFI:white:noise:sensitivity:LFI28:Rad:S:units}{285.3\muKs}

\setsymbol{LFI:white:noise:sensitivity:LFI18:Rad:M}{512.0}
\setsymbol{LFI:white:noise:sensitivity:LFI19:Rad:M}{581.4}
\setsymbol{LFI:white:noise:sensitivity:LFI20:Rad:M}{590.8}
\setsymbol{LFI:white:noise:sensitivity:LFI21:Rad:M}{455.2}
\setsymbol{LFI:white:noise:sensitivity:LFI22:Rad:M}{492.0}
\setsymbol{LFI:white:noise:sensitivity:LFI23:Rad:M}{507.7}
\setsymbol{LFI:white:noise:sensitivity:LFI24:Rad:M}{462.2}
\setsymbol{LFI:white:noise:sensitivity:LFI25:Rad:M}{413.6}
\setsymbol{LFI:white:noise:sensitivity:LFI26:Rad:M}{478.6}
\setsymbol{LFI:white:noise:sensitivity:LFI27:Rad:M}{277.7}
\setsymbol{LFI:white:noise:sensitivity:LFI28:Rad:M}{312.3}
\setsymbol{LFI:white:noise:sensitivity:LFI18:Rad:S}{465.7}
\setsymbol{LFI:white:noise:sensitivity:LFI19:Rad:S}{555.6}
\setsymbol{LFI:white:noise:sensitivity:LFI20:Rad:S}{623.2}
\setsymbol{LFI:white:noise:sensitivity:LFI21:Rad:S}{564.1}
\setsymbol{LFI:white:noise:sensitivity:LFI22:Rad:S}{534.4}
\setsymbol{LFI:white:noise:sensitivity:LFI23:Rad:S}{542.4}
\setsymbol{LFI:white:noise:sensitivity:LFI24:Rad:S}{399.2}
\setsymbol{LFI:white:noise:sensitivity:LFI25:Rad:S}{392.6}
\setsymbol{LFI:white:noise:sensitivity:LFI26:Rad:S}{418.6}
\setsymbol{LFI:white:noise:sensitivity:LFI27:Rad:S}{302.9}
\setsymbol{LFI:white:noise:sensitivity:LFI28:Rad:S}{285.3}

\setsymbol{LFI:knee:frequency:70GHz:units}{29.5\mHz}
\setsymbol{LFI:knee:frequency:44GHz:units}{56.2\mHz}
\setsymbol{LFI:knee:frequency:30GHz:units}{113.7\mHz}

\setsymbol{LFI:knee:frequency:70GHz}{29.5}
\setsymbol{LFI:knee:frequency:44GHz}{56.2}
\setsymbol{LFI:knee:frequency:30GHz}{113.7}

\setsymbol{LFI:knee:frequency:LFI18:Rad:M:units}{16.3\mHz}
\setsymbol{LFI:knee:frequency:LFI19:Rad:M:units}{15.1\mHz}
\setsymbol{LFI:knee:frequency:LFI20:Rad:M:units}{18.7\mHz}
\setsymbol{LFI:knee:frequency:LFI21:Rad:M:units}{37.2\mHz}
\setsymbol{LFI:knee:frequency:LFI22:Rad:M:units}{12.7\mHz}
\setsymbol{LFI:knee:frequency:LFI23:Rad:M:units}{34.6\mHz}
\setsymbol{LFI:knee:frequency:LFI24:Rad:M:units}{46.2\mHz}
\setsymbol{LFI:knee:frequency:LFI25:Rad:M:units}{24.9\mHz}
\setsymbol{LFI:knee:frequency:LFI26:Rad:M:units}{67.6\mHz}
\setsymbol{LFI:knee:frequency:LFI27:Rad:M:units}{187.4\mHz}
\setsymbol{LFI:knee:frequency:LFI28:Rad:M:units}{122.2\mHz}
\setsymbol{LFI:knee:frequency:LFI18:Rad:S:units}{17.7\mHz}
\setsymbol{LFI:knee:frequency:LFI19:Rad:S:units}{22.0\mHz}
\setsymbol{LFI:knee:frequency:LFI20:Rad:S:units}{8.7\mHz}
\setsymbol{LFI:knee:frequency:LFI21:Rad:S:units}{25.9\mHz}
\setsymbol{LFI:knee:frequency:LFI22:Rad:S:units}{15.8\mHz}
\setsymbol{LFI:knee:frequency:LFI23:Rad:S:units}{129.8\mHz}
\setsymbol{LFI:knee:frequency:LFI24:Rad:S:units}{100.9\mHz}
\setsymbol{LFI:knee:frequency:LFI25:Rad:S:units}{38.9\mHz}
\setsymbol{LFI:knee:frequency:LFI26:Rad:S:units}{58.9\mHz}
\setsymbol{LFI:knee:frequency:LFI27:Rad:S:units}{104.4\mHz}
\setsymbol{LFI:knee:frequency:LFI28:Rad:S:units}{40.7\mHz}

\setsymbol{LFI:knee:frequency:LFI18:Rad:M}{16.3}
\setsymbol{LFI:knee:frequency:LFI19:Rad:M}{15.1}
\setsymbol{LFI:knee:frequency:LFI20:Rad:M}{18.7}
\setsymbol{LFI:knee:frequency:LFI21:Rad:M}{37.2}
\setsymbol{LFI:knee:frequency:LFI22:Rad:M}{12.7}
\setsymbol{LFI:knee:frequency:LFI23:Rad:M}{34.6}
\setsymbol{LFI:knee:frequency:LFI24:Rad:M}{46.2}
\setsymbol{LFI:knee:frequency:LFI25:Rad:M}{24.9}
\setsymbol{LFI:knee:frequency:LFI26:Rad:M}{67.6}
\setsymbol{LFI:knee:frequency:LFI27:Rad:M}{187.4}
\setsymbol{LFI:knee:frequency:LFI28:Rad:M}{122.2}
\setsymbol{LFI:knee:frequency:LFI18:Rad:S}{17.7}
\setsymbol{LFI:knee:frequency:LFI19:Rad:S}{22.0}
\setsymbol{LFI:knee:frequency:LFI20:Rad:S}{8.7}
\setsymbol{LFI:knee:frequency:LFI21:Rad:S}{25.9}
\setsymbol{LFI:knee:frequency:LFI22:Rad:S}{15.8}
\setsymbol{LFI:knee:frequency:LFI23:Rad:S}{129.8}
\setsymbol{LFI:knee:frequency:LFI24:Rad:S}{100.9}
\setsymbol{LFI:knee:frequency:LFI25:Rad:S}{38.9}
\setsymbol{LFI:knee:frequency:LFI26:Rad:S}{58.9}
\setsymbol{LFI:knee:frequency:LFI27:Rad:S}{104.4}
\setsymbol{LFI:knee:frequency:LFI28:Rad:S}{40.7}

\setsymbol{LFI:slope:70GHz:units}{$-1.03$\mHz}
\setsymbol{LFI:slope:44GHz:units}{$-0.89$\mHz}
\setsymbol{LFI:slope:30GHz:units}{$-0.87$\mHz}

\setsymbol{LFI:slope:70GHz}{$-1.03$}
\setsymbol{LFI:slope:44GHz}{$-0.89$}
\setsymbol{LFI:slope:30GHz}{$-0.87$}

\setsymbol{LFI:slope:LFI18:Rad:M:units}{$-1.04$\mHz}
\setsymbol{LFI:slope:LFI19:Rad:M:units}{$-1.09$\mHz}
\setsymbol{LFI:slope:LFI20:Rad:M:units}{$-0.69$\mHz}
\setsymbol{LFI:slope:LFI21:Rad:M:units}{$-1.56$\mHz}
\setsymbol{LFI:slope:LFI22:Rad:M:units}{$-1.01$\mHz}
\setsymbol{LFI:slope:LFI23:Rad:M:units}{$-0.96$\mHz}
\setsymbol{LFI:slope:LFI24:Rad:M:units}{$-0.83$\mHz}
\setsymbol{LFI:slope:LFI25:Rad:M:units}{$-0.91$\mHz}
\setsymbol{LFI:slope:LFI26:Rad:M:units}{$-0.95$\mHz}
\setsymbol{LFI:slope:LFI27:Rad:M:units}{$-0.87$\mHz}
\setsymbol{LFI:slope:LFI28:Rad:M:units}{$-0.88$\mHz}
\setsymbol{LFI:slope:LFI18:Rad:S:units}{$-1.15$\mHz}
\setsymbol{LFI:slope:LFI19:Rad:S:units}{$-1.00$\mHz}
\setsymbol{LFI:slope:LFI20:Rad:S:units}{$-0.95$\mHz}
\setsymbol{LFI:slope:LFI21:Rad:S:units}{$-0.92$\mHz}
\setsymbol{LFI:slope:LFI22:Rad:S:units}{$-1.01$\mHz}
\setsymbol{LFI:slope:LFI23:Rad:S:units}{$-0.95$\mHz}
\setsymbol{LFI:slope:LFI24:Rad:S:units}{$-0.73$\mHz}
\setsymbol{LFI:slope:LFI25:Rad:S:units}{$-1.16$\mHz}
\setsymbol{LFI:slope:LFI26:Rad:S:units}{$-0.79$\mHz}
\setsymbol{LFI:slope:LFI27:Rad:S:units}{$-0.82$\mHz}
\setsymbol{LFI:slope:LFI28:Rad:S:units}{$-0.91$\mHz}

\setsymbol{LFI:slope:LFI18:Rad:M}{$-1.04$}
\setsymbol{LFI:slope:LFI19:Rad:M}{$-1.09$}
\setsymbol{LFI:slope:LFI20:Rad:M}{$-0.69$}
\setsymbol{LFI:slope:LFI21:Rad:M}{$-1.56$}
\setsymbol{LFI:slope:LFI22:Rad:M}{$-1.01$}
\setsymbol{LFI:slope:LFI23:Rad:M}{$-0.96$}
\setsymbol{LFI:slope:LFI24:Rad:M}{$-0.83$}
\setsymbol{LFI:slope:LFI25:Rad:M}{$-0.91$}
\setsymbol{LFI:slope:LFI26:Rad:M}{$-0.95$}
\setsymbol{LFI:slope:LFI27:Rad:M}{$-0.87$}
\setsymbol{LFI:slope:LFI28:Rad:M}{$-0.88$}
\setsymbol{LFI:slope:LFI18:Rad:S}{$-1.15$}
\setsymbol{LFI:slope:LFI19:Rad:S}{$-1.00$}
\setsymbol{LFI:slope:LFI20:Rad:S}{$-0.95$}
\setsymbol{LFI:slope:LFI21:Rad:S}{$-0.92$}
\setsymbol{LFI:slope:LFI22:Rad:S}{$-1.01$}
\setsymbol{LFI:slope:LFI23:Rad:S}{$-0.95$}
\setsymbol{LFI:slope:LFI24:Rad:S}{$-0.73$}
\setsymbol{LFI:slope:LFI25:Rad:S}{$-1.16$}
\setsymbol{LFI:slope:LFI26:Rad:S}{$-0.79$}
\setsymbol{LFI:slope:LFI27:Rad:S}{$-0.82$}
\setsymbol{LFI:slope:LFI28:Rad:S}{$-0.91$}

\setsymbol{LFI:FWHM:70GHz:units}{13\parcm01}
\setsymbol{LFI:FWHM:44GHz:units}{27\parcm92}
\setsymbol{LFI:FWHM:30GHz:units}{32\parcm65}

\setsymbol{LFI:FWHM:70GHz}{13.01}
\setsymbol{LFI:FWHM:44GHz}{27.92}
\setsymbol{LFI:FWHM:30GHz}{32.65}

\setsymbol{LFI:FWHM:LFI18:units}{13\parcm39}
\setsymbol{LFI:FWHM:LFI19:units}{13\parcm01}
\setsymbol{LFI:FWHM:LFI20:units}{12\parcm75}
\setsymbol{LFI:FWHM:LFI21:units}{12\parcm74}
\setsymbol{LFI:FWHM:LFI22:units}{12\parcm87}
\setsymbol{LFI:FWHM:LFI23:units}{13\parcm27}
\setsymbol{LFI:FWHM:LFI24:units}{22\parcm98}
\setsymbol{LFI:FWHM:LFI25:units}{30\parcm46}
\setsymbol{LFI:FWHM:LFI26:units}{30\parcm31}
\setsymbol{LFI:FWHM:LFI27:units}{32\parcm65}
\setsymbol{LFI:FWHM:LFI28:units}{32\parcm66}

\setsymbol{LFI:FWHM:LFI18}{13.39}
\setsymbol{LFI:FWHM:LFI19}{13.01}
\setsymbol{LFI:FWHM:LFI20}{12.75}
\setsymbol{LFI:FWHM:LFI21}{12.74}
\setsymbol{LFI:FWHM:LFI22}{12.87}
\setsymbol{LFI:FWHM:LFI23}{13.27}
\setsymbol{LFI:FWHM:LFI24}{22.98}
\setsymbol{LFI:FWHM:LFI25}{30.46}
\setsymbol{LFI:FWHM:LFI26}{30.31}
\setsymbol{LFI:FWHM:LFI27}{32.65}
\setsymbol{LFI:FWHM:LFI28}{32.66}

\setsymbol{LFI:FWHM:uncertainty:LFI18:units}{0.170\arcm}
\setsymbol{LFI:FWHM:uncertainty:LFI19:units}{0.174\arcm}
\setsymbol{LFI:FWHM:uncertainty:LFI20:units}{0.170\arcm}
\setsymbol{LFI:FWHM:uncertainty:LFI21:units}{0.156\arcm}
\setsymbol{LFI:FWHM:uncertainty:LFI22:units}{0.164\arcm}
\setsymbol{LFI:FWHM:uncertainty:LFI23:units}{0.171\arcm}
\setsymbol{LFI:FWHM:uncertainty:LFI24:units}{0.652\arcm}
\setsymbol{LFI:FWHM:uncertainty:LFI25:units}{1.075\arcm}
\setsymbol{LFI:FWHM:uncertainty:LFI26:units}{1.131\arcm}
\setsymbol{LFI:FWHM:uncertainty:LFI27:units}{1.266\arcm}
\setsymbol{LFI:FWHM:uncertainty:LFI28:units}{1.287\arcm}

\setsymbol{LFI:FWHM:uncertainty:LFI18}{0.170}
\setsymbol{LFI:FWHM:uncertainty:LFI19}{0.174}
\setsymbol{LFI:FWHM:uncertainty:LFI20}{0.170}
\setsymbol{LFI:FWHM:uncertainty:LFI21}{0.156}
\setsymbol{LFI:FWHM:uncertainty:LFI22}{0.164}
\setsymbol{LFI:FWHM:uncertainty:LFI23}{0.171}
\setsymbol{LFI:FWHM:uncertainty:LFI24}{0.652}
\setsymbol{LFI:FWHM:uncertainty:LFI25}{1.075}
\setsymbol{LFI:FWHM:uncertainty:LFI26}{1.131}
\setsymbol{LFI:FWHM:uncertainty:LFI27}{1.266}
\setsymbol{LFI:FWHM:uncertainty:LFI28}{1.287}

\setsymbol{HFI:center:frequency:100GHz:units}{100\,GHz}
\setsymbol{HFI:center:frequency:143GHz:units}{143\,GHz}
\setsymbol{HFI:center:frequency:217GHz:units}{217\,GHz}
\setsymbol{HFI:center:frequency:353GHz:units}{353\,GHz}
\setsymbol{HFI:center:frequency:545GHz:units}{545\,GHz}
\setsymbol{HFI:center:frequency:857GHz:units}{857\,GHz}

\setsymbol{HFI:center:frequency:100GHz}{100}
\setsymbol{HFI:center:frequency:143GHz}{143}
\setsymbol{HFI:center:frequency:217GHz}{217}
\setsymbol{HFI:center:frequency:353GHz}{353}
\setsymbol{HFI:center:frequency:545GHz}{545}
\setsymbol{HFI:center:frequency:857GHz}{857}

\setsymbol{HFI:Ndetectors:100GHz}{8}
\setsymbol{HFI:Ndetectors:143GHz}{11}
\setsymbol{HFI:Ndetectors:217GHz}{12}
\setsymbol{HFI:Ndetectors:353GHz}{12}
\setsymbol{HFI:Ndetectors:545GHz}{3}
\setsymbol{HFI:Ndetectors:857GHz}{4}

\setsymbol{HFI:FWHM:Maps:100GHz:units}{9\parcm88}
\setsymbol{HFI:FWHM:Maps:143GHz:units}{7\parcm18}
\setsymbol{HFI:FWHM:Maps:217GHz:units}{4\parcm87}
\setsymbol{HFI:FWHM:Maps:353GHz:units}{4\parcm65}
\setsymbol{HFI:FWHM:Maps:545GHz:units}{4\parcm72}
\setsymbol{HFI:FWHM:Maps:857GHz:units}{4\parcm39}
\setsymbol{HFI:FWHM:Maps:100GHz}{9.88}
\setsymbol{HFI:FWHM:Maps:143GHz}{7.18}
\setsymbol{HFI:FWHM:Maps:217GHz}{4.87}
\setsymbol{HFI:FWHM:Maps:353GHz}{4.65}
\setsymbol{HFI:FWHM:Maps:545GHz}{4.72}
\setsymbol{HFI:FWHM:Maps:857GHz}{4.39}

\setsymbol{HFI:beam:ellipticity:Maps:100GHz}{1.15}
\setsymbol{HFI:beam:ellipticity:Maps:143GHz}{1.01}
\setsymbol{HFI:beam:ellipticity:Maps:217GHz}{1.06}
\setsymbol{HFI:beam:ellipticity:Maps:353GHz}{1.05}
\setsymbol{HFI:beam:ellipticity:Maps:545GHz}{1.14}
\setsymbol{HFI:beam:ellipticity:Maps:857GHz}{1.19}

\setsymbol{HFI:FWHM:Mars:100GHz:units}{9\parcm37}
\setsymbol{HFI:FWHM:Mars:143GHz:units}{7\parcm04}
\setsymbol{HFI:FWHM:Mars:217GHz:units}{4\parcm68}
\setsymbol{HFI:FWHM:Mars:353GHz:units}{4\parcm43}
\setsymbol{HFI:FWHM:Mars:545GHz:units}{3\parcm80}
\setsymbol{HFI:FWHM:Mars:857GHz:units}{3\parcm67}

\setsymbol{HFI:FWHM:Mars:100GHz}{9.37}
\setsymbol{HFI:FWHM:Mars:143GHz}{7.04}
\setsymbol{HFI:FWHM:Mars:217GHz}{4.68}
\setsymbol{HFI:FWHM:Mars:353GHz}{4.43}
\setsymbol{HFI:FWHM:Mars:545GHz}{3.80}
\setsymbol{HFI:FWHM:Mars:857GHz}{3.67}

\setsymbol{HFI:beam:ellipticity:Mars:100GHz}{1.18}
\setsymbol{HFI:beam:ellipticity:Mars:143GHz}{1.03}
\setsymbol{HFI:beam:ellipticity:Mars:217GHz}{1.14}
\setsymbol{HFI:beam:ellipticity:Mars:353GHz}{1.09}
\setsymbol{HFI:beam:ellipticity:Mars:545GHz}{1.25}
\setsymbol{HFI:beam:ellipticity:Mars:857GHz}{1.03}

\setsymbol{HFI:CMB:relative:calibration:100GHz}{$\lsim 1\%$}
\setsymbol{HFI:CMB:relative:calibration:143GHz}{$\lsim 1\%$}
\setsymbol{HFI:CMB:relative:calibration:217GHz}{$\lsim 1\%$}
\setsymbol{HFI:CMB:relative:calibration:353GHz}{$\lsim 1\%$}
\setsymbol{HFI:CMB:relative:calibration:545GHz}{}
\setsymbol{HFI:CMB:relative:calibration:857GHz}{}

\setsymbol{HFI:CMB:absolute:calibration:100GHz}{$\lsim 2\%$}
\setsymbol{HFI:CMB:absolute:calibration:143GHz}{$\lsim 2\%$}
\setsymbol{HFI:CMB:absolute:calibration:217GHz}{$\lsim 2\%$}
\setsymbol{HFI:CMB:absolute:calibration:353GHz}{$\lsim 2\%$}
\setsymbol{HFI:CMB:absolute:calibration:545GHz}{}
\setsymbol{HFI:CMB:absolute:calibration:857GHz}{}

\setsymbol{HFI:FIRAS:gain:calibration:accuracy:statistical:100GHz}{}
\setsymbol{HFI:FIRAS:gain:calibration:accuracy:statistical:143GHz}{}
\setsymbol{HFI:FIRAS:gain:calibration:accuracy:statistical:217GHz}{}
\setsymbol{HFI:FIRAS:gain:calibration:accuracy:statistical:353GHz}{2.5\%}
\setsymbol{HFI:FIRAS:gain:calibration:accuracy:statistical:545GHz}{1\%}
\setsymbol{HFI:FIRAS:gain:calibration:accuracy:statistical:857GHz}{0.5\%}

\setsymbol{HFI:FIRAS:gain:calibration:accuracy:systematic:100GHz}{}
\setsymbol{HFI:FIRAS:gain:calibration:accuracy:systematic:143GHz}{}
\setsymbol{HFI:FIRAS:gain:calibration:accuracy:systematic:217GHz}{}
\setsymbol{HFI:FIRAS:gain:calibration:accuracy:systematic:353GHz}{}
\setsymbol{HFI:FIRAS:gain:calibration:accuracy:systematic:545GHz}{7\%}
\setsymbol{HFI:FIRAS:gain:calibration:accuracy:systematic:857GHz}{7\%}

\setsymbol{HFI:FIRAS:zero:point:accuracy:100GHz:units}{0.8\MJysr}
\setsymbol{HFI:FIRAS:zero:point:accuracy:143GHz:units}{}
\setsymbol{HFI:FIRAS:zero:point:accuracy:217GHz:units}{}
\setsymbol{HFI:FIRAS:zero:point:accuracy:353GHz:units}{1.4\MJysr}
\setsymbol{HFI:FIRAS:zero:point:accuracy:545GHz:units}{2.2\MJysr}
\setsymbol{HFI:FIRAS:zero:point:accuracy:857GHz:units}{1.7\MJysr}

\setsymbol{HFI:FIRAS:zero:point:accuracy:100GHz}{0.8}
\setsymbol{HFI:FIRAS:zero:point:accuracy:143GHz}{}
\setsymbol{HFI:FIRAS:zero:point:accuracy:217GHz}{}
\setsymbol{HFI:FIRAS:zero:point:accuracy:353GHz}{1.4}
\setsymbol{HFI:FIRAS:zero:point:accuracy:545GHz}{2.2}
\setsymbol{HFI:FIRAS:zero:point:accuracy:857GHz}{1.7}

\setsymbol{HFI:unit:conversion:100GHz:units}{0.2415\MJysrmK}
\setsymbol{HFI:unit:conversion:143GHz:units}{0.3694\MJysrmK}
\setsymbol{HFI:unit:conversion:217GHz:units}{0.4811\MJysrmK}
\setsymbol{HFI:unit:conversion:353GHz:units}{0.2883\MJysrmK}
\setsymbol{HFI:unit:conversion:545GHz:units}{0.05826\MJysrmK}
\setsymbol{HFI:unit:conversion:857GHz:units}{0.002238\MJysrmK}

\setsymbol{HFI:unit:conversion:100GHz}{0.2415}
\setsymbol{HFI:unit:conversion:143GHz}{0.3694}
\setsymbol{HFI:unit:conversion:217GHz}{0.4811}
\setsymbol{HFI:unit:conversion:353GHz}{0.2883}
\setsymbol{HFI:unit:conversion:545GHz}{0.05826}
\setsymbol{HFI:unit:conversion:857GHz}{0.002238}

\setsymbol{HFI:colour:correction:alpha=-2:V1.01:100GHz}{0.9893}
\setsymbol{HFI:colour:correction:alpha=-2:V1.01:143GHz}{0.9759}
\setsymbol{HFI:colour:correction:alpha=-2:V1.01:217GHz}{1.0007}
\setsymbol{HFI:colour:correction:alpha=-2:V1.01:353GHz}{1.0028}
\setsymbol{HFI:colour:correction:alpha=-2:V1.01:545GHz}{1.0019}
\setsymbol{HFI:colour:correction:alpha=-2:V1.01:857GHz}{0.9889}

\setsymbol{HFI:colour:correction:alpha=0:V1.01:100GHz}{1.0008}
\setsymbol{HFI:colour:correction:alpha=0:V1.01:143GHz}{1.0148}
\setsymbol{HFI:colour:correction:alpha=0:V1.01:217GHz}{0.9909}
\setsymbol{HFI:colour:correction:alpha=0:V1.01:353GHz}{0.9888}
\setsymbol{HFI:colour:correction:alpha=0:V1.01:545GHz}{0.9878}
\setsymbol{HFI:colour:correction:alpha=0:V1.01:857GHz}{1.0014}

\usepackage{graphicx}
\usepackage[breaklinks, colorlinks, citecolor=blue]{hyperref}
\usepackage{subfigure}
\usepackage{epsfig}
\usepackage{multirow}
\usepackage{color}
\usepackage{txfonts}

\usepackage{url}
\usepackage{breakurl}

\newcommand\apjl{ApJ}%
\newcommand\apjs{ApJS}%
\newcommand\aap{A\&A}%
\def\Oxford{1}
\def\ESOGarching{2}
\def\IAS{3}
\def\MPIfR{4}
\def\UChicago{5}
\def\UDP{6}
\def\CfA{7}
\def\Cambridge{8}
\def\KICPChicago{9}
\def\EFIChicago{10}
\def\PhysicsUChicago{11}
\def\Miss{12}
\def\AAUChicago{13}
\def\ANL{14}
\def\Dal{15}
\def\McGill{16}
\def\JPL{17}
\def\Caltech{18}
\def\Colorado{19}
\def\Toulouse{20}
\def\Berkeley{21}
\def\UFlorida{22}
\def\IFCA{23}
\def\SISSA{24}
\def\UCL{25}
\def\Stanford{26}
\def\LBNL{27}
\def\UCLA{28}
\def\Arizona{29}
\def\Michigan{30}
\def\IRAP{31}
\def\IPAC{32}
\def\IAP{33}
\def\STScI{34}
\def\CaseWestern{35}
\def\SAIC{36}
\def\UBC{37}
\def\TorontoDunlap{38}
\def\TorontoAstronomy{39}
\def\Illinois{40}
\def\Melbourne{41}
\def\Marseille{42}

\title[\vspace{-0.5cm}Star formation around $z\sim1$ SPT lensing halos]{\vspace{-0.5cm}Probing star formation in the dense environments of \textit{z}$\sim$1 lensing halos aligned with dusty star-forming galaxies detected with the South Pole Telescope}
\author[N. Welikala et al.]
{\parbox{\textwidth}{\vspace{-0.5cm}N.~Welikala$^{\Oxford}$\thanks{E-mail: niraj.welikala@astro.ox.ac.uk}, 
M.~B\'ethermin$^{\ESOGarching}$,
D.~Guery$^{\IAS}$,
M.~Strandet$^{\MPIfR}$\thanks{Member of the International Max Planck Research School (IMPRS) for Astronomy and Astrophysics at the Universities of Bonn and Cologne},
K.~A.~Aird$^{\UChicago}$,
M.~Aravena$^{\UDP}$,
M.~L.~N.~Ashby$^{\CfA}$,
M.~Bothwell$^{\Cambridge}$,
A.~Beelen$^{\IAS}$, 
L.~E.~Bleem$^{\KICPChicago,\PhysicsUChicago}$, 
C.~de~Breuck$^{\ESOGarching}$,
M.~Brodwin$^{\Miss}$,
J.~E.~Carlstrom$^{\KICPChicago,\EFIChicago,\PhysicsUChicago,\AAUChicago,\ANL}$, 
S.~C.~Chapman$^{\Dal}$,
T.~M.~Crawford$^{\KICPChicago,\AAUChicago}$, 
H.~Dole$^{\IAS}$,
O.~Dor\'e$^{\JPL,\Caltech}$,
W.~Everett$^{\Colorado}$,
I.~Flores-Cacho$^{\Toulouse,\IRAP}$, 
A.~H.~Gonzalez$^{\UFlorida}$, 
J.~Gonz\'alez-Nuevo$^{\IFCA,\SISSA}$, 
T.~R.~Greve$^{\UCL}$,
B.~Gullberg$^{\ESOGarching}$,
Y.~D.~Hezaveh$^{\Stanford}$,
G.~P.~Holder$^{\McGill}$, 
W.~L.~Holzapfel$^{\Berkeley}$, 
R.~Keisler$^{\Stanford}$, 
G.~Lagache$^{\Marseille,\IAS}$, 
J.~Ma$^{\UFlorida}$,
M.~Malkan$^{\UCLA}$,	
D.~P.~Marrone$^{\Arizona}$,   
L.~M.~Mocanu$^{\KICPChicago,\AAUChicago}$,
L.~Montier$^{\Toulouse,\IRAP}$,
E.~J.~Murphy$^{\IPAC}$,
N.~P.~H.~Nesvadba$^{\IAS}$,
A.~Omont$^{\IAP}$,
E.~Pointecouteau$^{\Toulouse,\IRAP}$, 
J.~L.~Puget$^{\IAS}$, 
C.~L.~Reichardt$^{\Berkeley,\Melbourne}$, 
K.~M.~Rotermund$^{\Dal}$,
D.~Scott$^{\UBC}$,
P.~Serra$^{\IAS}$,
J.~S.~Spilker$^{\Arizona}$,
B.~Stalder$^{\CfA}$, 
A.~A.~Stark$^{\CfA}$, 
K.~Story$^{\KICPChicago,\PhysicsUChicago}$, 
K.~Vanderlinde$^{\TorontoDunlap,\TorontoAstronomy}$, 
J.~D.~Vieira$^{\Illinois,\Caltech}$,
A.~Wei\ss$^{\MPIfR}$
}\vspace{1cm}\\
\parbox{\textwidth}{
$^{\Oxford}$ Department of Physics, University of Oxford, Denys Wilkinson Building, Keble Road, OX1 3RH, UK\\
$^{\ESOGarching}$ European Southern Observatory, Karl Schwarzschild Stra\ss e 2, 85748 Garching, Germany\\
$^{\IAS}$ Institut d'Astrophysique Spatiale, CNRS (UMR8617) Universit\'{e} Paris-Sud 11, B\^{a}timent 121, Orsay, France\\
$^{\MPIfR}$ Max-Planck-Institut f\"{u}r Radioastronomie, Auf dem H\"{u}gel 69 D-53121 Bonn, Germany\\
$^{\UChicago}$ University of Chicago, 5640 South Ellis Avenue, Chicago, IL 60637, USA\\
$^{\UDP}$ N\'{u}cleo de Astronom\'{\i}a, Facultad de Ingenier\'{\i}a, Universidad Diego Portales, Av. Ej\'{e}rcito 441, Santiago, Chile\\
$^{\CfA}$ Harvard-Smithsonian Center for Astrophysics, 60 Garden Street, Cambridge, MA 02138, USA\\ 
$^{\Cambridge}$ Cavendish Laboratory, University of Cambridge, JJ
Thompson Ave, Cambridge CB3 0HA, UK\\
$^{\KICPChicago}$ Kavli Institute for Cosmological Physics, University
of Chicago, 5640 South Ellis Avenue, Chicago, IL 60637, USA\\
$^{\EFIChicago}$ Enrico Fermi Institute, University of Chicago, 5640
South Ellis Avenue, Chicago, IL 60637, USA\\
$^{\PhysicsUChicago}$ Department of Physics, University of Chicago,
5640 South Ellis Avenue, Chicago, IL 60637, USA\\
$^{\Miss}$ Department of Physics and Astronomy, University of Missouri, 5110 Rockhill Road, Kansas City, MO 64110, USA\\
$^{\AAUChicago}$ Department of Astronomy and Astrophysics, University of Chicago, 5640 South Ellis Avenue, Chicago, IL 60637, USA\\
$^{\ANL}$ Argonne National Laboratory, 9700 S. Cass Avenue, Argonne,
IL, USA 60439, USA\\
$^{\Dal}$ Dalhousie University, Halifax, Nova Scotia, Canada\\
$^{\McGill}$ Department of Physics, McGill University, 3600 Rue University, Montreal, Quebec H3A 2T8, Canada\\
$^{\JPL}$ Jet Propulsion Laboratory, California Institute of Technology, 4800 Oak Grove Drive, Pasadena, California, U.S.A.\\
$^{\Caltech}$ California Institute of Technology, 1200 E. California Blvd., Pasadena, CA 91125, USA\\
$^{\Colorado}$ Department of Astrophysical and Planetary Sciences and Department of Physics, University of Colorado, Boulder, CO, 80309, USA\\
$^{\Toulouse}$ Universit\'{e} de Toulouse, UPS-OMP, IRAP, F-31028 Toulouse cedex 4, France\\
$^{\Berkeley}$ Department of Physics, University of California, Berkeley, CA 94720, USA\\
$^{\UFlorida}$ Department of Astronomy, University of Florida, Gainesville, FL 32611, USA\\
$^{\IFCA}$ Instituto de F\'{\i}sica de Cantabria (CSIC-Universidad de Cantabria), Avda. de los Castros s/n, Santander, Spain\\
$^{\SISSA}$ SISSA, Astrophysics Sector, via Bonomea 265, 34136, Trieste, Italy\\
$^{\UCL}$ Department of Physics and Astronomy, University College London, Gower Street, London WC1E 6BT, UK\\
$^{\Stanford}$ Kavli Institute for Particle Astrophysics and Cosmology, Stanford University, Stanford, CA 94305, USA\\
$^{\LBNL}$ Physics Division, Lawrence Berkeley National Laboratory, Berkeley, CA 94720, USA\\
$^{\UCLA}$ Department of Physics and Astronomy, University of California, Los Angeles, CA 90095-1547, USA\\
$^{\Arizona}$ Steward Observatory, University of Arizona, 933 North Cherry Avenue, Tucson, AZ 85721, USA\\
$^{\Michigan}$ Department of Physics, University of Michigan, 450 Church Street, Ann Arbor, MI, 48109, USA\\
$^{\IRAP}$ CNRS, IRAP, 9 Av. colonel Roche, BP 44346, F-31028 Toulouse cedex 4, France\\
$^{\IPAC}$ Infrared Processing and Analysis Center, California Institute of Technology, MC 314-6, Pasadena, CA 91125, USA\\
$^{\IAP}$ Institut d'Astrophysique de Paris, CNRS (UMR7095), 98 bis Boulevard Arago, F-75014, Paris, France\\
$^{\STScI}$ Space Telescope Science Institute, 3700 San Martin Dr., Baltimore, MD 21218, USA\\
$^{\CaseWestern}$ Physics Department, Center for Education and Research in Cosmology  and Astrophysics,  Case Western Reserve University, Cleveland, OH 44106, USA\\
$^{\SAIC}$ Liberal Arts Department, School of the Art Institute of Chicago,  112 S Michigan Ave, Chicago, IL 60603, USA\\
$^{\UBC}$Department of Physics \& Astronomy, University of British Columbia, 6224 Agricultural Road, Vancouver, British Columbia, Canada\\
$^{\TorontoDunlap}$ Dunlap Institute for Astronomy \& Astrophysics, University of Toronto, 50 St George St, Toronto, Ontario, M5S 3H4, Canada\\
$^{\TorontoAstronomy}$ Department of Astronomy \& Astrophysics, University of Toronto, 50 St George St, Toronto, Ontario, M5S 3H4, Canada\\
$^{\Illinois}$ Department of Astronomy and Department of Physics,
University of Illinois, 1002 West Green Street, Urbana, IL 61801, USA\\
$^{\Melbourne}$ School of Physics, University of Melbourne, Parkville,
VIC 3010, Australia\\
$^{\Marseille}$ Aix Marseille Universit\'e, CNRS, LAM (Laboratoire d'Astrophysique de Marseille) UMR 7326, 13388, Marseille, France
}}

\begin{document}

\maketitle

\label{firstpage}
\begin{abstract}
We probe star formation in the environments of massive
($\sim10^{13}\Msolar$) dark matter halos at redshifts of $z$$\sim$$1$. This star formation is linked to a sub-millimetre clustering signal which we detect in maps of the \textit{Planck} High Frequency
  Instrument that are stacked at the positions of a sample of high-redshift ($z$$>$$2$) strongly-lensed 
  dusty star-forming galaxies (DSFGs) selected from the South Pole Telescope
  (SPT) 2500 deg$^2$ survey. The clustering signal has sub-millimetre
  colours which are consistent with the mean redshift of the foreground lensing halos
($z$$\sim$$1$). We report a mean excess of star formation rate (SFR) compared to the field, 
 of $(2700\pm700)\,\Msolar\,{\rm yr}^{-1}$ from
all galaxies contributing to this clustering signal within a radius of
$3\parcm5$ from the SPT DSFGs. 
The magnitude of the \Planck\ excess is in broad agreement with
predictions of a current model of the cosmic infrared background. The
model predicts that 80$\%$  of the excess emission measured by \Planck\ originates from
galaxies lying in the neighbouring halos of the lensing halo. Using \textit{Herschel} maps of the
same fields, we find a clear excess, relative to the field, of
individual sources which contribute to the \Planck\ excess. The mean excess SFR compared to the
field is measured to be ($370\pm40$)$\,\Msolar\,{\rm yr}^{-1}$ per
resolved, clustered source. Our findings suggest that the environments around these massive $z$$\sim$$1$
lensing halos host intense star formation out to about $2$\,Mpc. 
The flux enhancement due to clustering should also be considered when measuring flux densities of galaxies in \Planck\ data.
\end{abstract}



\section{Introduction}
\label{sec:intro}

Although it is known that the local environment of a galaxy
impacts its star formation, the magnitude of
the effect is unclear, particularly at high redshifts. 
Studies in the low redshift ($z\sim0.1$) 
Universe show that star formation in
galaxies is suppressed in highly dense environments such as in the
centres of clusters, consistent with the effects of physical mechanisms such as
ram-pressure stripping \citep[e.g.,][]{hogg2004,blanton2005}. 
However, the high-redshift picture is murkier. 
Some studies -- for example, \citet{elbaz2007}, \citet{cooper2008} and
\citet{popesso2011} -- have found that the 
star formation rate (SFR)-density relation is either reversed or weaker at $z\sim1$ than
at $z\sim0$. 
The picture that has emerged from these studies is one of galaxies that are still actively
forming stars at $z\sim1$ in high density environments such as the centres of
groups. These may precede the formation of red, passive ellipticals
that are observed in the centres of clusters at
$z\sim0$. However, not all studies agree. 
\citet{feruglio2010} found no reversal of the SFR-density relation in
the Cosmic Evolution Survey (COSMOS), 
and \citet{ziparo2014} who investigated the evolution of the
SFR-density relation up to $z\sim1.6$ in the Extended Chandra Deep Field-South Survey
(ECDFS) and the Great Observatories Origins Deep Survey (GOODS), also
found no reversal. 


In this paper, we target dense environments associated with massive 
($M\gse10^{13}\Msolar$) dark matter lensing halos at $z$$\sim$$1$ and
probe star formation in these dense environments. 
Our study falls into the context of a known correlation between the
Cosmic Infrared Background (CIB, the thermal radiation from UV-heated
dust in distant galaxies) and gravitational lensing \citep[see,
e.g.,][]{blake2006,wang2011,hildebrandt2013,holder2013,planck2013-p13}. 
To select the dense environments, we start with a sample of high-redshift
($z>2$) strongly-lensed dusty star-forming galaxies (DSFGs) discovered with the South Pole Telescope
\citep[SPT,][]{carlstrom2011}. These DSFGs have been strongly lensed
by foreground, massive early-type galaxies at $z$$\sim$$1$ which trace
high-density environments \citep{hezaveh2013,vieira2013}. 
Our approach is to stack the \Planck\ maps at the positions of the SPT DSFGs and search for an excess of far-infrared emission, relative to the field, in the environments of these foreground 
halos. 

The  stacked image contains the sum of a number of astrophysical
components: (1) the parent sample of SPT DSFGs, (2) the mean
background from the CIB \citep{lagache2005, dole2006}, (3)
high-redshift sources clustered around the DSFGs, and (4) foreground
sources associated with and clustered around the lensing halo. The first component should be unresolved relative to
the  point spread function (PSF) of the  \Planck\ map, and the second
component should be a flat DC component in the map. 
The latter two clustered components would manifest themselves as a radially dependent
excess relative to the \Planck\ PSF. We use higher-resolution
\textit{Herschel} maps to isolate the emission from the background
DSFGs and from the clustered signal. \Planck\ is well suited to
characterising this clustering signal because the beam size of
\Planck\ is well matched to the angular scale of the excess signal
\citep[e.g.,][]{fernandez-conde2008,fernandez-conde2010,berta2011,bethermin2012a,viero2013a},
and its wide frequency coverage enables an estimate of its mean redshift.
At $z$$\sim$$1$, the \Planck\ beam probes physical scales of around
2\,Mpc. In the context of the halo model
\citep{mo1996,sheth1999,benson2000,sheth2001}, on these scales, we are
probing both the `one-halo term' (which is due to 
distinct baryonic mass elements that lie within the same dark matter
halo and which describes the clustering of galaxies on scales
smaller than the virial radius of the halo), and the `two-halo term' 
(due to pairs of galaxies in separate halos and which gives rise to galaxy clustering on larger scales).


The paper is structured as follows. In Sect.~\ref{sec:data}, we
  describe the SPT DSFG sample and the ancillary data that we use for
  the analysis. We describe our methods in
  Sect.~\ref{sec:methods}. We show the results in
  Sect.~\ref{sec:results}, which is split into two parts. The first
  part (Sect.~\ref{subsec:results_planck}) presents the excess of flux density we observe in the \Planck\
  stacks we construct relative to the flux densities from higher-resolution data at the same frequencies. We measure the clustered component from the
\Planck\ stacks, quantify the clustering contamination, obtain an SED and mean
photometric redshift of the clustered component, derive a
corresponding far-infrared (FIR) luminosity and SFR, and show the
radial profiles of the various components of the \Planck\ stack. In
the second part (Sect.~\ref{subsec:results_herschel}), we use
\textit{Herschel}/SPIRE observations to search for the individual
sources that are responsible for the \Planck\ excess and to constrain
the nature of these sources. In Sect.~\ref{sec:modeling}, we interpret the \Planck\ excess 
using a model of the CIB that relates infrared galaxies to dark matter halos. We discuss the implications of our
results in Sect.~\ref{sec:discussion} 
and present our conclusions in Sect.~\ref{sec:conclusions}. Some 
supporting analyses and descriptions are presented in the
Appendix. We refer to frequency rather than wavelength units
throughout this paper. We use a $\Lambda$CDM
cosmology with $H_0=70\,\mathrm{km}\,\mathrm{s}^{-1}\mathrm{Mpc^{-1}}$, $\Omega_{\mathrm{M}} = 0.27$ and $\Omega_{\Lambda} = 0.73$.


\begin{table*}
\caption{SPT survey parameters and the DSFG sample used in this analysis} 
\centering 
\begin{tabular}{l c} 
\hline

Sky coverage in SPT main survey & $2500\,{\rm deg}^2$\\
Spatial resolution at 220 GHz & 1$\arcmin$\\
Sensitivity at 220 GHz & $3.4-4.5\,{\rm mJy}\,{\rm beam}^{-1}$ rms\\[1ex]

Main sample: number of DSFGs with $S_{220} > 20$\,mJy& 65\\  
Number of DSFGs observed with APEX/LABOCA & 65\\
Number of DSFGs detected in APEX/LABOCA and with measured LABOCA flux densities & 61\\
Number of DSFGs observed with \textit{Herschel} SPIRE & 65\\
Number of DSFGs detected in \textit{Herschel} SPIRE and with measured
SPIRE flux densities & 62\\
Number of DSFGs detected in \textit{Herschel} SPIRE and with ALMA 100\,GHz positions & 26\\
\hline
\end{tabular} 
\label{table:sample} 
\end{table*}

\section{Data}
\label{sec:data}

\subsection{South Pole Telescope selection}
\label{subsec:data_spt}

The South Pole Telescope \citep[SPT,][]{carlstrom2011} is a 10-metre diameter millimetre/submillimetre
(mm/sub-mm) telescope located at the geographic South Pole and is
designed for low-noise observations of diffuse, low-contrast sources
such as primary and secondary anisotropies in the cosmic microwave
background \citep[CMB, e.g.,][]{reichardt2012,story2013}. The first generation SPT-SZ camera was a
960-element, three-band (95, 150 and 220\,GHz) bolometric
receiver. The sensitivity and angular resolution of the SPT make it an
excellent instrument for detecting extragalactic sources of emission  \citep{vieira2010}. 

The observations, data reduction, flux calibration, and generation of
the extragalactic millimetre-wave point source catalogue are described in 
\citet{vieira2010} and \citet{mocanu2013}. Sources detected in the SPT maps were classified as dust-dominated or
synchrotron-dominated based on the ratio of their 150\,GHz and
220\,GHz flux densities. 
Approximating the spectral behaviour of sources between 150\,GHz and
220\,GHz as a power law, $S_{\nu}\propto
\nu^{\alpha}$, we estimated the spectral index $\alpha$ for every
source. A spectral index $\alpha\simeq3$ is typical for sources
dominated by dust emission while $\alpha\simeq-$1 is typical for the
synchrotron-dominated population \citep[see][for
details]{vieira2010}. The sample of DSFGs used here is selected from
the full 2500 deg$^2$ SPT source catalog using a cut on the raw 220~GHz flux
density ($S_{220} > 20$~mJy) and on spectral index ($\alpha >
1.66$). In addition, sources also found in the Infrared Astronomy
Satellite Faint-Source Catalogue \citep[IRAS-FSC,][]{moshir1992},
which are typically at $z \ll 1$ (median $\langle z\rangle=0.003$), were removed from the sample, leaving a population of bright, dust-dominated galaxies without counterparts in IRAS.

In this work, our parent sample comprises 65 DSFGs discovered by SPT over 2500 ${\rm deg}^2$ \citep{vieira2010}. 
The 220\,GHz source selection in this work exploits the
nearly redshift-independent selection function of DSFGs at this
frequency \citep[e.g.,][]{blain2002}. The mean redshift of the SPT
sample is $\langle z\rangle=3.5$, as determined by \citet{weiss2013}
through a CO redshift survey conducted with ALMA for a sample of 26 of
these DSFGs. ALMA has now confirmed that the majority of the SPT DSFGs
are strongly lensed \citep{hezaveh2013,vieira2013}. The lensing dark matter halos which are
  aligned with the SPT DSFGs are empirically observed to lie in the redshift range
  $z\sim0.1$--$2.0$, in agreement with the theoretical prediction of
  $\langle z_{\rm lens}\rangle=1.15$ (with a $FWHM=1.53$) from
  \citet{hezaveh2011}. Table~\ref{table:sample} summarizes the SPT sample selection, the SPT
sky coverage and depths, and the number of sources with ancillary
observations that were used in this analysis. These include
\textit{Herschel}/SPIRE, APEX/LABOCA and ALMA imaging, the latter used to
obtain accurate positions of the SPT sources in the analysis. 
The ancillary observations are described more fully below.

\subsection{\textit{Planck}}
\label{subsec:data_planck}

\Planck\footnote{\textit{Planck} is a project of the European Space
  Agency - ESA - with instruments provided by two scientific Consortia
  funded by ESA member states (in particular the lead countries:
  France and Italy) with contributions from NASA (USA), and telescope
  reflectors provided in a collaboration between ESA and a scientific
  Consortium led and funded by Denmark.} \citep{tauber2010a, planck2011-1.1, planck2013-p01} is the third space
mission to measure the anisotropy of the CMB. It observed the 
sky in nine frequency bands covering $28.5-857$ GHz with 
high sensitivity and angular resolution
from 32\parcm24 to 4\parcm33. The
High Frequency Instrument \citep[HFI][]{lamarre2010, planck2011-1.5, planck2013-p03} 
covered the 100, 143, 217, 353, 545, and 857 GHz bands with
bolometers cooled to 0.1\,K. In the present work we use the public
\Planck\ HFI maps, which can be obtained from the \Planck\ Legacy
Archive\footnote{\tiny{\url{http://www.sciops.esa.int/index.php?page=Planck_Legacy_Archive\&project=planck}}}. 
The HFI data come from the nominal mission 
acquired between 13 August 2009 and 27 November 2010. These are converted 
from units of thermodynamic temperature to intensity units
\citep[${\rm MJy}\,{\rm sr}^{-1}$,][]{planck2013-p03d}. From the full-sky
\Planck\ {\tt HEALpix} maps \citep{gorski2005} with a resolution parameter $N_{\rm side}=2048$,
we extract \Planck\ patches (in the tangential plane, using a gnomic
projection) corresponding to each SPT field. The pixel scale in
these \Planck\ patches is $1\arcmin$. 
We then extract $1\deg\times1\deg$ cutouts around each SPT source,
centred on the SPT-derived position of the source. 

\subsection{IRIS}
\label{subsec:data_iras}

We combine the \Planck-HFI data with 3000\,GHz IRIS
photometry \citep{miville-deschenes2005}. IRIS is a reduction of the
IRAS 3000\,GHz data \citep{neugebauer1984} that benefits from an
improved zodiacal light subtraction, and from a calibration and zero
level which are compatible with the Diffuse Infrared Background
Experiment (DIRBE), and from better de-striping. 
At 3000\,GHz, IRIS maps are a significant improvement compared to the
\citet{schlegel1998} maps. The angular resolution of the maps is
4.3$\arcmin$. From the IRIS maps, we extract $1\deg\times\,1\deg$ cutouts of the SPT sources as in Sect.~\ref{subsec:data_planck}.

\subsection{APEX continuum imaging}
\label{subsec:data_apex}

All the SPT sources from the $2500\,{\rm deg}^2$ survey
data were imaged at 345\,GHz with the Large APEX BOlometer CAmera
(LABOCA) at APEX\footnote{Based on observations from MPI projects
  085.F-0008 (2010), 087.F-0015 (2011), 089.F-0009, 091.F-0031 (2013), and ESO
  project 089.A-0906A (2012)}. LABOCA \citep{siringo2009} is a 295-element
bolometer array with a field-of-view of $11\parcm4$ in diameter and an angular resolution of $19\parcs7$
(FWHM). The central frequency of LABOCA is $345\,{\rm GHz}$
($870\,\mu{\rm m}$), with a passband FWHM of approximately 60\,$\rm
  GHz$. The map size is approximately 12$\arcmin$. Observations were carried out under good weather conditions
(median precipitable water vapour value of 0.9\,mm, with a range of
0.3\,mm to 1.5\,mm) . The data reduction was performed in the same
manner as in \citet{greve2012}. Sixty one of the 65 SPT sources in 
this study were detected in the LABOCA maps and had measured flux
densities.

\subsection{\textit{Herschel}}
\label{subsec:data_herschel}

We  use \textit{Herschel} Spectral and Photometric Imaging Receiver (SPIRE) observations of the SPT DSFGs in order to: (a) look for a statistical excess (relative to the field) of bright, individually detected sources that
  contribute to the \Planck\ excess signal; (b) confirm that these
  bright, detected sources are associated with the $z\sim1$ SPT lensing halos; and (c) 
  estimate the mean contribution of these clustered sources to the
  excess of star formation that is observed in the environments around
  the lensing halos. The SPIRE instrument, its in-orbit performance and its scientific
capabilities are described in \citet{griffin2010}, while its
calibration methods and accuracy are outlined in \citet{swinyard2010}. 
We use two sets of SPIRE maps for this work. 

\begin{itemize}
\item \textit{SPIRE $10\arcmin\times10\arcmin$ maps}:
  The SPIRE maps at 1200\,GHz ($250\micron$), 857\,GHz
  (350$\micron$), and 545\,GHz (500$\micron$) used in this work were
  made from data taken during observing programmes OT1\_jvieira\_4, OT2\_jvieira\_5,
  DDT\_mstrande\_1 and DDT\_tgreve\_2 for the lensed SPT DSFGs that were
  selected from the 2500 deg$^2$ SPT survey. These maps had coverage complete to a radius of 5$\arcmin$ from the nominal
SPT-derived position. More accurate positions of the SPT DSFGs were then
obtained for the analysis on the SPIRE maps (see Sec.~\ref{subsec:data_alma}). The maps
were produced via the standard reduction pipeline {\tt HIPE} v9.0,
the SPIRE Photometer Interactive Analysis package v1.7, and
the calibration product v8.1. The median rms in these maps is 9.7\,mJy at
1200\,GHz, 8.9\,mJy at 857\,GHz and 9.9\,mJy at 545\,GHz. This is
dominated by confusion noise (approximately 6 mJy in each band). All
65 SPT sources were imaged with SPIRE and 62 were detected and had measured flux densities.
\item \textit{SPIRE observations of the Lockman--SWIRE field}: We use
  archival SPIRE data from the Herschel Multi-tiered Extragalactic
  Survey \citep[HerMES,][]{oliver2012} of the Lockman- SWIRE field
  centred on RA=10:48:00.00, Dec=$+$58:08:00.0 and 18.2\,deg$^2$ in area
  \footnote{\tiny{\url{http://hedam.oamp.fr/HerMES/release.php}}}. This
  data does not overlap with the SPT coverage but is used as a
  reference field in the analysis. The 5$\sigma$
  confusion noise is 27.5 mJy at 857\,GHz \citep{nguyen2010} and the
  total 5$\sigma$ noise (including instrumental noise) at 857\,GHz is
  approximately 40 mJy. 
\end{itemize}

\subsection{ALMA}
\label{subsec:data_alma}

When performing the analysis on the \textit{Herschel}/SPIRE images,
we use the positions of the SPT DSFGs that were derived from ALMA
100\,GHz (3\,mm) continuum observations whenever they are available. Thus for 26
galaxies, we use the ALMA positions and for the remainder, we use the
positions given by LABOCA. The ALMA positions used here were reported in \citet{weiss2013}.


\begin{figure}{}
\centerline{\rotatebox{0}{
\includegraphics[width=.25\textwidth]{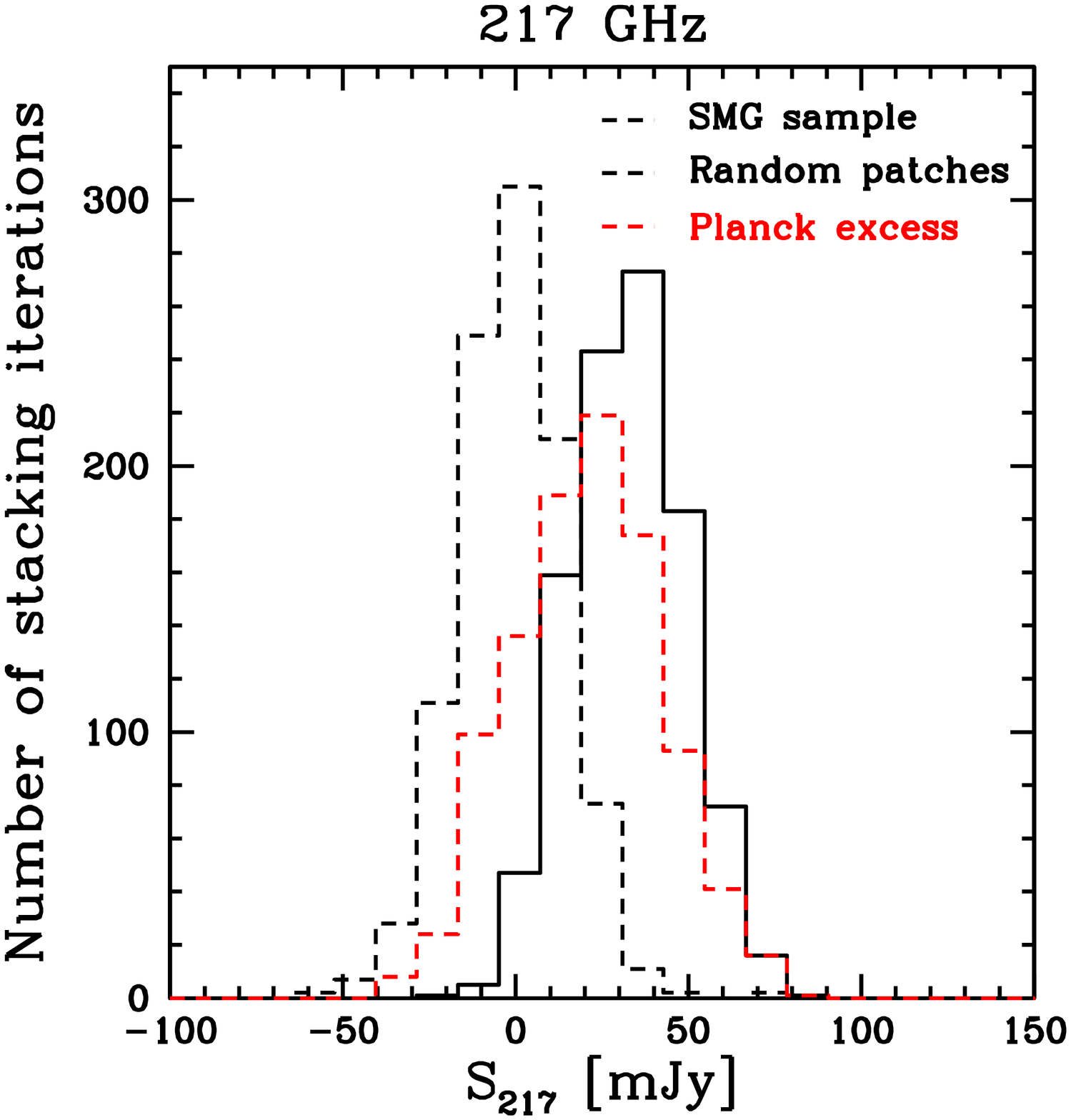}
\includegraphics[width=.25\textwidth]{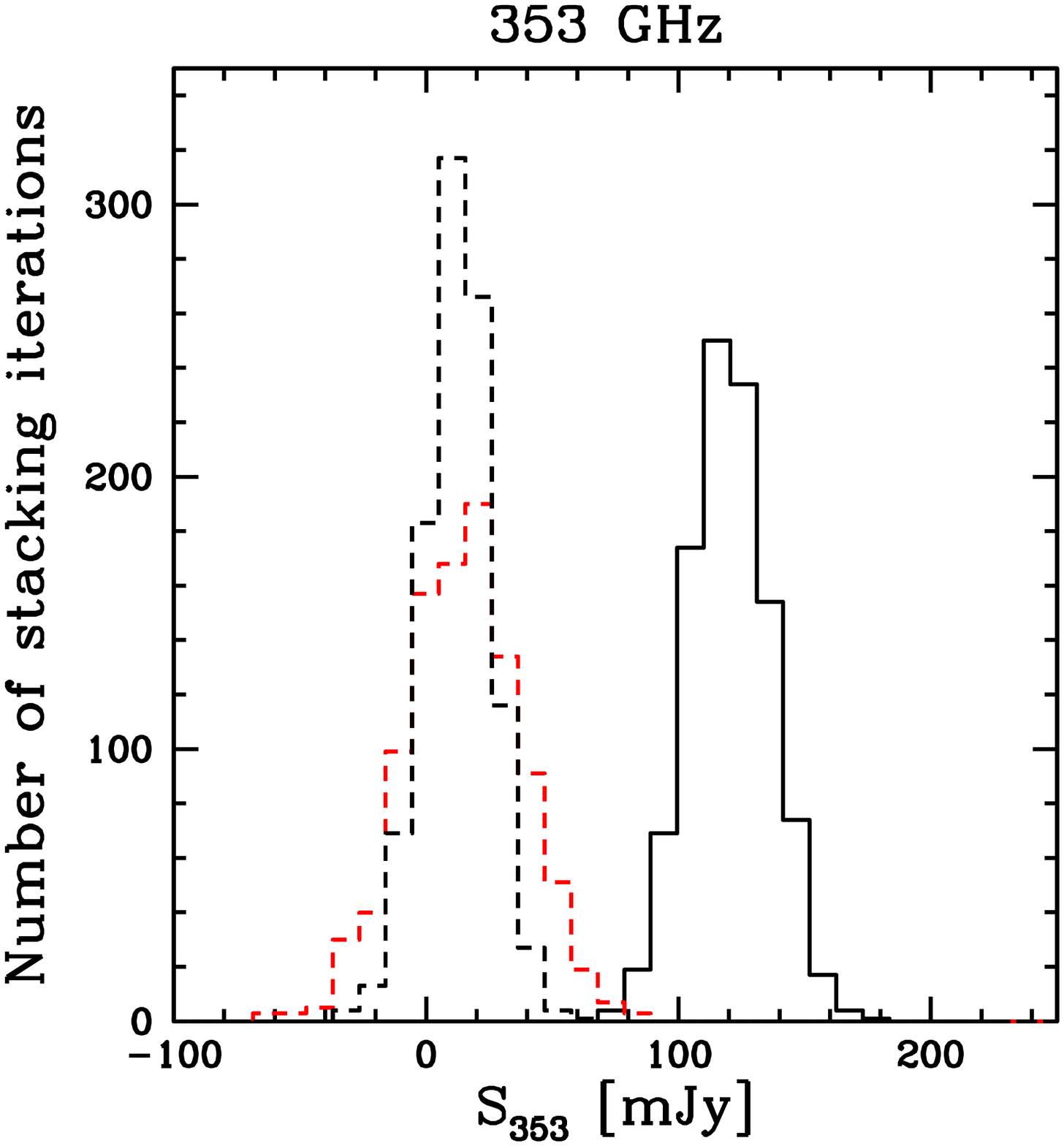}
}}   
\vspace{0.1cm}
\centerline{\rotatebox{0}{
\includegraphics[width=.25\textwidth]{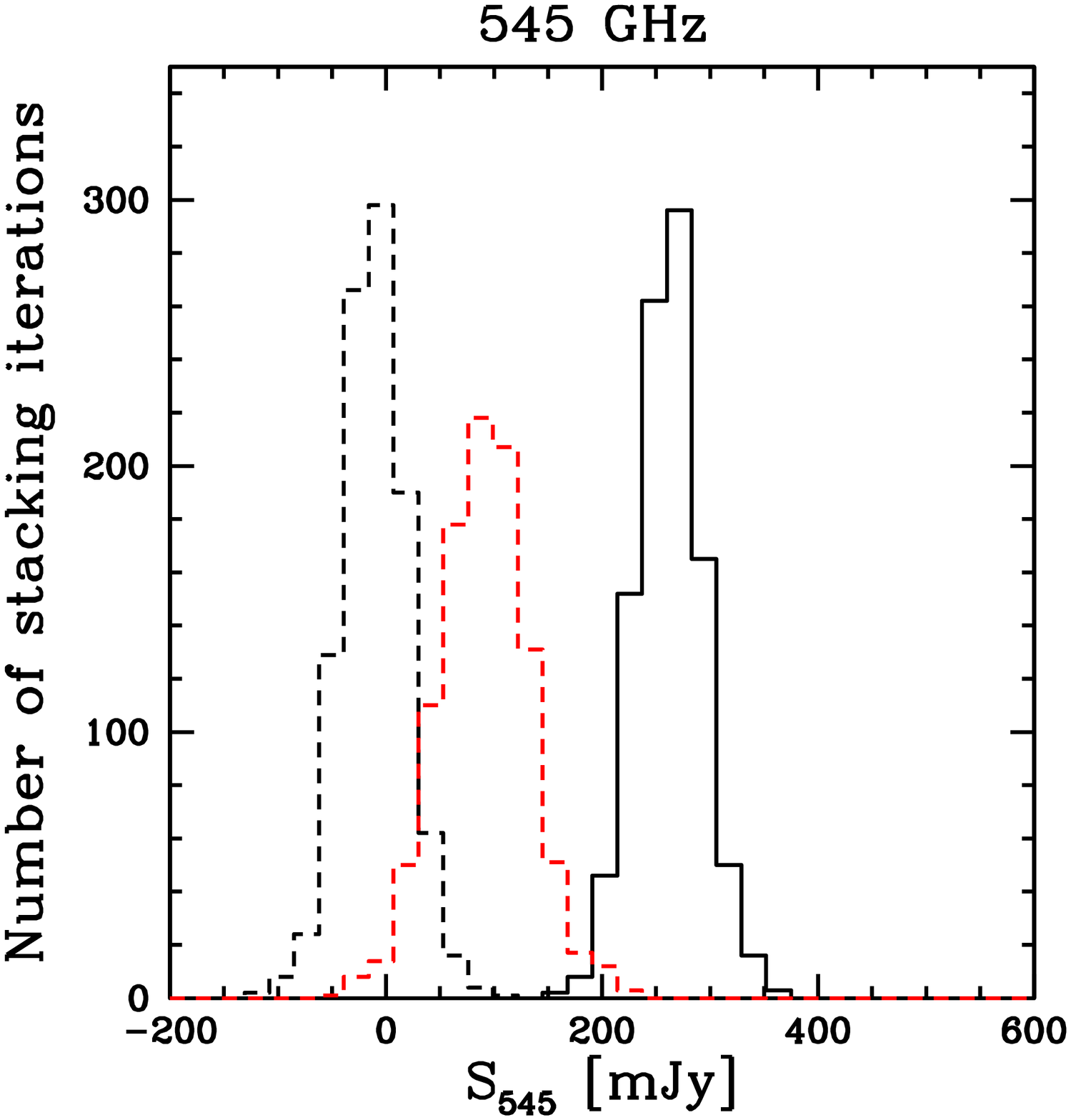}
\includegraphics[width=.25\textwidth]{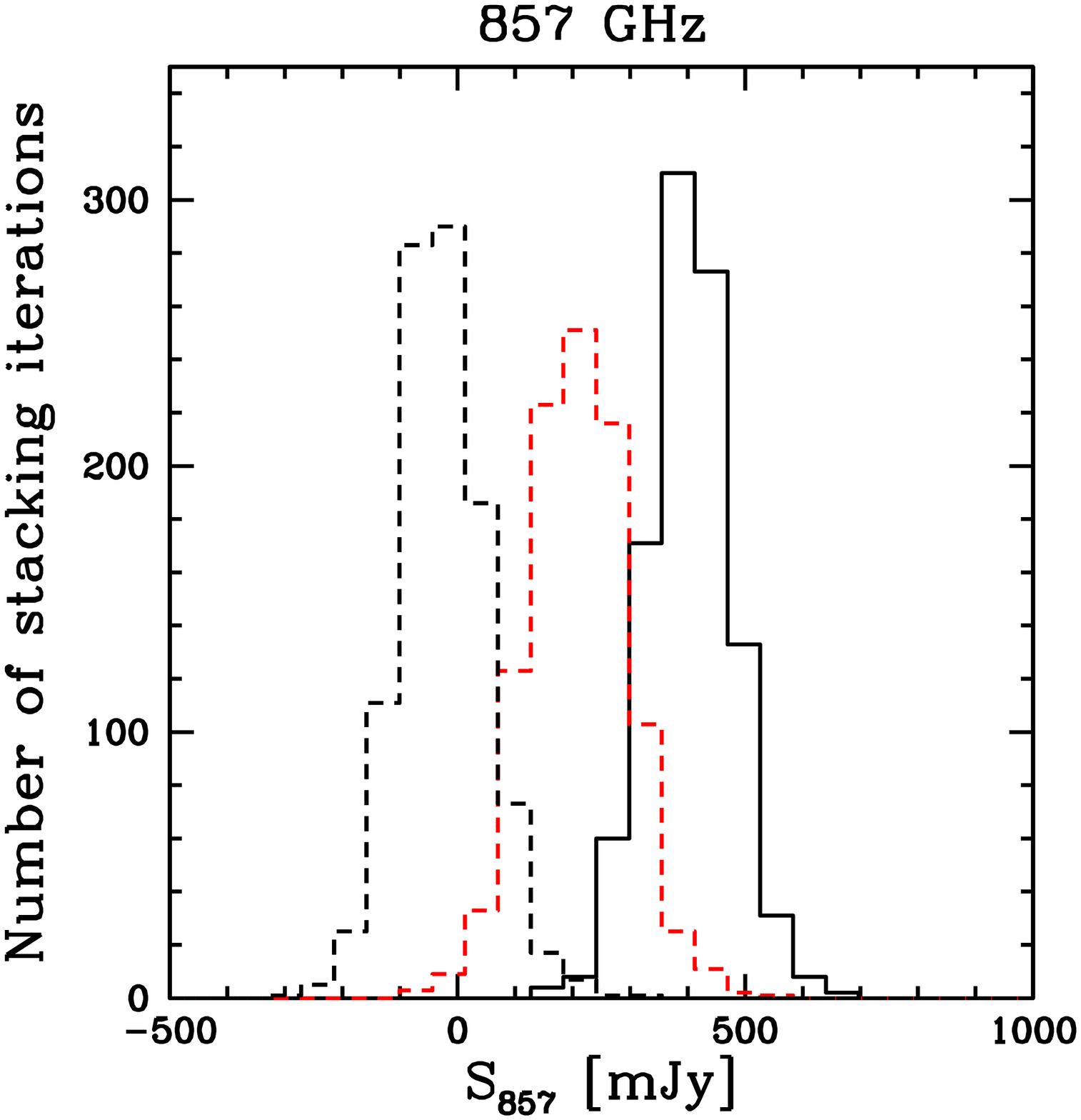}
}
}
\vspace{0.1cm}
\centerline{\rotatebox{0}{
\includegraphics[width=.25\textwidth]{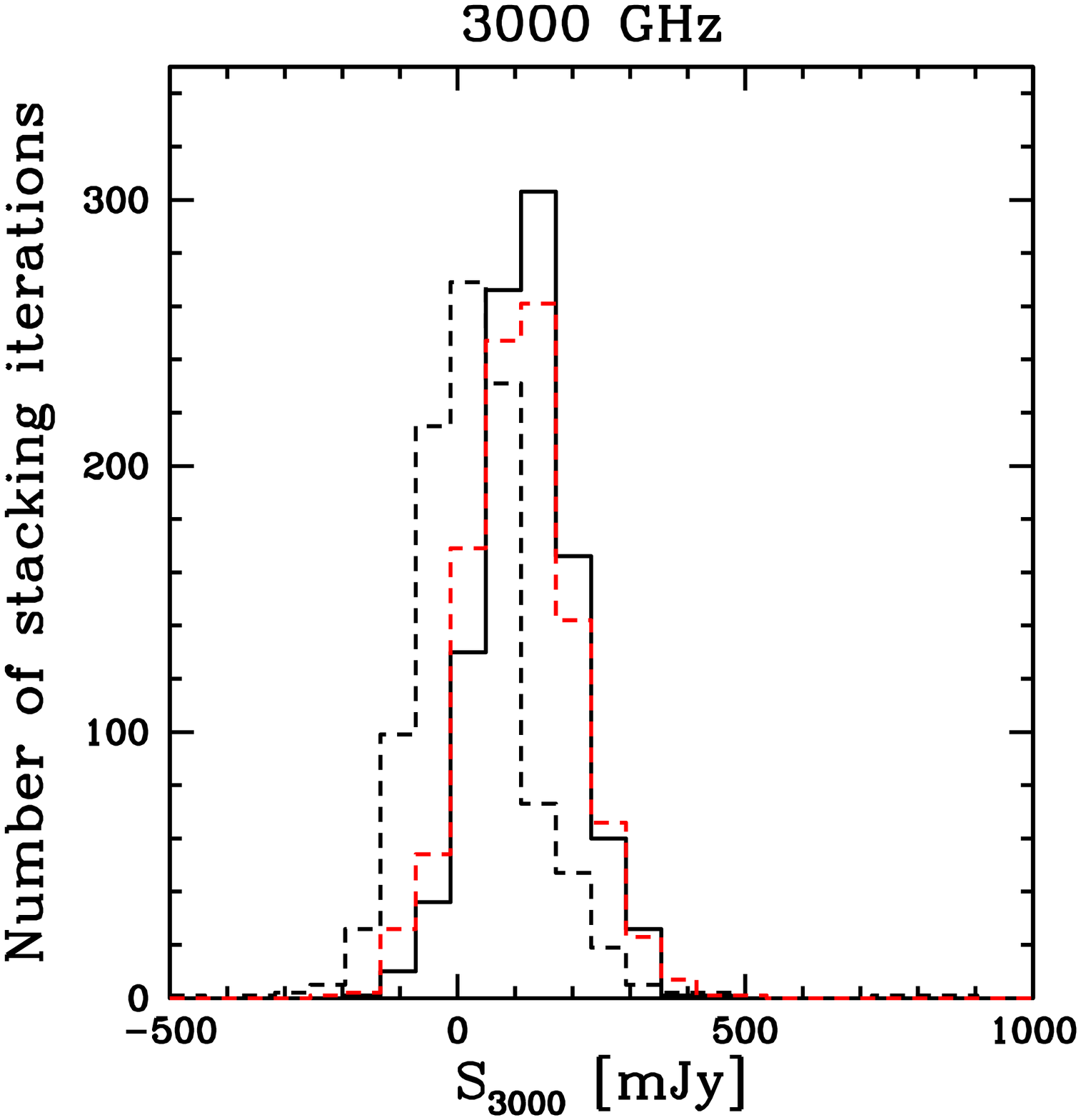}
}
}

\caption{Distribution of \Planck\ and IRIS flux densities from aperture
    photometry within a radius of $3\parcm5$ over: (1) 1000 bootstrap realisations of stacking 65
    $1\deg\times1\deg$ patches of the SPT DSFGs (black solid line); (2) 1000 iterations of stacking the same number (65) of $1\deg\times1\deg$ patches
    selected randomly from the \Planck\ maps covering the SPT
    fields at 217, 353, 545, 857\,GHz and from the IRIS maps at
    3000\,GHz (black dashed
    line); (3) 1000 bootstrap
      realisations of the \Planck\ and IRIS excess after removing the high
      redshift compact source (the SPT DSFGs) from the stacked map in
      each realisation using the formalism in
      Appendix~\ref{app:planck_stack_formalism} (red dashed line). At
    217 and 3000\,GHz, there is a much larger number of stacks on
    random locations (black dashed line) which have flux densities that are as high as the flux densities of the stacks on the
    SPT sources (black solid line), compared to the other frequencies. The 353, 545, and
    857\,GHz  channels are therefore cleaner.
\label{fig:bootstrap_bright}}
\end{figure}

\section{Methods}
\label{sec:methods}

In this section, we describe our methods for (1) stacking \Planck\ HFI maps at the
positions of the SPT DSFGs and performing photometry on the stacked
maps and (2) performing source detection and
photometry on the \textit{Herschel}/SPIRE maps.

\subsection{Stacking \textit{Planck} maps at the locations of SPT DSFGs}
\label{subsec:methods:stacking}

The noise at the high frequencies in \Planck\ is dominated by confusion noise from the 
CIB \citep{planck2011-6.6}. Stacking the 
\Planck\ maps at the locations of SPT sources enables us to go beyond the confusion
noise level that impacts individual detections of DSFGs
\citep[e.g.,][]{dole2006}. We also perform simulations to correct for
a positional offset of the SPT DSFGs due to the effect of pixelization in the {\tt HEALPix}
scheme (see Appendix~\ref{app:healpix_simus}).

We perform aperture photometry on the stacked maps at each \Planck\ HFI
frequency within a $3\parcm5$ radius of the SPT DSFG locations. 
This corresponds exactly to the radius of the region over
which we perform the \textit{Herschel} detection and photometry of
sources around the SPT DSFGs (see Sect.~\ref{methods:herschel}).  
We also investigated larger aperture sizes (up to a radius of
5$\arcmin$) and found that it produced no significant differences in
the results. 

We constrain the uncertainties on the average flux densities measured via stacking by performing 1000
bootstrap realizations of the stacked sample. Each bootstrap realization
is constructed by randomly selecting, with replacement, 65 SPT sources,
stacking their \Planck\ maps, and measuring the flux density in the resulting image. The scatter is determined by the $68\%$
confidence level in the resulting flux density distribution.
Fig.~\ref{fig:bootstrap_bright} shows the distribution of flux densities obtained after
doing aperture photometry on bootstrap realizations of these stacked maps at each \Planck\
frequency and at the IRIS frequency. Also shown, for the same frequencies, are the flux
density distributions (again after doing aperture photometry with a
3.5$\arcmin$ aperture radius) for 1000 iterations of stacking the same number (65) of
$1\deg\times1\deg$ maps which are selected randomly in the \Planck\
sky of the SPT fields. The flux density distributions that result from this
null test are all peaked around zero, as expected, and at 353, 545,
and 857\,GHz, are quite distinct from the distribution of flux
densities obtained from the 1000 bootstrap realizations of stacking
maps at the positions of the 65 SPT DSFGs. However, at 217 and
3000\,GHz, there is a much larger number of stacks in the null test
which have flux densities that are as high as those derived
from the bootstrap realizations on the SPT sources, compared to the
other frequencies. 
This is due to fluctuations of the Galactic cirrus at 3000 GHz and of
the CMB at 217 GHz in the stacked \Planck\ and IRIS maps. 

Our paper therefore focuses on the signal from 857, 545, and
353\,GHz. In Appendix~\ref{app:stack_uncertainties}, we show that the bootstrap
and photometric uncertainties in the \Planck\ flux densities are similar and that the
uncertainty due to inhomogeneity in the SPT sample is negligible. We
will use the bootstrap uncertainties throughout the analysis.

\subsection{\textit{Herschel} source detection and photometry}
\label{methods:herschel}

We create $10\arcmin$-by-$10\arcmin$ maps centred on the SPT DSFGs in each SPIRE band. 
Due to the short size of the scan pass ($10\arcmin$), the mapmaker
does not accurately recover angular scales as large as several
arcminutes. This means that these maps are poorly suited to recovering the
clustering signal on $3\parcm5$ scales (as was done with
\Planck). Therefore we focus on individually detected sources in the SPIRE maps.

We extract the resolved sources in the SPIRE
maps as well as in the blank HerMES Lockman-SWIRE field (which was
used as a reference field) in order to verify that there is indeed an
excess of resolved sources that contribute to the large-scale clustering signal observed by \Planck. We use the {\tt STARFINDER} algorithm \citep{diolaiti2000} which was developed to blindly extract sources
from confused maps, for this purpose. In order to avoid an extraction bias (which can
vary with position in the maps), we consider
only high significance detections: $S_{857}>50\,$mJy, approximately
$6\sigma$ in the HerMES Lockman SWIRE field and in the SPIRE maps of the SPT sources. 

The coverage of the maps of the SPT sources is not homogeneous. We only extract
sources within $3\parcm5$ of the SPT DSFG in order to minimize the effect of inhomogeneity. 
We have also verified that small changes to this radius (between
2.5$\arcmin-$3.5$\arcmin$) do not impact our results. We do not use the $S_{545}/S_{857}$ colours in the analysis because the
600\,GHz ($500\micron$) maps (beam FWHM$=$36$\arcsec$) suffer from a
larger degree of source confusion than the 1200\,GHz (FWHM$=$18$\arcsec$) and 857\,GHz
(FWHM$=$25$\arcsec$) maps. Hence we focus on the $S_{857}$/$S_{1200}$ colours in this work.

We compute $S_{857}/S_{1200}$ colours of these 857\,GHz-flux-selected galaxies using two different methods,
depending on whether or not they are detected independently at 1200\,GHz. 
For objects detected at both frequencies, we take the flux densities
reported by {\tt STARFINDER} at each frequency. Some red objects are not detected at
1200\,GHz. For these galaxies, we measure the 1200\,GHz flux
density at the 857\,GHz position using {\tt FASTPHOT}
\citep{bethermin2010b}, which is designed to deblend sources with known positions. To obtain
the most accurate flux densities possible, we also add the other
sources in the same field, which are detected at
1200 and 857\,GHz, to the list of positions used by
{\tt FASTPHOT}. In general, we recover source flux densities at
3$-$6\,$\sigma$ (which is just below the blind detection threshold),
and the precision on the colours is between 16.5 and 33.0$\%$. 
The same algorithm is applied to the maps of the SPT sources and the control field
so as to have the same potential residual biases, since our goal is not to
obtain an absolute measurement of the colour distribution, but to
detect potential differences between the environment of SPT sources
and blank fields. In order to check the quality of our source
extraction we perform Monte Carlo simulations
(Appendix~\ref{app:montecarlo_herschel}), injecting sources into both
the maps of the SPT sources and the larger HerMES field. We check the output
against input flux densities at each frequency. We also 
examine the completeness as a
function of flux density, where completeness is defined as the
fraction of recovered sources. For the rather conservative flux
density cut at $S_{857}>50\,$mJy, the completeness is higher than
95$\%$ and flux boosting (due to Malmquist and Eddington bias and from
source confusion) is below 5$\%$ in both the maps of the
SPT sources and the control field.


\begin{figure}{}
\centerline{\rotatebox{0}{
\includegraphics[width=.165\textwidth]{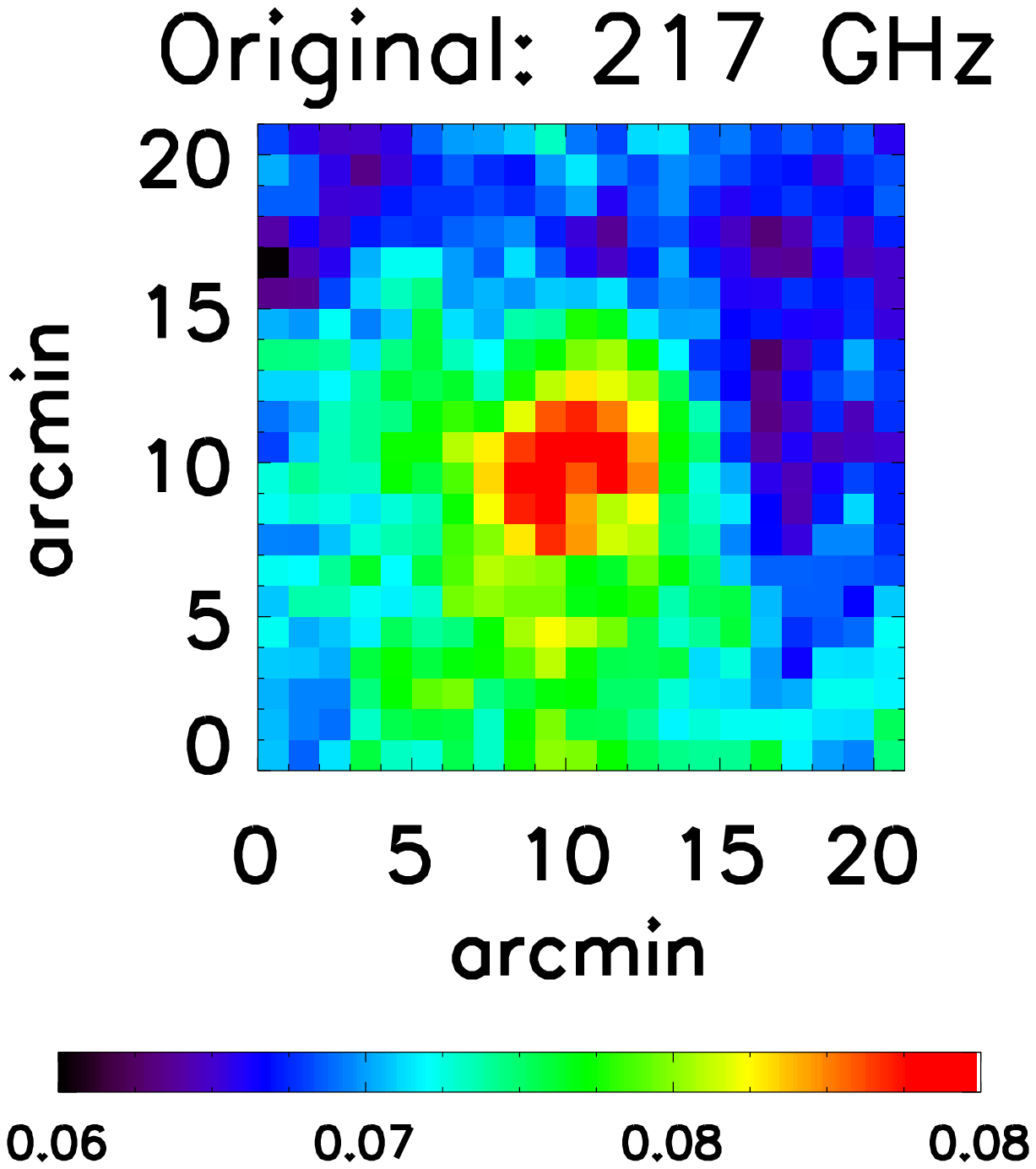}
\includegraphics[width=.165\textwidth]{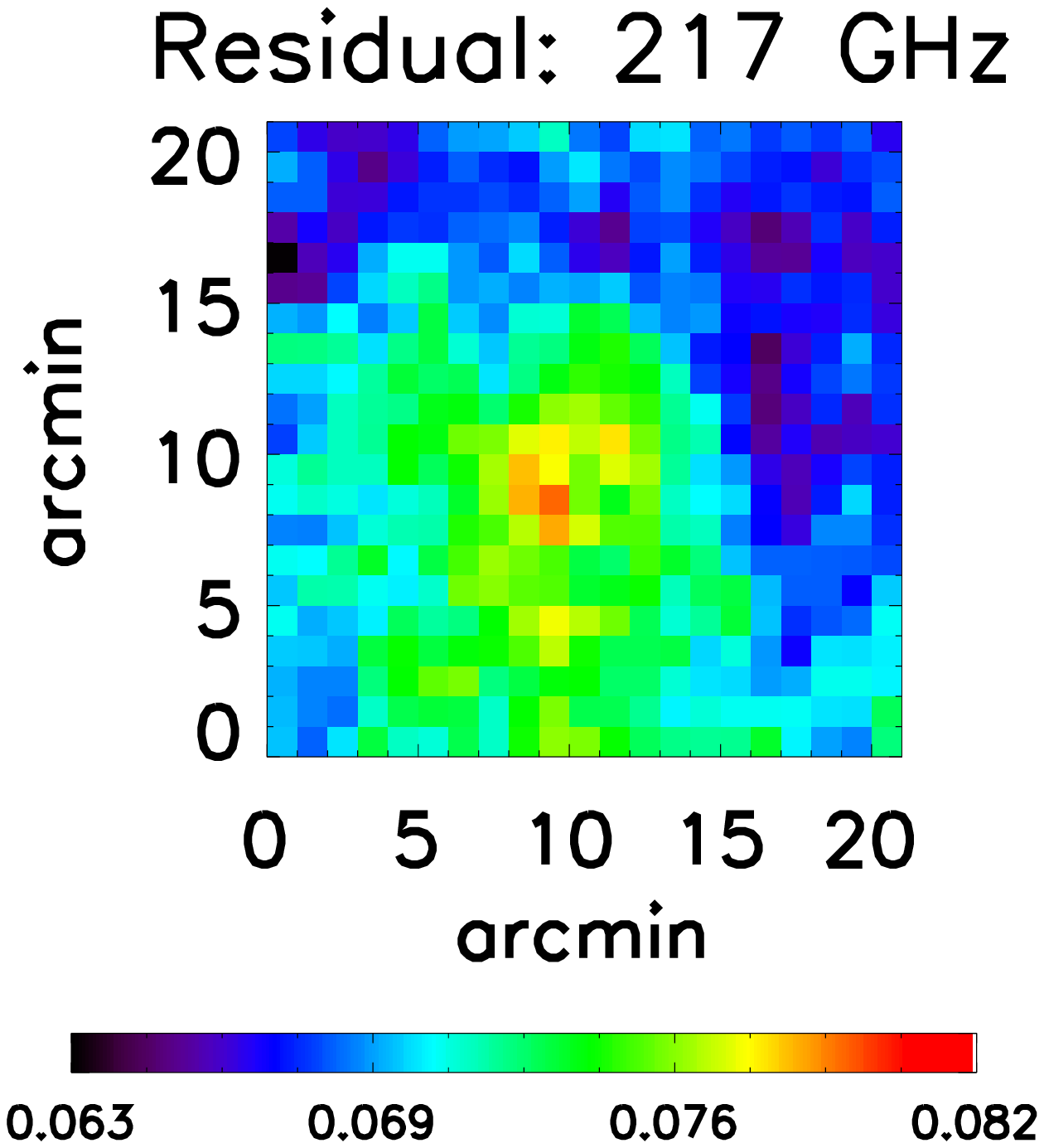}
}}   
\vspace{0.1cm}
\centerline{\rotatebox{0}{
\includegraphics[width=.165\textwidth]{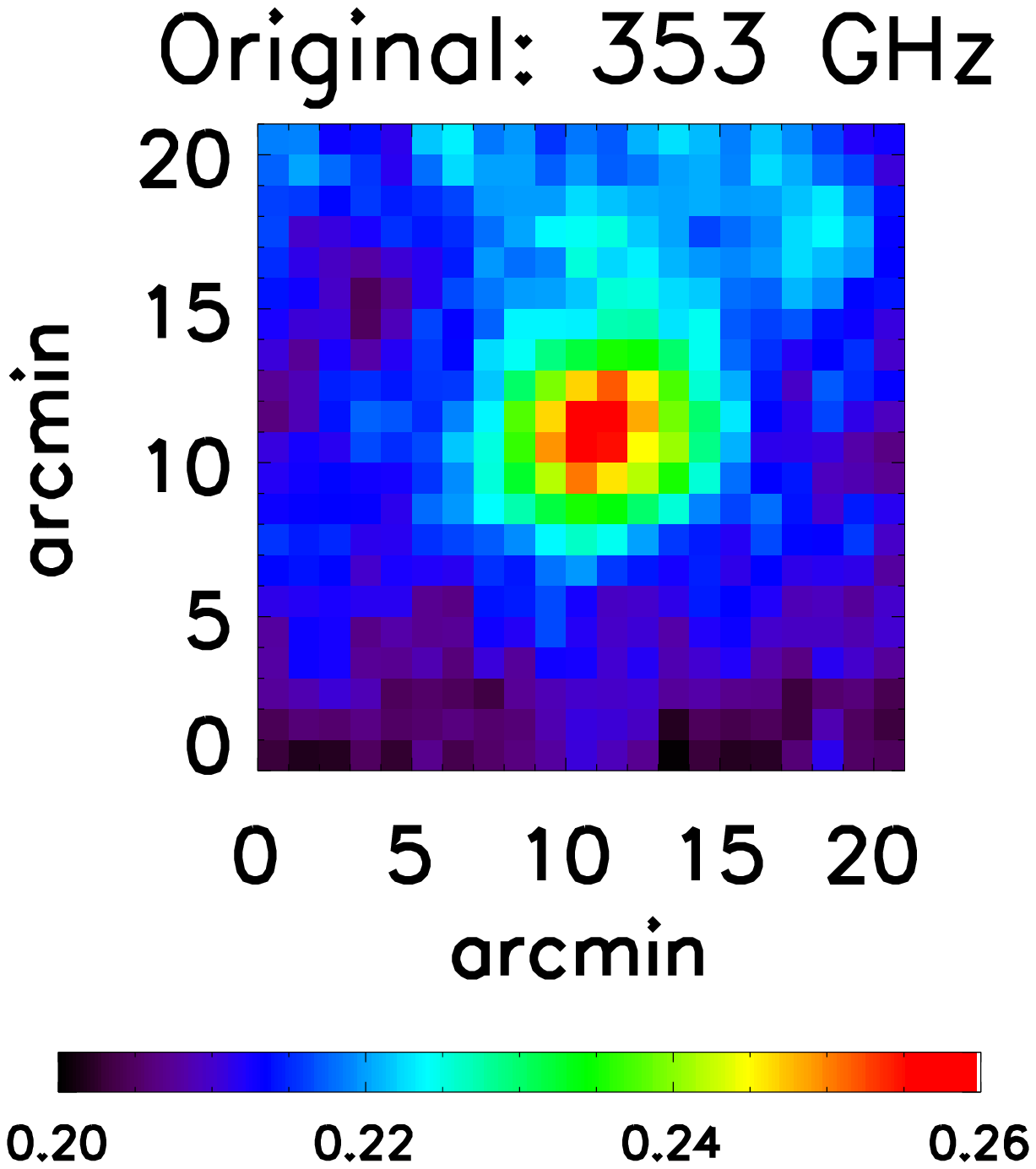}
\includegraphics[width=.165\textwidth]{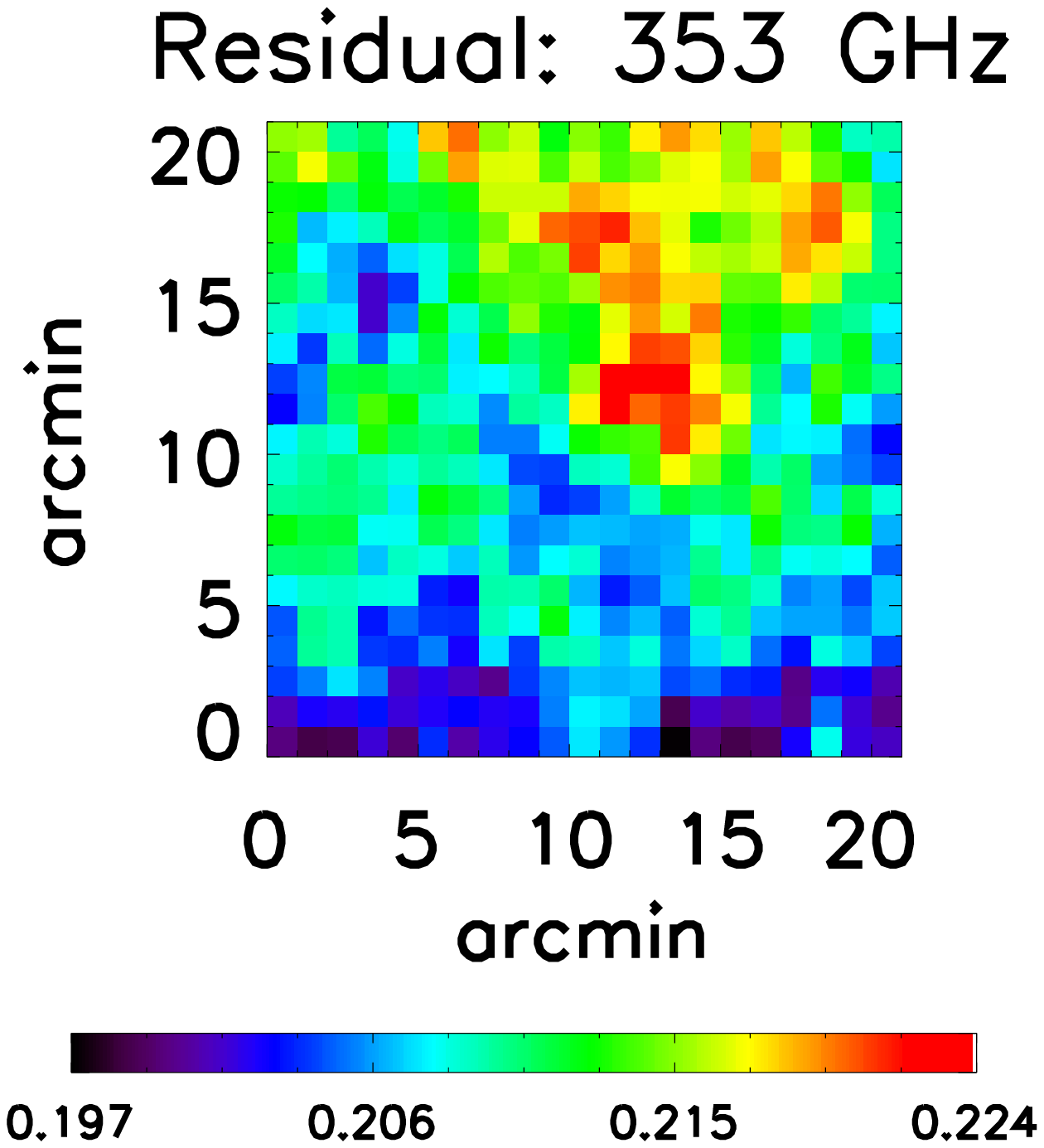}
}
}
\vspace{0.1cm}
\centerline{\rotatebox{0}{
\includegraphics[width=.165\textwidth]{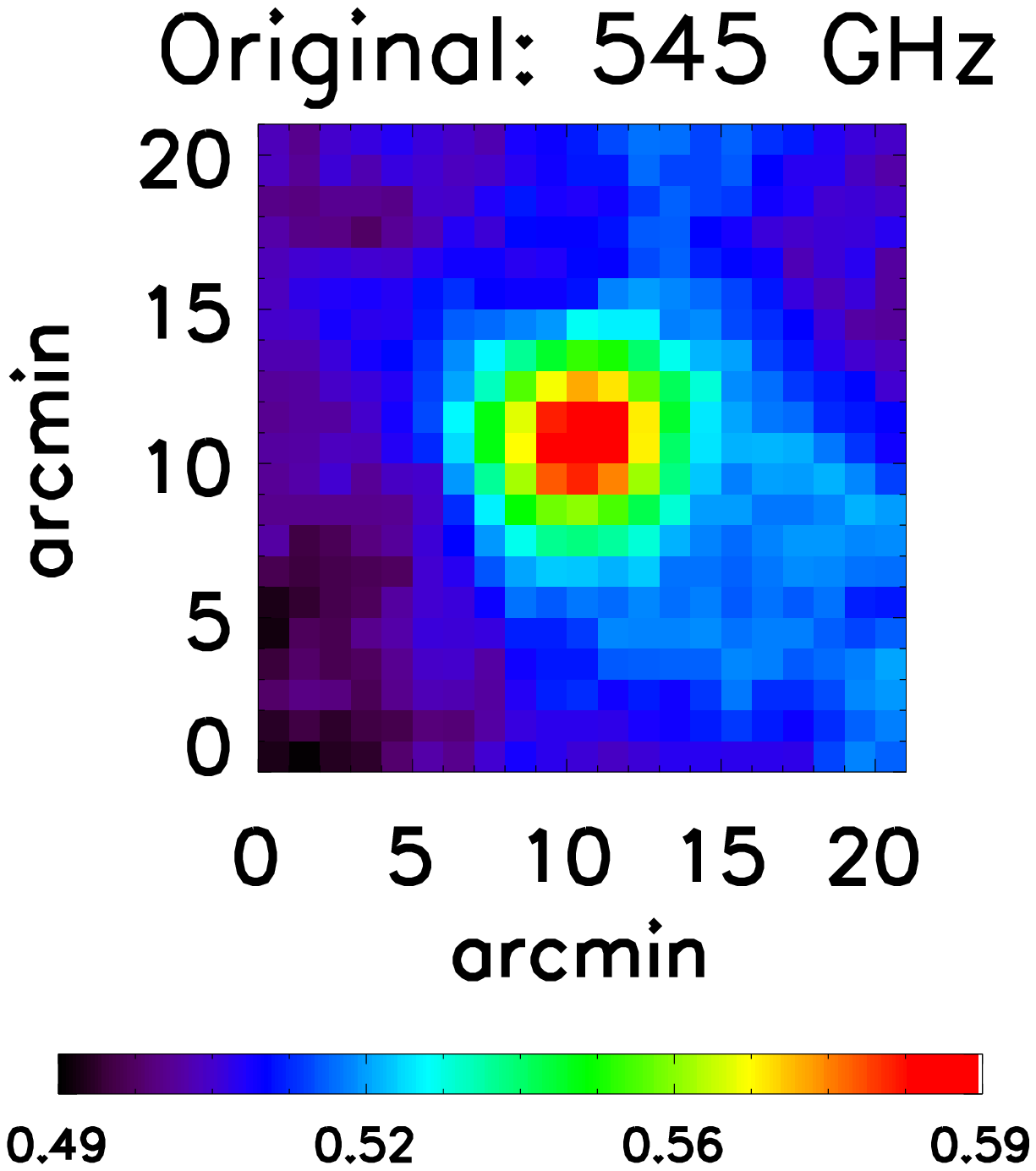}
\includegraphics[width=.165\textwidth]{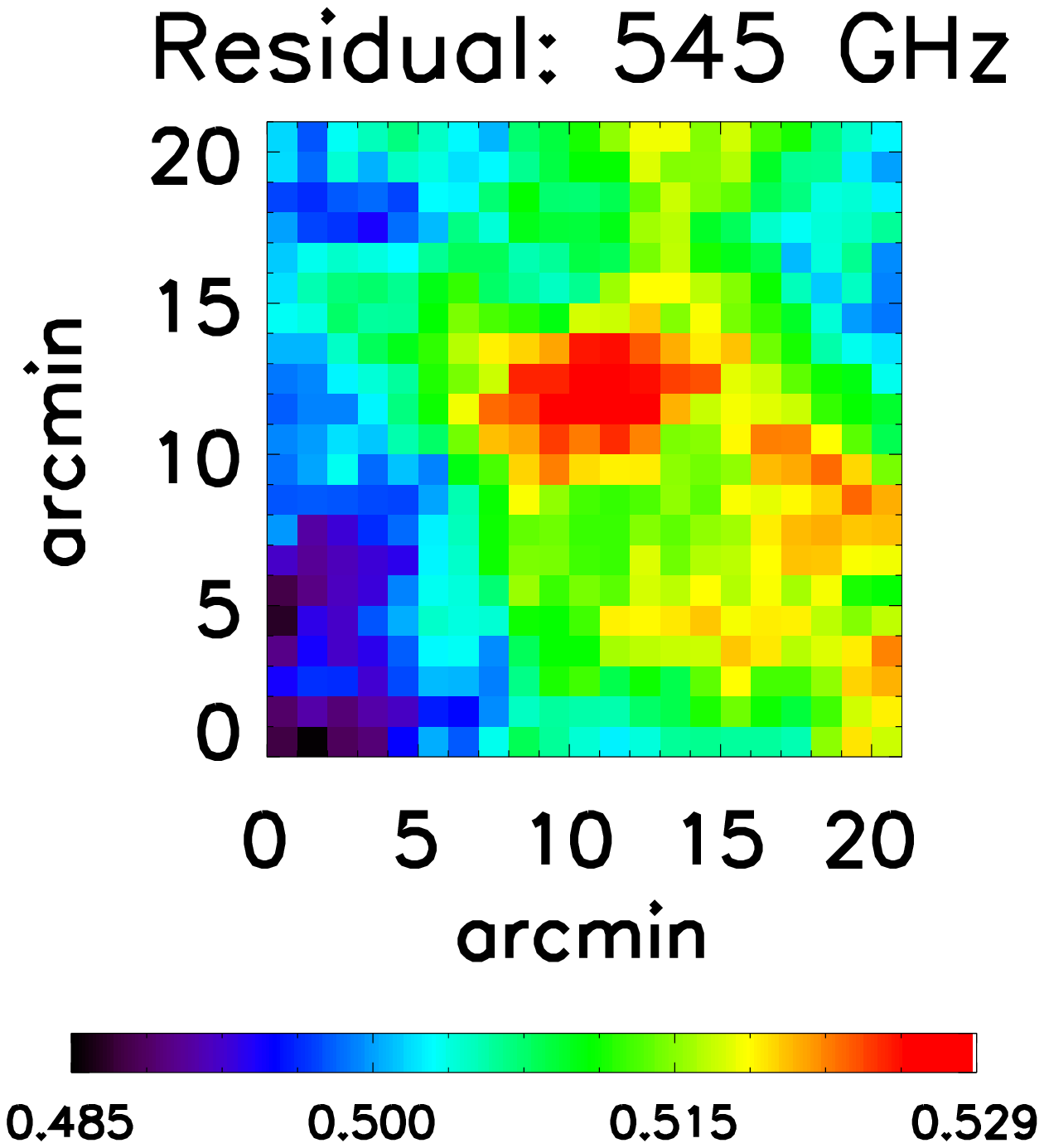}
}
}
\vspace{0.1cm}
\centerline{\rotatebox{0}{
\includegraphics[width=.165\textwidth]{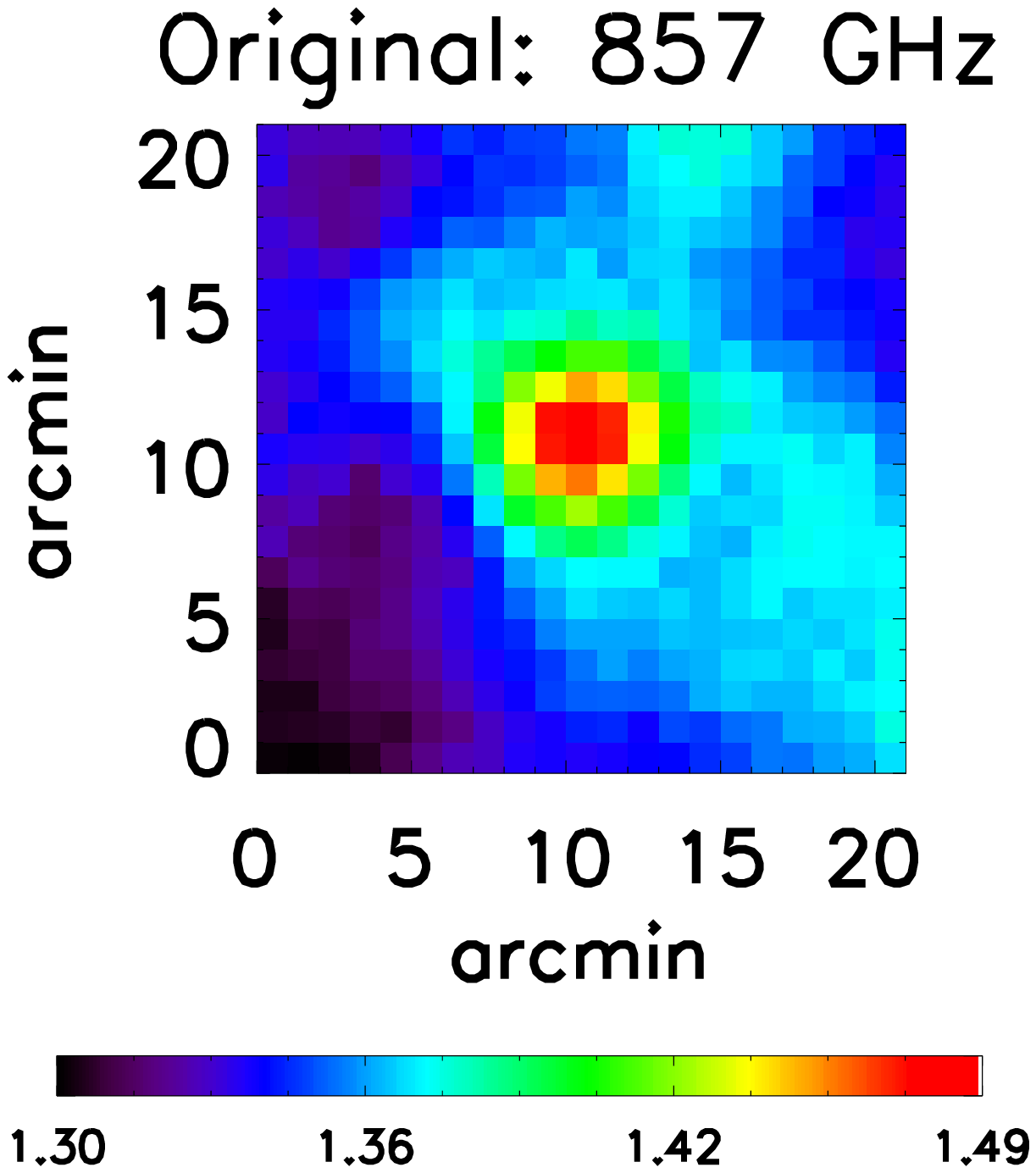}
\includegraphics[width=.165\textwidth]{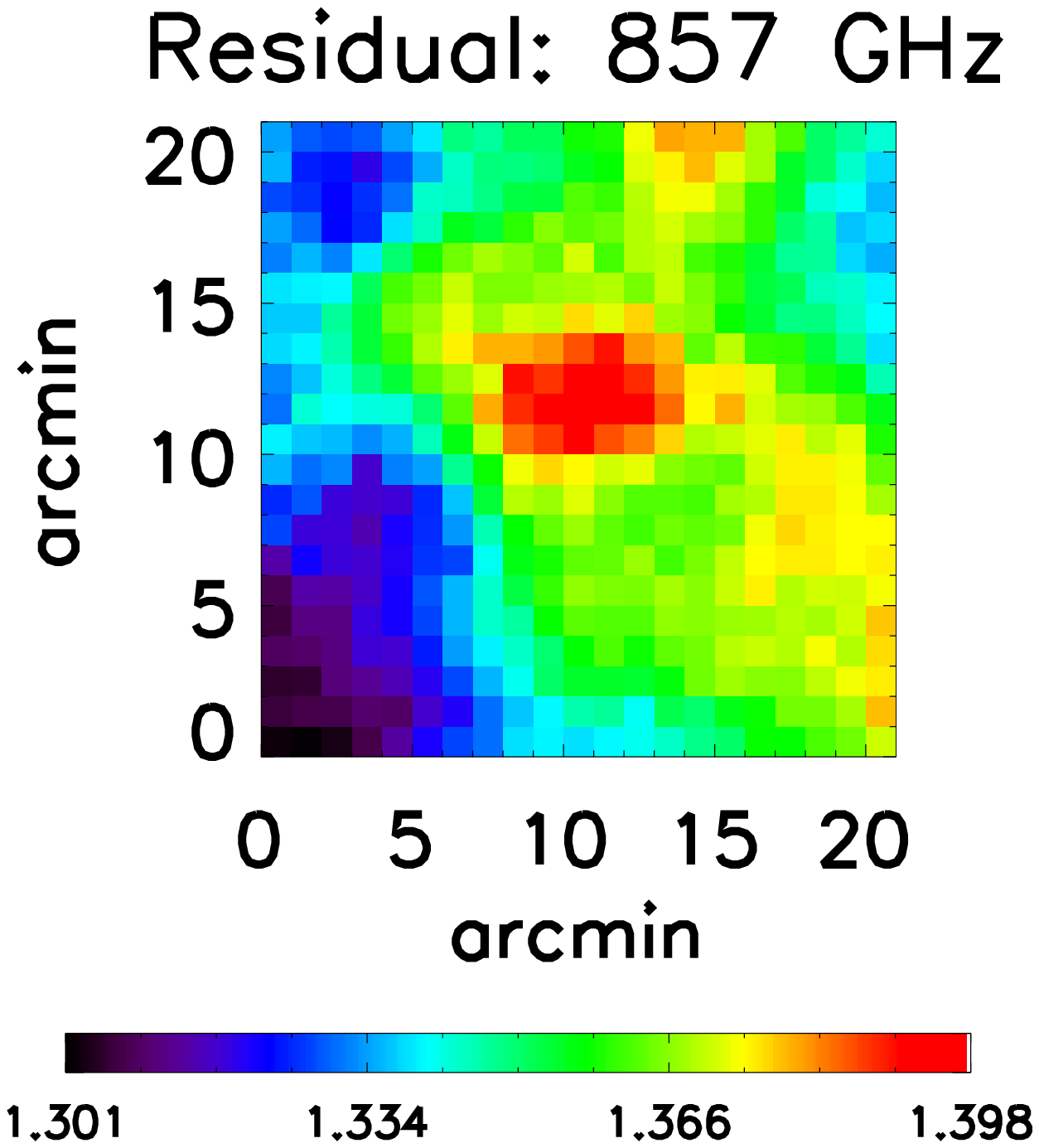}
}
}
\vspace{0.1cm}
\centerline{\rotatebox{0}{
\includegraphics[width=.165\textwidth]{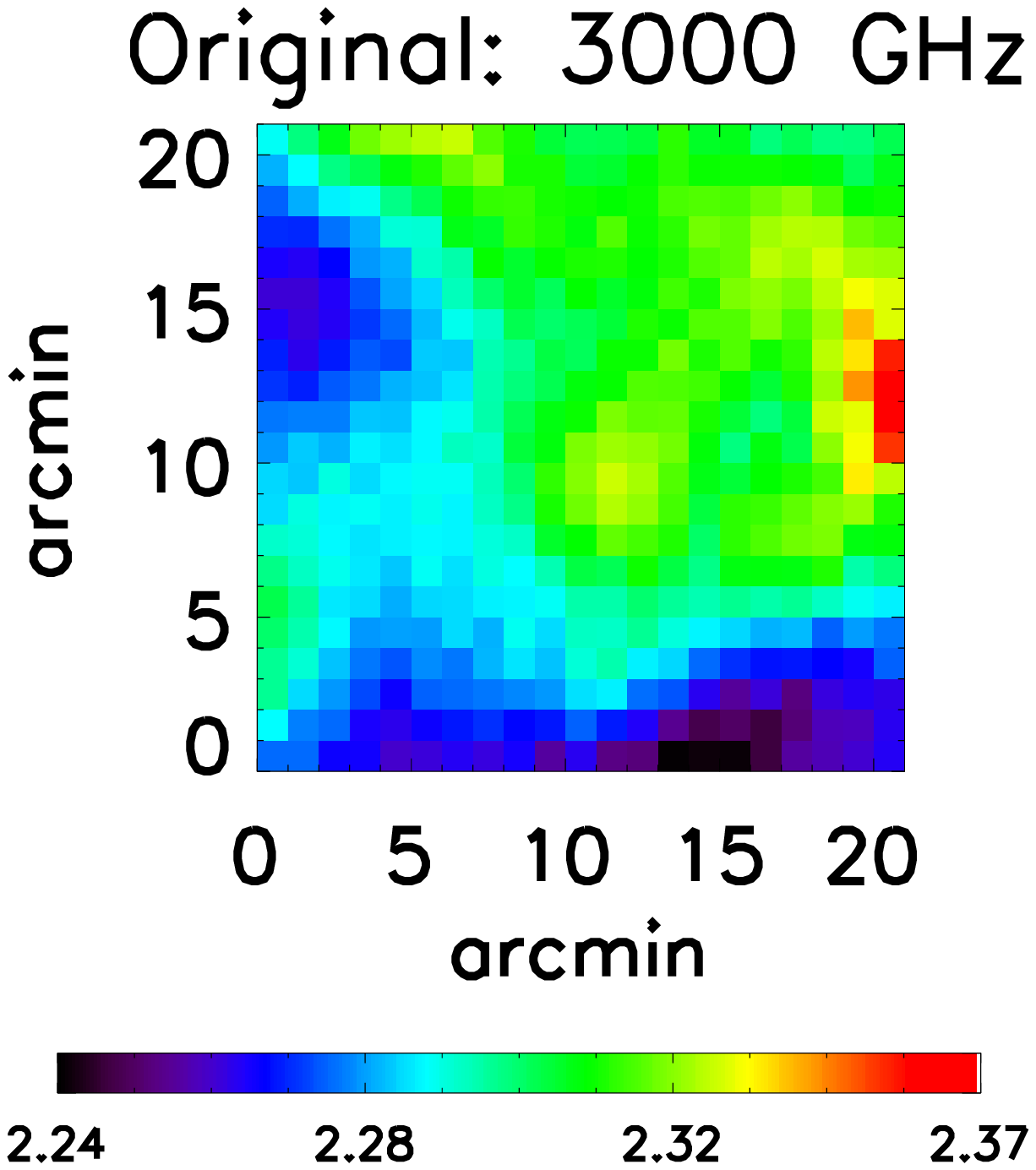}
\includegraphics[width=.165\textwidth]{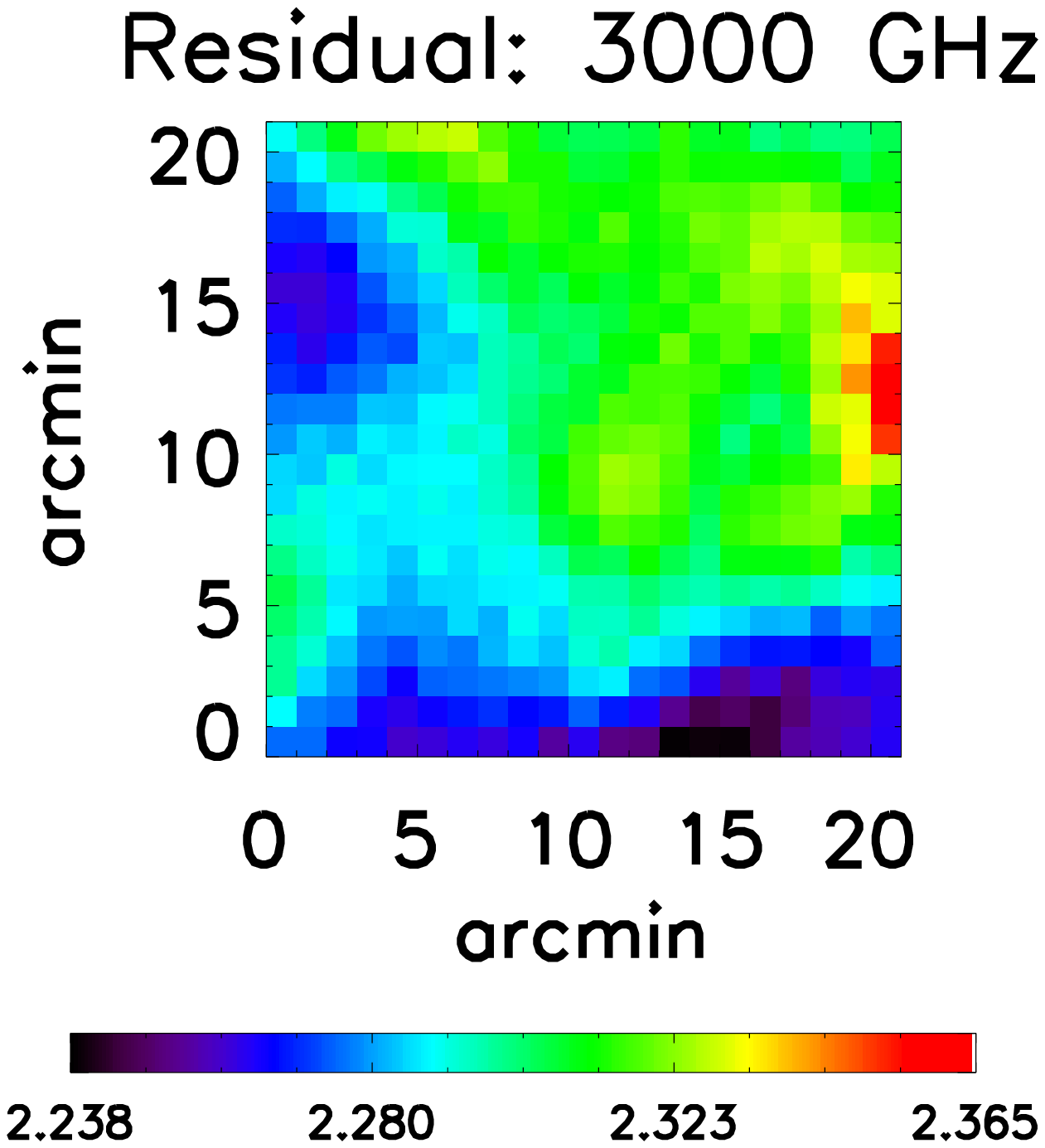}
}
}

\caption{Left panel: \Planck\ and IRIS maps, in units of MJy$sr^{-1}$ which are obtained by stacking individual maps at the
  positions of the SPT DSFGs. Each map in the stack is centred on the
    SPT-derived position of the DSFG. The original size of the stacked
    maps is $1\deg\times1\deg$. Here, we zoom into the central
    $20\arcmin\times20\arcmin$ region in order to show
    structure more clearly. The signal from the DSFGs is strong at 353, 545, and 857
  GHz. Right panel: residual maps obtained after removing the central
  compact source from each stacked map using the formalism in
  Appendix~\ref{app:planck_stack_formalism}. These residual maps show an
  extended but isolated structure at 545 and 857\,GHz.
\label{fig:stack_maps}}
\end{figure}

\section{Results}
\label{sec:results}

Here, we present our results in two broad divisions: 
(1) the measurement and analysis of the clustered component 
from stacking the \Planck\ HFI maps at the locations of the SPT DSFGs; and (2) the
confirmation, using \textit{Herschel} observations, of the clustering signal 
and the nature of the sources contributing to this clustering signal.

\subsection{The \textit{Planck} excess}
\label{subsec:results_planck}

We present the results of the stacking analysis, including the
measurement of the clustered component, its SED and
photometric redshift, and we estimate the SFR of all the galaxies
contributing to the signal. Finally, we present azimuthally-averaged profiles of
the different components in the \Planck\ stack.

\subsubsection{Measuring the clustered component}
\label{subsubsec:measure_clustering}

\begin{figure}
\includegraphics[width=0.5\textwidth]{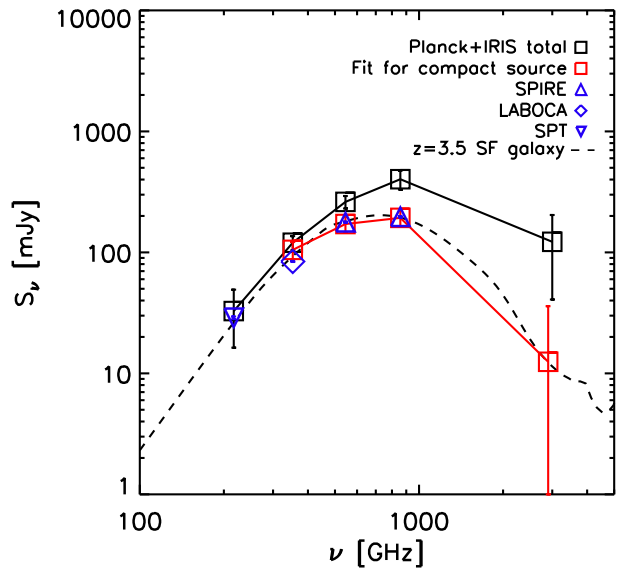}
\caption{Comparison of the mean \Planck\ (217$-$857\GHz) and
    IRIS (3000\,GHz) flux densities
    of the SPT sample after stacking the
    \Planck\ and IRIS maps (at the positions of the SPT DSFGs) with: (a)
    the mean SPT 220\,GHz flux density of the sample (blue inverted triangle); (b) the mean APEX/\textit{LABOCA} flux
    density at 345\,GHz (blue diamond); and (c) the mean \textit{Herschel}/SPIRE
    flux density at 857\,GHz and 545\,GHz (blue triangles). The mean
    \Planck\ and IRIS flux densities are estimated from: (i) aperture
    photometry (black squares and line); and (ii) after fitting simultaneously for the source,
clustering and background in the stacked \Planck\ and IRIS maps using
the formalism given in Appendix~\ref{app:planck_stack_formalism} (red
squares and line). There is no fitted flux measurement of the compact
source component shown at 217 GHz because we have SPT flux
measurements for the full SPT sample and we use the mean SPT flux
density at 220 GHz to constrain the fitting to the clustered term, as described in
Sec.~\ref{subsubsec:measure_clustering}. \Planck\ and IRIS photometric
uncertainties are obtained by bootstraping ($N_{\rm boot}=1000$ over
the stack). Also shown is an SED of a $z=3.5$ star-forming galaxy
generated from the \citet{magdis2012} effective templates (dashed line). The SED
derived from aperture photometry in the stack (black line) is wider than this typical SED of a
star-forming galaxy, because it is a superposition of the SEDs of a high redshift compact component and a low redshift clustered component. Subtracting the best-fit clustered term from the \Planck\ flux
densities brings them into agreement with the SPIRE and LABOCA flux densities.}
\label{fig:planck_sed}
\end{figure}

The left panel of Figure~\ref{fig:stack_maps} shows the \Planck\ and IRIS maps
which are stacked at the positions of the 65 SPT DSFGs. 
Figure~\ref{fig:planck_sed} shows the mean spectral energy
distribution (SED) of the sample that is derived from
\Planck\ and IRIS data after performing aperture
photometry on the stacked maps (black squares and line). The dashed line in
Fig.~\ref{fig:planck_sed} is a model galaxy SED at $z=3.5$ generated
from the SED library of \citet{magdis2012}. We observe that the mean SED of
the sample that is derived from doing aperture photometry on the
stacked maps is not simply a rescaling of a typical star-forming galaxy
SED at $z=3.5$. As a comparison with the \Planck\ flux density measurements, we also show the mean flux density measurements of the DSFGs (with the same selection in $S_{220}$) at
higher resolution, at 220\,GHz (the SPT measurement), 345\,GHz
(LABOCA), 545\,GHz and 857\,GHz (SPIRE). The LABOCA and SPIRE
measurements shown in Fig.~\ref{fig:planck_sed} are the mean flux
densities for all SPT sources which were detected in the LABOCA and SPIRE
maps respectively and which had measured flux densities (see Table~\ref{table:sample}).
We observe an excess in the \Planck\ flux density particularly at the
highest frequencies, compared to the flux density from other
observations at the same frequencies (albeit with relatively
high uncertainties): $206\pm73\,$mJy
at 857\,GHz, $84\pm31\,$mJy at 545\,GHz, and $36\pm16\,$mJy at
353\,GHz. At 220\,GHz, the excess is statistically not significant: $4\pm16\,$mJy.

One possible source of the excess in the \Planck\ maps is sub-mm emission
from sources clustered within the \Planck\ beam. The stacked signal
can therefore be decomposed into two components, a DSFG contribution
and a clustered component. We consider two scenarios here:

\begin{itemize}
\item If the clustered component is at the same redshift as the DSFGs
  and consists itself primarily of DSFGs, the SEDs of both components
  should be very similar. In particular, the peaks of the SEDs will be
  at approximately the same frequencies. The excess will thus be constant in frequency modulo some noise due to dust temperature and emissivity variations.
\item If the clustered component is at a lower redshift than the DSFGs, then the
  SED of the clustered component would be expected to peak at a higher frequency than the
  stacked DSFGs. 
\end{itemize}


The trend of the measured excess signal with frequency is
more consistent with the second scenario. 
This implies that the clustered signal within the \Planck\ beam has a much larger 
contribution from low redshift sources than from any clustered sources in the
neighborhood of the DSFGs. Given the fact that the majority of SPT DSFGs are
lensed, their positions are correlated with massive dark matter halos
at $z\sim1$, so we expect to detect sub-mm emission from galaxies in the lensing halos. 

We next test the hypothesis that there is a clustered signal within the $3\parcm5$ radius aperture. We fit the stacked \Planck\ maps to
a model following the formalism of
\citet{bethermin2010b,bethermin2012a} and \citet{heinis2013}. The model has 3
components:
(1) the compact source,
(2) the clustered component, and
(3) the background.

The method is described fully in Appendix~\ref{app:planck_stack_formalism}.
We use this formalism to extract the mean flux density of the compact
source (red points and line in Fig.~\ref{fig:planck_sed}) by
fitting simultaneously for all three components in the stack. 
The right panel of Figure~\ref{fig:stack_maps} shows the residual maps
after the compact source has been removed from the stacked maps using
this formalism. The residual images at 545 and 857\,GHz in particular 
show an extended but isolated structure around the centre of each map. 
The \Planck\ excess is now defined as the difference between the
compact source's flux density and the total flux density within the
$3\parcm5$ radius aperture. The same excess is recovered if we 
perform aperture photometry on the residual maps at each frequency (see also
Sect.~\ref{subsubsec:radial_profiles}, where we measure radial
profiles of the different components).

In addition, at 217 GHz, since we have measured
SPT flux densities for the full SPT DSFG sample, we remove a compact
source from the \Planck\ stack where the normalization of that compact
source in the fit is fixed by the mean SPT flux density, and then
perform aperture photometry on the residual map. This results in a statistical uncertainty in
  the mean \Planck\ excess measured at 220 GHz that is lower than if
  we did not use this prior. At 353\,GHz, as seen in
  Fig.~\ref{fig:planck_sed}, the total flux density in the stack and
  the flux density from the compact source that is obtained from the model fits, are $0.6\sigma$ apart, and
we find no significant evidence for an excess. However, at higher frequencies, a
clustered component is needed to reconcile the \Planck\ flux densities
with those obtained from the higher resolution observations in Fig.~\ref{fig:planck_sed}.

In Appendix~\ref{app:clustering_tests}, we describe three tests to
verify that the clustered component is real and not simply an
artefact of the stacking procedure. In the first test (see Appendix~\ref{app:stacking_simulations}), we perform stacking
simulations, with artificial compact source components and clustering
components generated using the same model as in
Appendix~\ref{app:planck_stack_formalism} and injected into blank
\Planck\ maps before they are stacked. 
We find no significant bias arising from the stacking procedure in the
mean flux densities obtained from either aperture photometry or from 
fitting to the source and clustered components. In Appendix~\ref{app:random_rotations}, we also test whether the extended
component seen in the residual maps at 545 and 857\,GHz around the
central compact source in Figure~\ref{fig:stack_maps} is actually part of the structure in the
background, by creating many realisations of the stacked maps where
the individual \Planck\ maps are rotated randomly by 90$\deg$ before
they are stacked. The clustered component appears consistently at 545
and 857\,GHz as an isolated structure around the compact source and is
therefore not simply part of the structure in the background.

In Appendix~\ref{app:blazars}, we show that the clustering
component does not appear at 545 and 857\,GHz if there are no lensing
halos in the foreground. We stack \Planck\ and IRIS maps at the
positions of a sample of 65 SPT synchrotron
sources \citep{vieira2010}. These sources are not angularly correlated
with foreground structure and we find no extended component in the residual maps after removing the central
compact source (the synchrotron source itself) from the stacked maps
using the same fitting formalism. This suggests that the
clustered component found in this study is specific to the foreground
lensing halos of the STP DSFGs.

\begin{figure}
\centerline{\rotatebox{0}{
\includegraphics[width=0.5\textwidth]{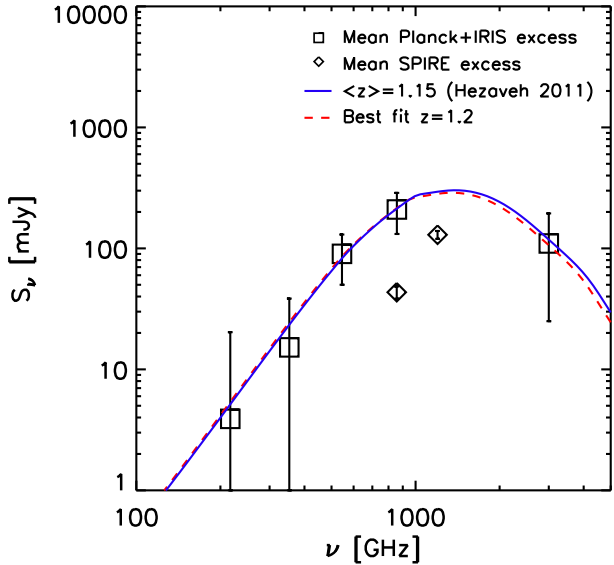}
}
}
\caption{SED of the \Planck\ excess (black squares), which
    is derived from the difference between the total flux density
    within a $3\parcm5$ radius (black squares in
    Fig.~\ref{fig:planck_sed}) and the flux density of the compact
    source in the stack (red squares in
    Fig.~\ref{fig:planck_sed}). There is no significant evidence of an
    excess at 220 GHz or 353 GHz; the data are consistent with zero
    at 1$\sigma$. We also compare the \Planck\ excess SED with two star-forming galaxy SEDs that
    are generated from the B12 library \citep{bethermin2012b} and
    redshifted to: (1) the predicted mean redshift ($z\sim1.15$) of the SPT
    lensing halos in \citet{hezaveh2011} (blue line); (2) the best fitting redshift
    ($z\sim1.2$) found by maximizing the probability distribution for
    the redshift $p(z)$ (red dashed line). The data requires $T_d>50$\,K at 
95$\%$ confidence if we assume the excess emission originates from
   $z=3.5$. On the other hand, if we assume $z=1.15$, we obtain
$T_{\rm d}=(32\pm19)$\,K (in addition, $T_{\rm d}=(33\pm20)$\,K for
$z=1.2$ from the best fit to the \Planck\ excess) which is within the range of
expected dust temperatures of galaxies (see Sect.~\ref{subsubsec:excess_pz}). Finally, we show the mean excess
    of flux density $S_{\rm excess}$ at 857\,GHz and 1200\,GHz from sources that are detected in
    \textit{Herschel}/SPIRE within $3\parcm5$ of the SPT DSFGs. This
    excess in flux density is computed relative to all other sources that have been detected at the same
    flux density threshold in a larger control field (see
    Sect.~\ref{subsec:results_herschel} and
    Eq.~\ref{eq:herschel_Sexcess}). It is expected that the detected SPIRE sources
    account for a fraction (approximately 20\% at 857\,GHz) of the
    \Planck\ excess \citep{bethermin2012a}.
\label{fig:excess_sed}}
\end{figure}


\begin{figure}
\centerline{\rotatebox{0}{
\includegraphics[width=0.5\textwidth]{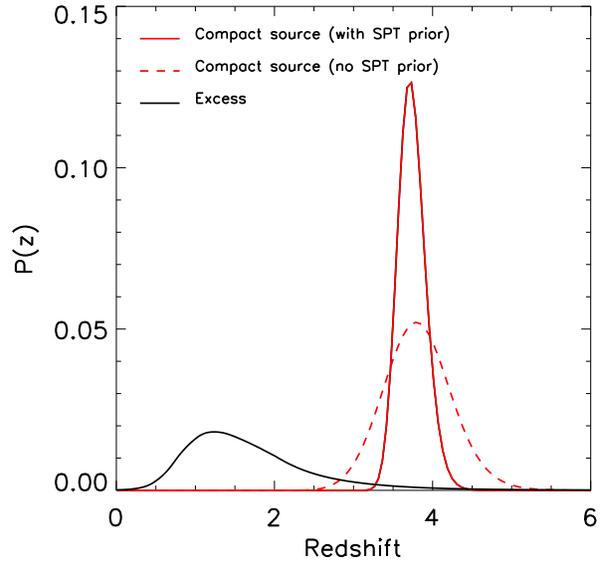}
}
}
\caption{Probability distribution for the mean redshift, $p(z)$
    for two components of the \Planck\ stack. The dashed red line
    shows $p(z)$ for the compact source, where the SED is given by the
    red line in Fig.~\ref{fig:planck_sed} which is obtained from the fit described in
    Appendix~\ref{app:planck_stack_formalism} using only the 353, 545,
    857 and 3000 GHz data. The solid red line is the result of fitting to an SED where we
    also use a 217 GHz data point, assuming the compact
    source has the same mean flux density at 217 GHz as
    the SPT mean flux density of the sample. The black line shows
    $p(z)$ for the \Planck\ excess, the SED for which is shown in
    Fig.~\ref{fig:excess_sed} (see also the fifth row of
    Table~\ref{table:clusteringbias}). The quantity $p(z)$ for each component is derived by fitting SED
  templates from the \citet{magdis2012} library in a range of redshifts, to the measured SED of
  that component, using Eqs.~\ref{eq:pz_chi2}$-$\ref{eq:pz_final} (see
  Sect.~\ref{subsubsec:excess_pz} and Appendix~\ref{app:prob_z} for
  details). The distribution $p(z)$ for the compact source component peaks
  near the mean of the redshift distribution for SPT sources
  $z\sim2$--$6$ found in \citet{weiss2013}, whereas the $p(z)$ for the \Planck\ excess has a maximum at $z=1.2$.
\label{fig:pz_excess}}
\end{figure}

\subsubsection{Clustering contamination in the stacked flux densities of the DSFGs}
\label{subsec:bias}

We quantify the contribution of the clustered component associated with the
foreground lensing halos relative to the measured stacked flux
densities of the high redshift lensed galaxies.
The enhancement introduced by the clustering signal 
\citep{bethermin2010b,kurczynski2010,bethermin2012a,bourne2012,viero2013b}
needs to be taken into account in order to obtain a correct estimate of
the mean flux density of the background lensed galaxies in the stack. 
In this study, in particular, the clustering contamination is
significant, because the beam size of \Planck\ is comparable to the
angular scale of the clustering signal. Our aim is therefore to quantify the
clustering contamination in the different frequency channels of \Planck\ HFI.

The relative clustering contamination can be expressed as the ratio of the flux
density of the clustered component to the flux density of the compact source
component in the stack. In Table~\ref{table:clusteringbias}, we list the mean
flux densities of the clustered component and compact source
component in the stack, as well as the relative clustering contamination for the
217, 353, 545, and 857 GHz channels. The flux densities of the compact
source component and the clustered component are obtained from
the fits. When fitting the clustered component at 217 GHz, however, we exploit the fact that we have
measured SPT flux densities for the full SPT DSFG sample and introduce
the mean SPT flux density in the fitting in order to compute the
strength of the clustered term, as described in
Sec.~\ref{subsec:results_planck}. At 217 GHz, therefore, the strength of
the clustered term is defined as the flux density of the residual component
obtained after removing a compact source (through the same fitting procedure) whose normalization is
given by the mean SPT flux density itself.


We find that the relative clustering contamination has a large
uncertainty at 220 GHz but thereafter increases with frequency in the \Planck\ HFI channels (the beam FWHM is relatively
stable among the HFI frequencies, so we focus on the frequency dependence here).
This flux density contribution from sources clustered around the foreground
lensing halos adds to the stacked flux density of the background
lensed galaxies. This boosts the flux density estimates of the
background galaxies that are derived from aperture photometry performed on
\Planck\ data. The clustering contamination should therefore be taken into account in order to
obtain the correct flux densities of galaxies (both
ensemble-averaged flux densities from stacking but also flux densities
of individual galaxies) in \Planck\ data.

\begin{table*}
\caption{Mean flux densities of the components in the \Planck\ stack and
  the relative clustering contamination values as a function of frequency. The
  latter is expressed as the ratio of the flux
  density of the \Planck\ excess to the flux density of the compact
  source component. The flux density of
  the compact source component is expressed in two different ways: (1) from the
  high-resolution measurements (SPT, LABOCA and SPIRE: third row) assuming that
  there is negligible clustering of sources in the SPT, LABOCA and
  SPIRE beams; (2) from the fits to the components in the \Planck\
  stack, as described in Appendix~\ref{app:planck_stack_formalism}
  (fourth row). The flux density of the clustered
  component (fifth row) is then computed from the difference between the total flux density
  within a $3\parcm5$ aperture and the fit to the compact source component. In
  addition, at 217 GHz, since we have measured
  SPT flux densities for the full SPT DSFG sample, we use the mean SPT flux
  density in order to constrain the strength of
  the clustered component at 217 GHz: we remove a compact source from
  the \Planck\ stack where the normalization of that compact source in
  the fit is fixed by the mean SPT flux density, and then perform aperture
  photometry on the residual map. As we employ this prior based on the
  SPT flux density, we do not quote a value for the flux density of
  the compact source component at 217 GHz obtained from the fits. Finally, the relative clustering
  contamination is expressed as the ratio of the flux density of the
  clustered component to that of the compact source component, which
  are both obtained from the fits. At 217 GHz, this is computed as the
  ratio of: (1) the clustered component computed with the prior on the
  SPT flux density and (2) the SPT flux density itself.
} 
\centering 
\begin{tabular}{l c c c c} 
\hline
Frequency & 217\,GHz & 353\,GHz & 545\,GHz & 857\,GHz \\
Total flux density from aperture photometry [mJy] & 32.7$\pm$16.4 & 120.1$\pm$16.1  & 261.6$\pm$30.9 & 402.4$\pm$72.5  \\ 
Flux density of the compact source component (high resolution measurements)
[mJy] & 28.8$\pm$0.7 & 84.1$\pm$0.9  & 177.5$\pm$2.0 & 196.7$\pm$2.4  \\ 
Flux density of the compact source component (from fit)
[mJy] & - & 104.9$\pm$16.9  & 171.4$\pm$25.5  & 192.8$\pm$28.9 \\   
Flux density of the clustered component (from fit) [mJy] & 3.9$\pm$16.4 & 15.2$\pm$23.3  &
90.1$\pm$40.1   &  209.6$\pm$78.0  \\
Relative clustering contamination & 0.1$\pm$0.6 &  0.2$\pm$0.2  & 0.5$\pm$0.2    &  1.1$\pm$0.4       \\
\hline
\end{tabular} 
\label{table:clusteringbias} 
\end{table*}

\subsubsection{SED and photometric redshift of the clustered component}
\label{subsubsec:excess_pz}

In Fig.~\ref{fig:excess_sed}, we show the SED of the excess signal.
In order to derive redshifts from the sub-mm SEDs, we use
the effective SED library of \citet{bethermin2012b}\footnote{\tiny{\url{http://irfu.cea.fr/Sap/Phocea/Page/index.php?id=537}}}, which is
based on the \citet[][hereafter M12]{magdis2012} SED libraries and the
\citet[][hereafter B12]{bethermin2012b} model. 
These templates are the luminosity-weighted average SED of all the galaxies
described by the B12 model at a given redshift. 
There are two families of templates included -- ``main-sequence'' (MS) and
``starburst'' (SB) galaxies -- and both evolve with redshift. 
We also assume a scatter in the mean radiation field $\langle U\rangle$  of 0.2 dex
(about 0.05 dex in the dust temperature) 
at fixed redshift for a given family of templates.

We fit the template SEDs as a function of redshift to the SED of: (1)
the compact source; and (2) the \Planck\ excess (after subtracting the
contribution from the compact source). We derive the probability distribution for
the redshift, $p(z)$, for these two components (see Appendix~\ref{app:prob_z} for
a full description of how $p(z)$ was computed), as shown in Fig.~\ref{fig:pz_excess}. 
The $p(z)$ of the compact source component is narrower than the redshift
distribution from $z\sim2$--$6$ found by \citet{weiss2013} for a subset
of the sources analysed here, but has a
consistent central value at $z\sim4$. 
The $p(z)$ of the excess is quite different and
peaks at $z\sim1.2$, with a tail to higher redshifts. In Fig.~\ref{fig:excess_sed}, we show the 
template SED redshifted to: (a) the best-fit redshift $z=1.2$; and (b) the theoretical mean redshift of the lensing halos ($z=1.15$)
predicted by \citet{hezaveh2011}. Although still uncertain, the agreement supports the hypothesis that the
clustered sources are primarily associated with the foreground lenses
rather than the DSFGs. In addition, we estimate the dust
temperatures of sources contributing to 
the \Planck\ excess by fitting a modified blackbody with spectral index $\beta=2.0$, to the
Rayleigh-Jeans part of the spectrum in Fig.~\ref{fig:excess_sed} ($\nu\le857$\,GHz) and assuming: (1)
$z=3.5$, consistent with the mean redshift of the DSFGs
\citep{weiss2013}, and (2) $z=1.15$ for the foreground lenses
\citep{hezaveh2011}. The data requires $T_d>50$\,K at 
95$\%$ confidence if we assume the excess emission originates from
the environments around the high-redshift DSFGs. This is incompatible with
what is known of high redshift galaxies \citep[see
e.g.,][]{hwang2010a,magnelli2010}. On the other hand, if we assume $z=1.15$, we obtain
$T_{\rm d}=(32\pm19)$\,K (in addition, $T_{\rm d}=(33\pm20)$\,K for
$z=1.2$ from the best fit to the \Planck\ excess in
Fig.~\ref{fig:excess_sed}) which is within the range of
expected dust temperatures of galaxies. This is a further indication that the sources contributing to the
\Planck\ excess are associated with the foreground lenses rather than the high-redshift DSFGs themselves.

\subsubsection{Far-infrared luminosity and SFR of the clustered component}

Assuming a mean redshift of $z=1.15$ for the lenses \citep[consistent
with the estimate for SPT DSFG lens redshifts in][]{hezaveh2011}, 
the total far-infrared luminosity $L_{\rm IR}$ (computed between 8 and
1000$\micron$ in the rest frame) for the sources contributing to the
excess within the \Planck\ beam is
$(1.5\pm0.4)\times10^{13}\,\Lsolar$. Using the relation between SFR
computed in the IR and $L_{\rm IR}$ in \citet{kennicutt1998}, ${\rm
  SFR} (\Msolar\,{\rm yr}^{-1}) = 1.7\times10^{-10}(L/\Lsolar)$, we
obtain a total SFR of $(2700\pm700)\,\Msolar\,{\rm yr}^{-1}$ from all
galaxies contributing to the clustering signal within a radius of $3\parcm5$ from the positions of the SPT DSFGs. In
Sect.~\ref{subsec:results_herschel}, we derive the contribution to
this overall SFR from galaxies that are resolved by \textit{Herschel}.


\begin{figure*}{}
\centerline{\rotatebox{0}{
\includegraphics[width=.325\textwidth]{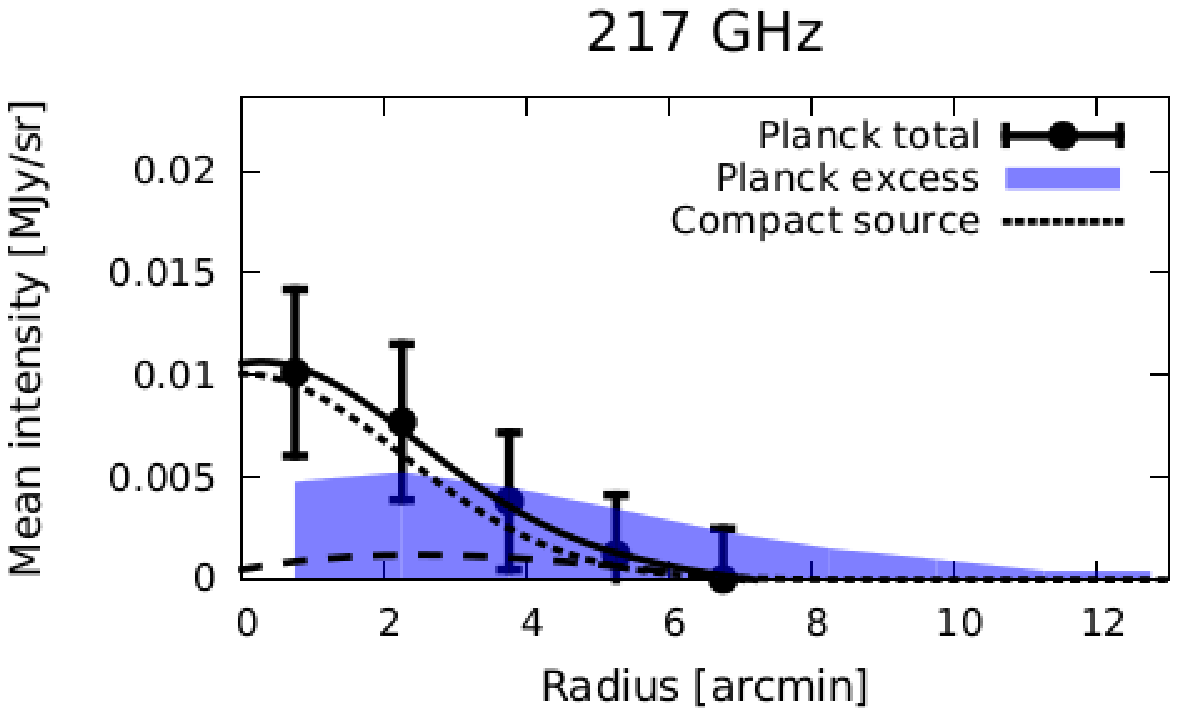}
\includegraphics[width=.325\textwidth]{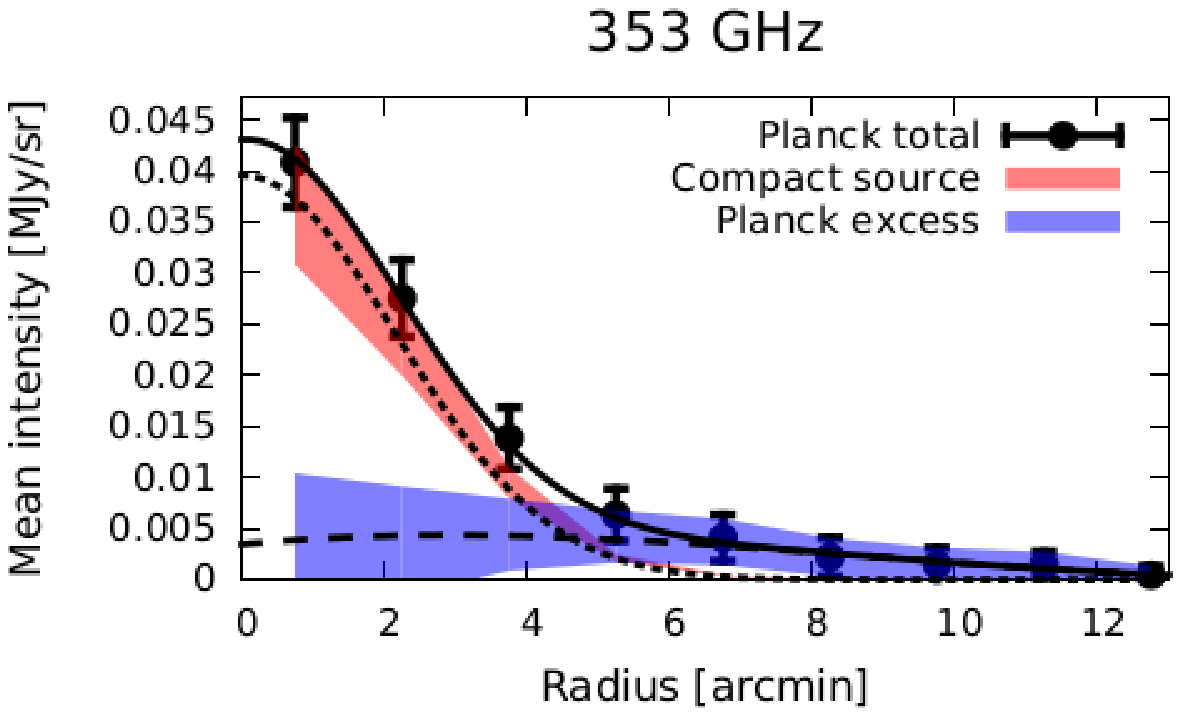}
}}
\vspace{0.1cm}
\centerline{\rotatebox{0}{
\includegraphics[width=.325\textwidth]{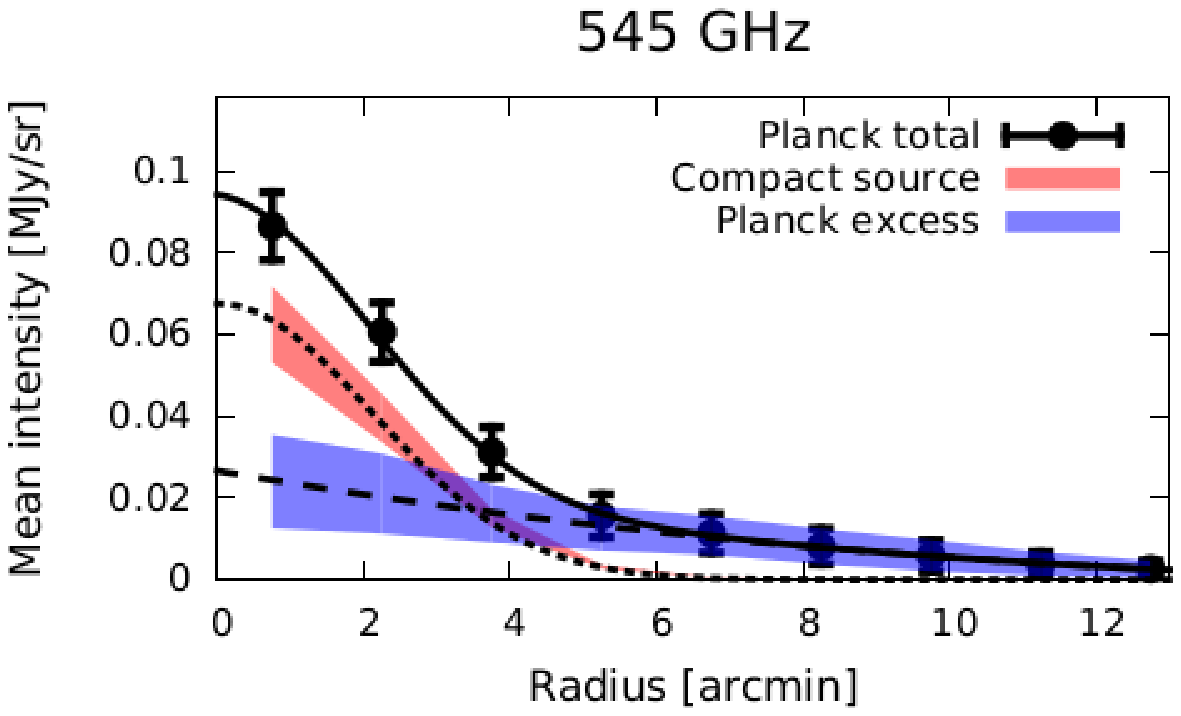}
\includegraphics[width=.325\textwidth]{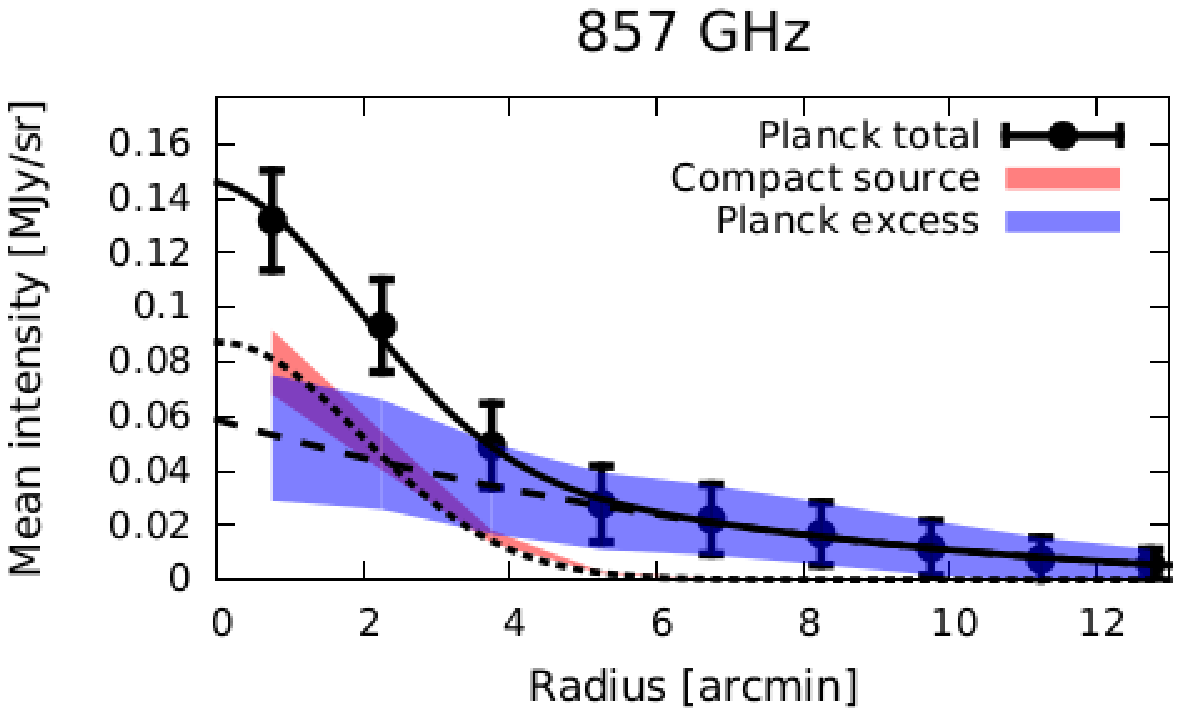}
}}
\vspace{0.1cm}
\centerline{\rotatebox{0}{
\includegraphics[width=.325\textwidth]{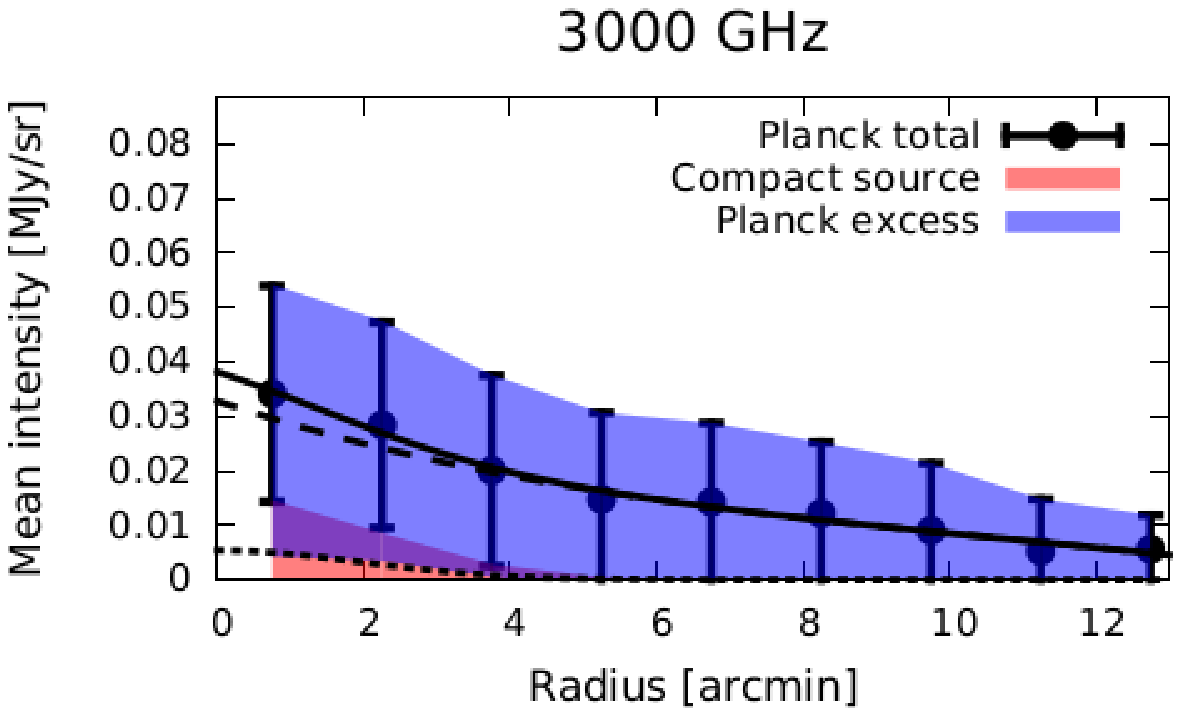}
}
}
\caption{Radial profiles of the different components in the \Planck\
  and IRIS maps stacked at the positions of the SPT DSFGs. The panels show the azimuthally-averaged mean
  intensity, at each frequency, of: (a) the stacked map of the DSFGs (filled
  circles with error bars) -- the cumulative flux densities obtained
  from this profile within a $3\parcm5$ radius aperture are shown in Fig.~\ref{fig:bootstrap_bright}
  and in the black line in Fig.~\ref{fig:planck_sed}; (b) the compact source component after fitting a Gaussian profile with a FWHM that is fixed by the effective \Planck\ (or IRIS, bottom panel) beam width ($1\sigma$ uncertainty, red shaded region); and (c) the excess
  obtained by removing the compact source component from the stack ($1\sigma$ uncertainty, blue
  shaded region). For each component of the stack, the uncertainties
  are derived from the bootstraps at each frequency. The short dashed line is a Gaussian fit (FWHM fixed
  by the beams) to the compact source profile. The long dashed line is
  a fit, using a cubic polynomial, to the mean intensity of the
  \Planck\ (or IRIS, bottom panel) excess. The solid line is a sum of the fit to the compact
  source and the fit to the excess. At 217 GHz, since we have SPT flux
    density measurements for the full DSFG sample, we obtain the
    clustered component by fixing the normalization of the compact
    source component in the fit to the mean SPT flux density of the sample, as described in
    Sec.~\ref{subsec:results_planck}.
\label{fig:radial_profiles}}
\end{figure*}

\subsubsection{Components of the \Planck\ stack: radial profiles}
\label{subsubsec:radial_profiles}

In Fig.~\ref{fig:radial_profiles}, we show the azimuthally-averaged intensity profiles 
(centred at the position of the compact source) of: (1) the original
stacked map; (2) the compact source after fitting to the source using the formalism in
Appendix~\ref{app:planck_stack_formalism}; and (3) the \Planck\ excess
after removing the source from the stacked map. The aperture
photometry flux densities we quote in this work (e.g., Fig.~\ref{fig:bootstrap_bright}
  and the black line in Fig.~\ref{fig:planck_sed}) are in fact the cumulative flux
densities obtained by integrating profile (1) within a $3\parcm5$
radius aperture. For each component of the stack, the uncertainties come from the bootstraps at each frequency.

If the excess emission measured by \Planck\ is indeed associated with
the SPT lensing halos at $z\sim1$ that are along the line of sight to
the high redshift compact source and if that
excess emission originates from only the lensing halos, we would only
detect this emission within the FWHM of the compact source
profile (corresponding to a radius of $\sim2.5\arcmin$ at 857 GHz). Instead, the
radial profiles suggest that the excess emission is extended on a
larger angular scale than that of the high redshift compact
source. It follows that the excess emission would, in this case, also extend beyond 
the foreground lensing halo that is between the observer and the high
redshift compact source. In particular, at 857\,GHz, where we observe the largest magnitude of
the excess emission (Fig.~\ref{fig:excess_sed}), we detect that
emission out to a radius of 3.5$\arcmin$ from
the compact source, at $2\sigma$ significance (beyond this radius, the
significance of the detection decreases with increasing radius). This suggests
that the excess emission could have a significant contribution from galaxies in neighbouring
halos that surround the lensing halos. A theoretical prediction of the
\Planck\ excess should therefore take the contribution of these 
neighbouring halos into account (as we will do in Sec.~\ref{sec:modeling}).

\begin{figure}
\centerline{\rotatebox{0}{
\includegraphics[width=0.55\textwidth]{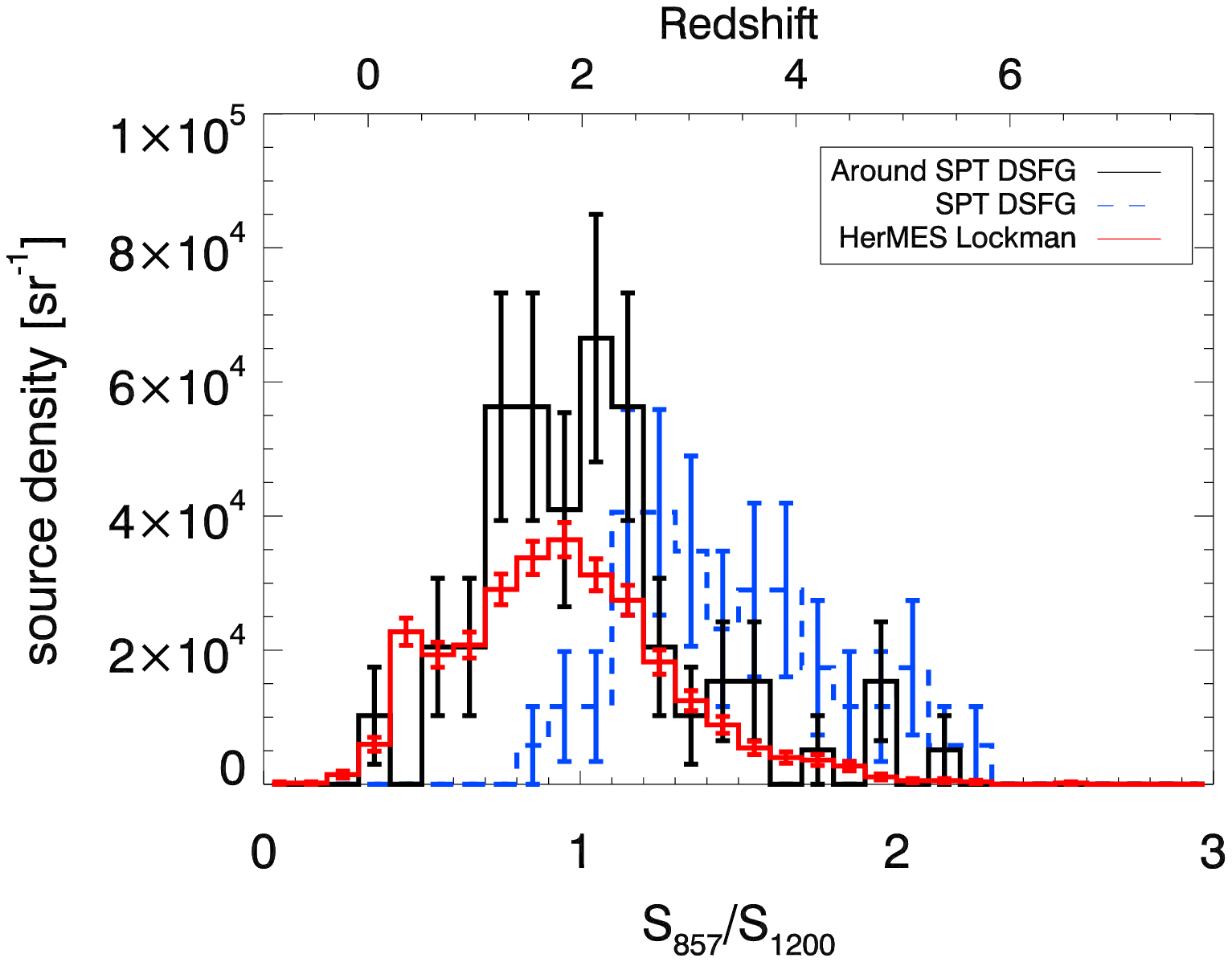}
}}
\caption{Resolving the excess with \textit{Herschel}/SPIRE: Number density
  (within a $3\parcm5$ radius from the position of the SPT DSFG)  of detected sources by $S_{857}/S_{1200}$
  colour bin. The sources considered are (a) sources detected at $S_{857}>50\,{\rm mJy}$ 
  around the SPT DSFGs (black); (b) all sources detected at $S_{857}>50\,{\rm mJy}$ in the HerMES Lockman-SWIRE field 
  (red); and (c) the SPT DSFGs themselves (blue). The horizontal axis on top
  shows the estimated redshift derived from the colours using a set of
  star-forming templates from \citet{magdis2012} (it should be noted that this
  photometric redshift estimate is model-dependent).
\label{fig:herschel_numberdensity}}
\end{figure}

\subsection{The sources contributing to the \Planck\ excess}
\label{subsec:results_herschel}

We use the \textit{Herschel}/SPIRE observations 
to probe the sources of the excess signal measured by \Planck. 
The source detection and photometry are described in
Sect.~\ref{methods:herschel} and Appendix~\ref{app:montecarlo_herschel}.
We first investigate if there is a statistical excess of such sources around the
SPT DSFGs relative to a Poisson distribution of sources.

We focus on only high significance detections ($S_{857}>50\,{\rm mJy}$),
measuring the number densities of three types of sources: (1)
$n_{\rm neighbours}$ for sources within $3\parcm5$ of the DSFG; (2) $n_{\rm null}$ for sources detected at the same significance
($S_{857}>50\,{\rm mJy}$) in the larger HerMES Lockman-SWIRE field; and
(3) $n_{\rm DSFG}$ for the DSFGs themselves.
 
The computation of the source densities is described fully in
Appendix~\ref{app:herschel_source_densities}. 
In order to determine if such a clustering of sources is associated
with the SPT DSFGs or with foreground structures along the line of sight to
the DSFGs, we measure the variation of the
number density of these three types of detected sources (DSFG neighbours,
HerMES Lockman-SWIRE sources and the DSFGs themselves) as a function
of their $S_{1200}/S_{857}$ colours. The result is shown in
Fig.~\ref{fig:herschel_numberdensity}. The top horizontal axis of the
same figure represents the photometric redshifts estimated from the sub-mm colours using 
the B12 effective template SEDs described in
Sect.~\ref{subsubsec:excess_pz}. We make the following observations:

\begin{itemize}
\item There is a significant excess of sources within $3\parcm5$ of the DSFG, compared to the null test (using all other
  sources in the HerMES Lockman-SWIRE field which are detected at the same
  significance). The excess can also be expressed as the ratio of the
  mean density of the DSFG neighbours within $3\parcm5$ of the DSFGs to the mean
  density of the sources in the entire HerMES Lockman-SWIRE field. We
  obtain a ratio of $2.18\pm0.15$ at 1200\,GHz and $1.76\pm0.19$ at
  857\,GHz. The excess extends over a broad range of photometric redshifts from $z\sim1$ to
  $z\sim2$. This is consistent with the combined spectroscopic and
  photometric $n(z)$ for the lens galaxies. For the lens galaxies themselves, multi-wavelength
  imaging and spectroscopy has been obtained for more than 50 of the
  lensed SPT DSFGs \citep{rotermund2014}.  Spectroscopic
  redshifts are complete for $\sim70\%$ of the sample, suggesting the
  median redshift of the lensing halos is at least $\langle z
  \rangle=0.6$, and with photometric redshifts for the remainder of
  the (optically fainter) sample, the median is close to the estimated
  SPT lens redshift of $z\sim1$ in \citet{hezaveh2011}.
\item On average, the sources clustered around the SPT DSFGs are
  significantly bluer (in sub-mm colours) than the DSFGs themselves. Our SED fits suggest that these sources
  are at $z\sim1-2$ whereas the DSFGs themselves are at $z>2$,
  consistent with $n(z)$ of the DSFG sample reported in \citet{weiss2013}. 
\end{itemize}

We also estimate the mean colours of the three types of sources in Fig.~\ref{fig:herschel_numberdensity}. We compute the mean colour of the sources
responsible for the \textit{Herschel} excess $C_{\rm excess}$ according to:

\begin{equation}
C_{\rm excess} = \frac{\sum{C_{{\rm X},i}(N_{{\rm X},i} - N_{{\rm null},i} )} }{\sum (N_{{\rm X},i} - N_{{\rm null},i} ) },
\end{equation}
 
\noindent where $C_{{\rm X},i}$ is the $S_{857}/S_{1200}$ colour of the sources around
the DSFG in each interval of colour $i$ in
Fig.~\ref{fig:herschel_numberdensity}, 
$N_{{\rm X},i}$ is the number of such
sources in that same colour interval, and $N_{{\rm null},i}$ is the number of HerMES Lockman-SWIRE
sources in that same colour interval. The mean colours are $\langle
S_{857}/S_{1200}\rangle=0.98\pm0.01$\,for the sources in the
Lockman-SWIRE field, $\langle S_{857}/S_{1200}\rangle=1.10\pm0.13$ for
$C_{\rm excess}$ and $\langle S_{857}/S_{1200}\rangle=1.47\pm0.05$ for
the DSFGs. We check that cosmic variance has a negligible effect on the uncertainties in the number densities in each bin of colour in Fig.~\ref{fig:herschel_numberdensity} by performing bootstrap
  realisations over the SPIRE fields around each SPT DSFG. The
  median ratio of the standard deviation in the
number density over the bootstrap realisations to the Poisson
uncertainty is 0.96. The mean colours are also dominated by the
Poisson errors and not the cosmic variance. The sources responsible for the excess observed by \textit{Herschel}
thus have the same mean colour, and hence probably the same redshift, 
as the low redshift sources in HerMES Lockman-SWIRE. However,
those sources clustered around the DSFGs are significantly bluer
(by $\langle S_{857}/S_{1200}\rangle=0.4$ on average) compared to the DSFGs.


We also estimate a mean excess in flux density, $S_{\rm excess}$, of the
detected sources around the DSFGs relative to all the other
detected sources in the HerMES Lockman-SWIRE field, according to

\begin{equation}
\label{eq:herschel_Sexcess}
S_{\rm excess} = \langle S_{\rm neighbours}\rangle - \langle S_{\rm null}\rangle, 
\end{equation}

\noindent where $\langle S_{\rm neighbours}\rangle$ is the mean flux density of the detected sources that are
within $3\parcm5$ of the SPT DSFGs and $\langle S_{\rm null}\rangle$ is the mean flux
density of all the sources detected within an aperture of $3\parcm5$
radius in the HerMES Lockman-SWIRE field, with:

\begin{equation}
\langle S_{\rm neighbours}\rangle = \frac{\sum{S_{\rm neighbours}}}{N_{\rm DSFG}},
\end{equation}

\noindent where $N_{\rm DSFG}$ is the number of SPIRE maps of the SPT
DSFGs (62 in practice, see Table~\ref{table:sample}) and 

\begin{equation}
\langle S_{\rm null}\rangle = \frac{\sum{S_{\rm null}} \times \pi \times
  (3\parcm5)^2}{A_{\rm L}}, 
\end{equation}

\noindent where $A_{\rm L}$ is the total area of the Lockman-SWIRE field
in square arcminutes.

We obtain $S_{\rm excess}$ of $130\pm10$\,mJy and
$43\pm5$\,mJy at 1200\,GHz and 857\,GHz, respectively. 
It is important to note that the \textit{Herschel} observations (with
$S_{857}>50\,{\rm mJy}$) thus recover
approximately 20\% of the \Planck\ excess we measure at
857\,GHz, and about 45\% at 1200\,GHz (see
Fig.~\ref{fig:excess_sed}). If we assume $z=1.15$ for the lenses
\citep{hezaveh2011}, this resolved excess emission at 857\,GHz
translates into a mean $L_{\rm IR}$ of 
$(2.2\pm0.2)\times10^{12}\,\Lsolar$ and a mean excess SFR of $(370\pm40)\,\Msolar\,{\rm yr}^{-1}$ per resolved
source. This suggests that the environments around these massive $z\sim1$
lensing halos host active star formation and that the galaxies in
these environments that are responsible for this excess FIR emission are ultra-luminous infrared galaxies (ULIRGs). 

To recover the full excess, we would require deeper imaging at a
higher angular resolution (e.g., with ALMA).  
It is expected that \textit{Herschel} detects this fraction of the
extragalactic sources contributing to the CIB \citep{bethermin2012a} and the excess we measure with
SPIRE (relative to random regions in the Universe) arises from bright,
star-forming galaxies which are associated mainly with the foreground
lensing halos of the SPT DSFGs. Finally, it should be noted that
neither in the \Planck\ nor \textit{Herschel} analysis is it possible to
pinpoint the sub-mm contribution from the lens galaxy itself. However,
the lens galaxies are largely passive elliptical galaxies with no
strong star formation \citep{hezaveh2013} and their contribution to $S_{\rm excess}$ is expected to be quite small.


\section{Modeling the \textit{Planck} excess}
\label{sec:modeling}

We have shown a large-scale excess of sub-mm emission 
that is detected out to a distance of $\sim3.5\arcmin$ from the SPT DSFGs. We cannot interpret it as a classical
clustering signal between the high redshift sources and their neighbours \citep{bethermin2010b,bethermin2012a}, because the colour of this excess
indicates that the signal corresponds to objects at $z<2$ (see
Sect.~\ref{subsec:results_planck}) whereas the SPT DSFGs lie mostly at
$z\sim2$--$6$ \citep{vieira2013,weiss2013}. However, both
theoretical models \citep{negrello2007,bethermin2011a,hezaveh2011}
and observations \citep{vieira2013} predict that the large
majority of bright SPT DSFGs are lensed. Consequently, there
must be relatively massive dark matter halos along the line of sight
to the SPT sources. \citet{hezaveh2011} predict a median mass of the
lensing halos of  $10^{13.3}\,\Msolar$. These massive halos are
also strongly clustered \citep{mo1996,sheth1999,sheth2001}. 
The excess we measure with \Planck\ could thus be 
the infrared emission coming mostly from galaxies which are in the neighbouring halos of the lenses.

The exact computation of the excess from a model of galaxy evolution
that links the star formation process to the dark matter halos is beyond the scope of this paper.
However, an estimate of the expected \textit{Planck} excess can be
performed with a more simplified computation. 
We use the halo model which assumes that all dark matter is bound in halos and provides a
formalism for describing the clustering statistics of halos and
galaxies \citep[see][and references therein]{cooray2002}. In this
model, the one-halo term (due to distinct baryonic mass
elements that lie within the same dark matter halo) dominates the
correlation function on scales smaller than the virial radii of halos,
while the two-halo term (due to baryonic mass elements in distinct pairs of
halos) dominates the correlation function on larger scales. 
The halo occupation distribution \citep[HOD, see][]{berlind2003} 
describes the clustering of galaxies within the halos -- it is the probability that a
halo of fixed virial mass hosts $N_{gal}$ galaxies. 
A standard approach to the HOD is to consider two populations of
galaxies in the halos: central galaxies located at the centre of the host halo,
and satellite galaxies distributed throughout the halo. 
In the context of the SPT lenses and their environments, 
the one-halo term thus takes into account the
excess signal coming from the satellite galaxies within the lensing
halo. The two-halo term accounts for the excess signal arising
from clustering with galaxies in neighbouring halos. 
The use of the two-halo term is justified here because the \Planck\ excess
emission we observe extends out to $3\parcm5$ from the DSFG,
corresponding to a physical distance of 1.7\,Mpc from the lensing halo at
$z\sim1$. 

We start by computing the angular auto-correlation function $w_{\rm
  lens}(\theta)$ of $10^{13.3}\,\Msolar$ halos assuming the redshift
distribution given by the \citet{hezaveh2011} model (median $z = 1.15$,
${\rm FWHM}= 1.53$). The computation is performed using the {\tt PMClib} tools
\citep{kilbinger2011,coupon2012}. We first estimate the two-halo term
contribution by computing the HOD assuming no satellites. 
The cross-correlation function $\Psi$($\theta$) between the lensing
 halo and the halo hosting the neighbouring galaxies is then
 $\Psi(\theta) = b_{\rm CIB}/b_{\rm lens}\,w_{\rm lens}(\theta)$,
 where $b_{\rm CIB}$ is the effective bias of sources responsible for
 the CIB, thus tracing galaxies in the neighbouring halos, and has a
 value of 2.4 at 857\,GHz \citep{viero2009}, $w_{\rm lens}(\theta)$
 has a typical value of 0.029 at $\theta=5\arcmin$, and
$b_{\rm lens}$ is the mean bias of the lensing halos. A mean bias of $b_{\rm lens}=3.6$ is
 used for the median halo mass at the median redshift of the lenses as
 predicted by \citet{hezaveh2011}. The simple conversion above comes from the fact that $\Psi \propto b_{\rm CIB} \times b_{\rm lens}$ when $w_{\rm lens} \propto b_{\rm lens}^2$ \citep{cooray2002}, in the approximation that the redshift distributions of the two components are similar. This is a fair assumption here as \citet{bethermin2012a} showed that the median redshift of the CIB at 857\,GHz is 1.2.

From the auto-correlation function, we can compute the mean number excess, $e$, of infrared galaxies around the lensing halos:
\begin{equation}
e = \int_{\theta=0}^{3\parcm5} \Psi(\theta) \theta d\theta.
\end{equation}

We find an excess in the number density of galaxies of 2.3\%. The total
flux density of all galaxies at 857\,GHz in a $3\parcm5$ radius can be
computed from the total contribution of galaxies to the CIB within this
area, which is estimated in \citet{bethermin2012a} to be 4300\,mJy --
the measured \Planck\ excess at 857\,GHz corresponds to 6$\%$ of this
total contribution to the CIB within the same radius. The expected \textit{Planck} signal from neighbouring halos (the 2-halo term) is thus $0.023\times4300=99$\,mJy.\\

Having computed the contribution from galaxies hosted by neighbouring halos of
the lensing halos, we then compute the one-halo term contribution from galaxies inside the lensing halo
itself, using a different formalism. We assume a standard
halo-mass-to-infrared-light ratio estimated from abundance matching
\citep{bethermin2012c,bethermin2012b} and the satellite mass function of \citet{tinker2010}. 
By contrast with the two-halo term computation, here we consider both
central and satellite galaxies in the lensing halo. 
For a halo of $10^{13.3}\,\Msolar$  at $z=1.15$, we find a total flux density from the central and satellite galaxies
in the lensing halo of 20\,mJy. These predictions are upper limits because the model neglects the
environmental quenching of satellites around massive galaxies. The
total expected contribution of both the one-halo and two-halo
terms is thus 119\,mJy at 857\,GHz. The prediction from this
relatively simple model is in broad agreement with the
\Planck\ measurement of the excess ($210\pm78$\,mJy at 857\,GHz). In
fact, there is a weak indication that the measured value is higher
than the model prediction, due to, perhaps, enhanced star formation that could originate from the dense
environments around the lensing halos, but the \Planck\ signal does
not have sufficient signal-to-noise to confirm this. 
Finally, we also determine how sensitive the predicted amplitude of
the emission is to the assumed halo mass. We obtain 50\,mJy (one-halo term) and 148\,mJy
(two-halo term) for a halo mass of $10^{13.8}\,\Msolar$, giving a total
predicted excess of 200\,mJy for $10^{13.8}\,\Msolar$ halos.   
We obtain 8\,mJy (one-halo term) and 70\,mJy
(two-halo term) for a halo mass of $10^{12.8}\,\Msolar$, giving a total
predicted excess of 80\,mJy for $10^{12.8}\,\Msolar$ halos.

\section{Discussion}
\label{sec:discussion}

Our results support the picture of active star
formation proceeding in dense environments at $z\sim1$. 
Using a simple model that connects star formation to dark matter
halos, we predict that most of this excess emission (around $80\%$) that is
detected by \Planck\ should arise from galaxies in the neighboring halos 
of the foreground lensing halos (the two-halo term in the context of
the halo model). A proportion of the excess emission measured by \Planck\ ($20\%$ at 
857\,GHz and $45\%$ at 1200\,GHz) is associated with individual 
sources detected by \textit{Herschel}. The sources that contribute to
this resolved excess are consistent with being ULIRGs ($L_{\rm
  IR}>10^{12}\,\Lsolar$). The remainder of the excess FIR emission measured by \Planck\ which is
not resolved by \textit{Herschel} must therefore come from an excess
of fainter infrared galaxies ($L_{\rm IR}<10^{12}\,\Lsolar$) at $z\sim1$ that are in these dense environments. 

Several studies \cite[e.g.,][]{noble2012} report 
an excess in the number densities of sub-mm galaxies in mass-biased regions of the $z\gtrsim1$
  Universe, relative to blank fields. Although the number statistics
  are low, surveys towards $z\sim1$ clusters
  \citep[e.g.,][]{best2002,webb2005} suggest that the optical Butcher–-Oemler
  effect (where a population of blue, star-forming galaxies appears in
  many $z>0.3$ clusters) is also
observed at sub-mm wavelengths. These studies also suggest 
that if the DSFGs responsible for this excess are confirmed to be at
the same redshift as the $z\sim1$ clusters, their SFRs would be
consistent with those of ULIRGs.

Our results are qualitatively consistent with other studies
that find active star formation proceeding in dense environments at
$z$$\sim$$1$. \citet{brodwin2013} investigated star-forming properties of galaxy
clusters at $1<z<1.5$ and found extensive star formation increasing toward the centres of clusters. 
\citet{alberts2014} showed that the SFR in clusters grows more rapidly
with increasing redshift than it does in the field, and surpasses the
field values around $z\sim1.4$. \citet{feruglio2010} found that although the ULIRG$+$LIRG fraction decreases
with increasing galaxy density up to $z$$\sim$$1$, the dependence on
density flattens from $z=0.4$ to $z=1$. They observed that a large
fraction of highly star-forming LIRGs is still present in the most dense environments at
$z$$\sim$$1$. The dense environments at $z$$\sim$$1$, including those
associated with the SPT lensing halos that we probe in this study, 
may well be the progenitors of the massive galaxies found in the
centres of clusters at $z\sim0$. 

An optical follow-up study of the lens environments will investigate the LIRG hypothesis in more detail. \citet{rotermund2014}
have already used spectroscopic and photometric studies to
constrain the $N(z)$ of the SPT lensing halos ($\langle z \rangle>0.6$), and
have studied the relative overdensities surrounding the lensing
galaxies. However, an analysis of star forming galaxies in these environments has yet to
be carried out. Finally, we note that the \Planck\
survey itself will be able to find overdensities at $z\gtrsim2$ across the full sub-mm
sky by selecting the coldest fluctuations of the CIB \citep{dole2014}.

\section{Conclusions}
\label{sec:conclusions}

We stack \Planck\ HFI maps at the locations of DSFGs identified in SPT
data. The stack provides an ensemble average of the flux density of
the background DSFGs, the foreground lensing halos at $z\sim1$, and
the surrounding environments. Though the SPT DSFGs lie at much higher redshift ($z \sim
2-6$), they are angularly correlated with massive ($\sim10^{13}\Msolar$) dark matter  halos  at $z \sim 1$
through strong gravitational lensing. We isolate a clustered component which extends to large
  angular scales in the stack and demonstrate that it originates from sub-mm emission from star
  formation in these environments. We exploit \Planck's wide
  frequency coverage to estimate a photometric redshift for the
  clustered component from the far-infrared colours. We then use
  higher resolution \textit{Herschel}/SPIRE observations in order to
  study the sources in these dense environments that contribute to the clustering signal. Our results can be summarized
  as follows.

\begin{itemize}

\item We find a mean excess of star
formation rate (SFR) compared to the field, 
 of $(2700\pm700)\,\Msolar\,{\rm yr}^{-1}$ from
all galaxies contributing to the clustering signal within a radius of
$3\parcm5$ from the positions of the SPT DSFGs. The sources responsible for the clustering signal are galaxies
  clustered within about $2$\,Mpc around the foreground lensing halo at
  $z\sim1$. The magnitude of the measured \Planck\ excess due to the
  clustered component ($210\pm78$\,mJy at 857\,GHz) broadly agrees with the prediction of a model of the CIB that links
  infrared luminosities with dark matter halos. The measured excess at
  857\,GHz corresponds to approximately $5\%$ of the total contribution of all galaxies to the
  CIB within a $3\parcm5$ radius. The model predicts
  that the excess emission (and hence star formation) should be
  dominated (around $80\%$) by the two-halo term contribution, due to
  galaxies in the neighbouring halos which are clustered around the
  lensing halo itself. 

\item A fraction (approximately $20\%$ at 857\,GHz with $S_{857}>50$\,mJy) of
  the excess emission from these dense $z\sim1$ environments is
  resolved by \textit{Herschel}. The sources contributing to this
  resolved excess are highly star-forming ULIRGs
  ($L\sim10^{12.5}\,\Lsolar$). The mean excess of SFR, relative to the
  field, due to these detected sources is $370\pm40\,\Msolar\,{\rm yr}^{-1}$ per resolved source. 
The remainder of excess star formation could originate from fainter LIRGs that
are in highly dense regions within the neighbouring halos. The overall
picture therefore suggests that these dense environments at $z\sim1$ are still actively
forming stars. This is qualitatively consistent with the SFR-density relation reversing at $z\sim1$ when compared to $z\sim0$.

\item Our work shows that in an experiment where the beam FWHM is
  comparable or larger than the angular scale of the clustering signal, 
  the stacked flux density estimates of high redshift lensed DSFGs will have significant contributions from galaxies clustered around the
  lensing halos that are along the line-of-sight to the background
  lensed galaxies. The relative clustering contamination has a clear dependence on
  frequency: in \Planck\ data, we measure it to be 0.1$\pm$0.6 at 217 GHz, 0.2$\pm$0.2 at 353 GHz,
  0.5$\pm$0.2 at 545 GHz, and 1.1$\pm$0.4 at 857 GHz. This
  contamination should be taken into account in order to obtain the
  correct flux densities of the background galaxies with \Planck\ data. 

\end{itemize}

\section*{Acknowledgments}

We thank the anonymous referee for their valuable comments. 
The South Pole Telescope is supported by the National Science Foundation through
grant PLR-1248097. Partial support is also provided by the NSF Physics
Frontier Center grant PHY-1125897 to the Kavli Institute of
Cosmological Physics at the University of Chicago, the Kavli
Foundation and the Gordon and Betty Moore Foundation grant GBMF 947. 
This material is based on work supported by the U.S. National Science
Foundation under grant No. AST-1312950. Based on observations obtained with \Planck\
(\url{http://www.esa.int/Planck}), an ESA science mission with
instruments and contributions directly funded by ESA Member States,
NASA, and Canada. The development of \Planck\ has been supported by: ESA; CNES
and CNRS/INSU-IN2P3-INP (France); ASI, CNR, and INAF (Italy); NASA and
DoE (USA); STFC and UKSA (UK); CSIC, MICINN and JA (Spain); Tekes, AoF
and CSC (Finland); DLR and MPG (Germany); CSA (Canada); DTU Space
(Denmark); SER/SSO (Switzerland); RCN (Norway); SFI (Ireland);
FCT/MCTES (Portugal); and PRACE (EU). A description of the Planck
Collaboration and a list of its members, including the technical or
scientific activities in which they have been involved, can be found
at \url{http://www.rssd.esa.int/index.php?project=PLANCK&page=PlanckCollaboration}. This
paper makes use of the following ALMA data: ADS/JAO.ALMA\#2011.0.00957.S. ALMA is a
partnership of ESO (representing its member states), NSF (USA) and NINS (Japan), together
with NRC (Canada) and NSC and ASIAA (Taiwan), in cooperation with the
Republic of Chile. The Joint ALMA Observatory is operated by ESO,
AUI/NRAO and NAOJ. APEX is a collaboration between the
Max-Planck-Institut f{\"u}r Radioastronomie, the European Southern
Observatory, and the Onsala Space Observatory. This work is based in part on
observations made with \textit{Herschel}, a European Space Agency
Cornerstone Mission with significant participation by NASA, and
supported through an award issued by JPL/Caltech for
OT2\_jvieira\_5. NW acknowledges support from the Beecroft
Institute for Particle Astrophysics and Cosmology and previous support
from the Centre National d'{\'E}tudes Spatiales (CNES). 
Part of the research described in this paper was carried out at the Jet
Propulsion Laboratory, California Institute of Technology, under a
contract with the National Aeronautics and Space Administration. 
MS was supported for this research through a stipend
from the International Max Planck Research School (IMPRS) for
Astronomy and Astrophysics at the Universities of Bonn and
Cologne. IFC acknowledges the support of grant ANR-11-BS56-015. 
JGN acknowledges financial support from the Spanish CSIC for a JAE-DOC
fellowship, co-funded by the European Social Fund, by the Spanish
Ministerio de Ciencia e Innovacion, AYA2012-39475-C02-01, and
Consolider-Ingenio 2010, CSD2010-00064, projects. NW thanks
B.~Partridge, J.~Delabrouille, D.~Harrison, P.~Vielva and S.~Alberts for useful comments.

\bibliographystyle{mn2e}
\bibliography{Planck_bib}

\appendix

\section{Uncertainties in the \textit{Planck}
  and IRIS stacked maps of the DSFGs}
\label{app:stack_uncertainties}

\begin{table*}
\caption{Uncertainties in the stacked \Planck\ (at
$217-857$\,GHz) and IRIS (at 3000\,GHz) maps which are co-added at
the locations of the SPT DSFGs:
(1) photometric uncertainties $\sigma_{phot}$ estimated from the
standard deviation of flux densities over 1000 iterations of stacking 65 randomly chosen
patches in the sky (dashed lines in Fig~\ref{fig:bootstrap_bright});
and (2) bootstrap uncertainties $\sigma_{\rm boot}$ computed from the standard
deviation of flux densities over 1000 bootstrap realizations of the
stacked maps of the 65 DSFGs (solid lines in Fig~\ref{fig:bootstrap_bright}).
The sample heterogeneity $\sigma_{\rm pop}$ is the intrinsic
dispersion in the DSFG population. It is estimated at each \Planck\
frequency by extrapolating the flux density dispersion at the SPT
frequency to the \Planck\ frequencies. $S_{\nu}$ is the mean flux
density from performing aperture photometry on the bootstrap realizations
of the stacked maps of the DSFGs, and $S_{\nu,compact}$ is the mean flux density of the compact source
component in the stack.
}
\centering
\begin{tabular}{c c c c c c}
\hline
Type of variance & 217\,GHz & 353\,GHz & 545\,GHz & 857\,GHz &
3000\,GHz \\
$\sigma_{\rm phot}$ [mJy] & 13$\pm$0.4 & 13$\pm$0.4 & 30$\pm$0.9 & 75$\pm$2.4 & 106$\pm$3.4\\
$\sigma_{\rm phot}/S_{\nu}$ & 0.40 & 0.11 & 0.11 & 0.19 & 0.87\\

$\sigma_{\rm pop}$ [mJy] & 1.5 & 5.4 & 8.9 & 10.0 & 0.6 \\
$\sigma_{\rm pop}/S_{\nu,compact}$& 0.05 & 0.05 & 0.05 & 0.05 & 0.05\\


$\sqrt{\sigma_{\rm phot}^2 + \sigma_{\rm pop}^2}$ & 13 & 14 & 31 & 76 & 106\\

$\sigma_{\rm boot}$ [mJy] & 16$\pm$2 & 16$\pm$2 & 31$\pm$4 & 73$\pm$10 & 85$\pm$11\\
$\sigma_{\rm boot}/S_{\nu}$ & 0.51 & 0.13 & 0.12 & 0.18 & 0.67 \\
\hline
\end{tabular}
\label{table:heterogeneity}
\end{table*}

In Table~\ref{table:heterogeneity}, we compare the uncertainties from
bootstrapping, $\sigma_{\rm boot}$, with the photometric uncertainties,
$\sigma_{\rm phot}$, derived from performing aperture photometry in
the random patches of the SPT fields. In order
to compare how close $\sigma_{\rm boot}$ and $\sigma_{\rm phot}$ are to each other, we
also compute an uncertainty on them -- these scale as $1/\sqrt(N_{sources})$ for
$\sigma_{\rm boot}$ and $1/\sqrt(N_{iter})$ for
$\sigma_{\rm phot}$, where $N_{sources}$ is the number of sources in the
stack (65) and $N_{iter}$ is the number of stacking iterations (1000).
$\sigma_{\rm phot}$ includes both the
instrumental and confusion noise. We estimate the
standard deviation of the average flux density of the stacked
population, $\sigma_{\rm pop}$ assuming that the relative scatter on
the flux density of the SPT sources does not depend on wavelength:

\begin{equation}
\label{eq:heterogeneity}
\sigma_{\rm pop} = \frac{\sigma_{\rm 220}}{\sqrt{N_{sources}}} \times \frac{S_{\nu,compact}}{S_{\rm 220}}
\end{equation}

where $S_{\rm 220}$ is the mean flux density of the DSFG sample
measured by SPT at 220 GHz, $\sigma_{\rm 220}$ is the standard
deviation of the SPT flux densities (sources detected individually) and $S_{\rm \nu, compact}$ is the mean flux
density of the compact source component in the stack. 
Table~\ref{table:heterogeneity} shows that the bootstrap uncertainties
are very close to the photometric uncertainties at 217$-$857\,GHz. 
The bootstrap uncertainties combine the photometric noise and the heterogeneity of the population \citep{bethermin2012a}:
\begin{equation}
\sigma_{\rm boot} = \sqrt{\sigma_{\rm phot}^2 + \sigma_{\rm pop}^2}
\end{equation}

Table~\ref{table:heterogeneity} shows that this intrinsic dispersion
as characterized by $\sigma_{\rm pop}$ is very small
compared to the photometric uncertainties. At 3000\,GHz, $\sigma_{\rm phot}$ is somewhat higher than $\sigma_{\rm
boot}$ (although still within $2\sigma$) due to possible complex
effects of Galactic cirrus. In general, the 353, 545 and 857\,GHz channels are cleaner than the 3000 and
217 GHz channels and allow better constraints on the properties of the DSFGs.

\section{Determining the effect of pixelisation on the FWHM in the {\tt HEALPix} maps}
\label{app:healpix_simus}

The \Planck\ HFI maps are pixelized using the HEALPix scheme at
resolution $N_\mathrm{grid} = 2048$, corresponding to $5 \times
10^7$ pixels over the full sky. This pixelation can lead to positional
offsets as large as $0\parcm5$ and can enlarge the effective beam. 
We calculate the magnitude of this effect using simulations of the
stacking analysis. Since the offsets depend on the sky position, we
begin by inserting simulated sources with the measured \Planck\ beam \citep{planck2013-p03c} at the known source locations. 
We then extract $1\deg\times1\deg$ maps centred at each source
location, stack these maps, and measure the beam FWHM in the stacked
map. The final FWHMs are $4\parcm64$, $4\parcm97$, $5\parcm10$, and
$5\parcm30$ for the \Planck\ 857, 545, 353, and 217 GHz bands
respectively, and $4\parcm61$ for the IRIS 3000\,GHz band.

\begin{figure}
\includegraphics[width=0.5\textwidth]{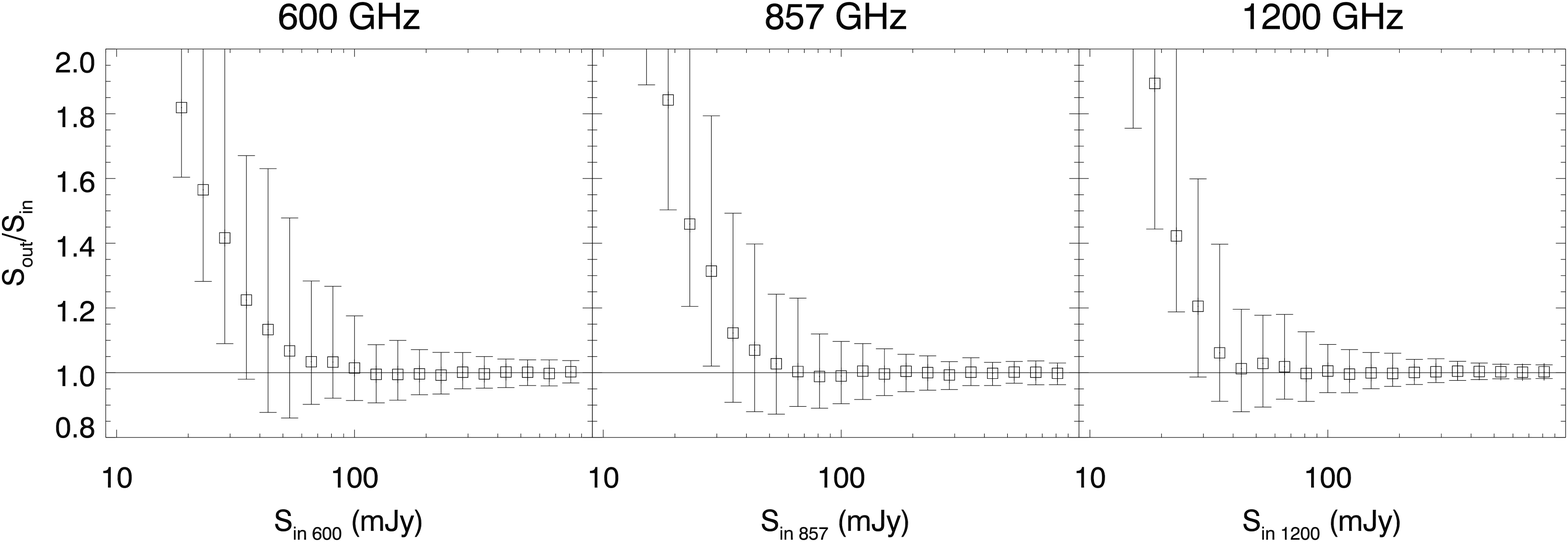}
\vspace{0.1cm}
\includegraphics[width=0.5\textwidth]{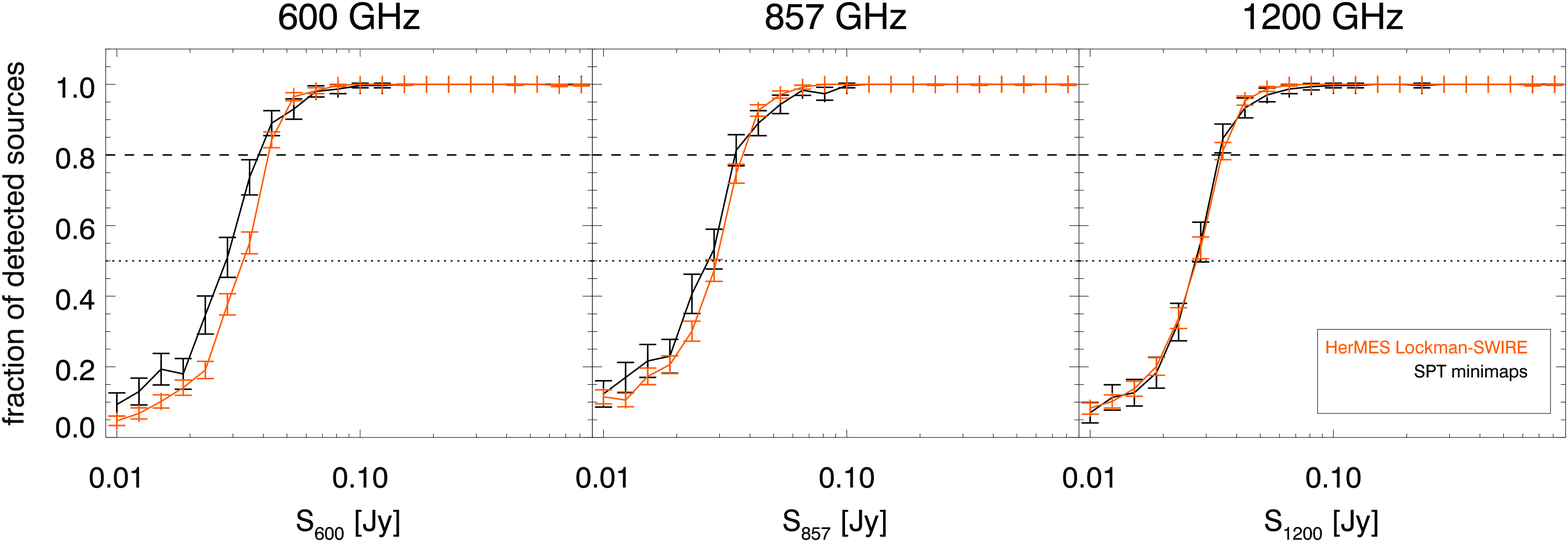}
\caption{Monte Carlo simulations with sources injected into
    the 10$\arcmin\times$10$\arcmin$ SPIRE maps of SPT DSFGs at
    600, 857, and 1200\,GHz. Top panel: ratio of output to input flux
    densities as a function of the input flux density at the each frequency. Bottom panel: fraction of
    recovered sources as a function of the input flux density of the
    sources at each frequency. This is plotted for: (a) the maps containing the
    SPT sources (black); (b) sources in the entire HerMES Lockman SWIRE
    field (orange).
\label{fig:montecarlo_herschel}}
\end{figure}

\section{Monte Carlo simulations on \textit{Herschel} maps}
\label{app:montecarlo_herschel}

We perform Monte Carlo simulations to test the robustness of the
source detection and photometry in both the
$10\arcmin\times10\arcmin$ SPIRE maps of the SPT DSFGs
and the larger HerMES Lockman-SWIRE field.
We inject sources of known flux densities at random positions into
the maps. We inject 5 sources of a given flux
density into each $10\deg\times10\deg$ map, and record the fraction of
sources that are detected. This process is repeated for source flux
densities from 10 to 1000 mJy. The same process is applied to the
larger HerMES Lockman-SWIRE field, however the number of sources is
increased to 1000. Fig.~\ref{fig:montecarlo_herschel} shows: (1) a comparison of
the input and output flux densities and (2) the completeness,
defined as the fraction of recovered sources, as a function of the
input flux densities, for both the SPIRE maps
of the SPT DSFGs and the HerMES Lockman-SWIRE field. 
The simulations for completeness show that the fraction of injected
sources that are recovered becomes $\gtrsim0.8$ at flux densities
above around $50$\,mJy for both the 1200 and 857\,GHz bands.

\section{Formalism to measure the flux density of the DSFG and excess in
  the stacked \textit{Planck} maps}
\label{app:planck_stack_formalism}

We use the formalism of \citet{bethermin2010a} to disentangle source and clustering contributions to the total flux within the \Planck\ beam. 
\citet{bethermin2012a} used this method to estimate the level of contamination due to
clustering in the deep number counts at 1200, 857 and 600\,GHz in HerMES. 
They fitted stacked images of the SPIRE sources with an
auto-correlation function (ACF) $w(\theta)$ which is convolved with
the beam function. \citet{heinis2013} have also applied this method to UV
stacking. In particular:

\begin{itemize}
\item We fit the compact source component with a 2-dimensional Gaussian profile whose width is determined by the PSF FWHM of the \Planck\ beams as described in Appendix~\ref{app:healpix_simus}.
\item We fit the clustered component around the source using an angular
  correlation function $w(\theta)\propto{\theta}^{-0.8}$ that is
  first convolved with the \Planck\ PSF FWHM at each frequency \citep[the
  exponent comes from measurements of the
  angular correlation function of galaxies, e.g.,][]{baugh1996, connolly1998}.
\item We assume a constant background level.
\end{itemize}

We define the quantity $s^2$ as the difference between the fluxes of the raw stacked \Planck\ maps and a linear combination of the above 3 profiles that are fitted to the stacked map $m_{i{,}j}$:
\begin{equation}
s^2 = \sum_{i{,}j}{(m_{ij}  -   \alpha p_{ij} + \beta c_{ij} + \gamma\times\textbf{1}_{ij}   )^2 },
\end{equation}

\noindent where $p_{ij}$ is an array containing the PSF in 2
dimensions ($i{,}j$); $c_{ij}$ is an array containing the clustering
signal and $\textbfss{1}_{ij}$ is an array containing only 1s and
represents the background (assumed to be constant). The sum runs over
all the pixels $N_{\rm pix}$ in the map. The quantities $\alpha$, $\beta$ and
$\gamma$ are normalisation constants for the flux density of the
compact source component, clustered component and the background
component, respectively. Minimizing $s^2$ with respect to $\alpha,
\beta$ and $\gamma$ leads to a simple matrix equation



\begin{equation}
\bmath{c} = \textbfss{A}\, \bmath{b},
\end{equation}

\noindent where \textbfss{A} is defined as:
\begin{equation}
\textbfss{A}  =  \left( 
\begin{array}{ccc}
\sum_{i,j}{p_{ij}^2} & \sum_{i,j}{p_{ij} c_{ij}} & \sum_{i,j}{p_{ij}} \\
\sum_{i,j}{p_{ij}c_{ij}} & \sum_{i,j}{c_{ij}^2} & \sum_{i,j}{c_{ij}} \\
\sum_{i,j}{p_{ij}} & \sum_{i,j}{c_{ij}} & N_{\rm pix} 
\end{array}
 \right) \, , 
\end{equation}

\noindent with $\bmath{b}$ defined as:

\begin{equation}
\bmath{b}  = \left( 
\begin{array}{c}
 \alpha \\ 
 \beta \\
 \gamma
\end{array}
\right) \, , 
\end{equation}

\noindent and $\bmath{c}$ defined as:

\begin{equation}
\bmath{c}  = \left( 
\begin{array}{c}
\sum_{i,j}{m_{ij} p_{ij}} \\ 
\sum_{i,j}{m_{ij} c_{ij}}  \\
\sum_{i,j}{m_{ij}}
\end{array}
\right).
\end{equation}

\noindent By inverting \textbfss{A} and solving this equation, we obtain flux
densities of each component $\alpha$, $\beta$, and $\gamma$.

\section{Tests of the \textit{Planck} clustered component}
\label{app:clustering_tests}

\subsection{Stacking simulations}
\label{app:stacking_simulations}

We generate 1000 realisations of 65 $1\deg\times1\deg$ maps (the same number
  as in the SPT DSFG sample) where each map contains a compact source component (which is at the
  centre of the map and modeled as a Gaussian with a FWHM given by the \Planck\ beams) and a
  clustered component, according to the model given in
  Appendix~\ref{app:planck_stack_formalism}. The input flux densities
  of the two components are chosen to match the measured mean flux densities given in
  Table~\ref{table:clusteringbias}. In each individual simulated map, the source and clustering
  components are added to one of
  the randomly chosen blank maps in the \Planck\ sky (discussed in
  Sect~\ref{subsec:methods:stacking} and in
  Figure~\ref{fig:bootstrap_bright}). For each of 1000 realisations of
  the stacked maps, we measure the total flux density within a
  3.5$\arcmin$ radius of the central compact source using aperture
  photometry, and we compare this flux density to the total input flux density at
  each frequency. We also apply the formalism
in Appendix~\ref{app:planck_stack_formalism} to each realisation of
the stacked maps in order to recover the flux densites of the compact
source component and the clustered component at each frequency, and we compare these with
their input flux densities. In Table~\ref{table:stacksims}, we report
the difference between the recovered mean flux density (over 1000
realisations) and the input flux density at each frequency, as a fraction of the statistical
uncertainty (given by the photometric uncertainty $\sigma_{phot}$ in each stacked map).

\begin{table*}
\caption{The systematic bias, arising from the stacking procedure, in the measured flux density of
    the high redshift compact component and the clustering
  component (obtained from the fitting method in
  Appendix~\ref{app:planck_stack_formalism}) and in the measured total
  flux density (obtained from aperture photometry). Artificial compact source
  components and clustered components are injected into blank \Planck\ and IRIS
  maps before these maps are stacked (see
  Appendix~\ref{app:stacking_simulations} for details). The difference between the mean
recovered flux density of each component (over 1000 stacking
realisations) and their true flux density at each frequency is expressed in terms
  of the photometric uncertainties $\sigma_{phot}$ in the stacked maps.
} 
\centering 
\begin{tabular}{l c c c c c} 
\hline
Frequency & 217\,GHz & 353\,GHz & 545\,GHz & 857\,GHz & 3000\,GHz \\
Systematic bias in mean flux density of all components (from aperture photometry) & 0.31$\sigma$ & 0.38$\sigma$ & 0.39$\sigma$ & 0.33$\sigma$ & 0.19$\sigma$ \\ 
Systematic bias in mean flux density of compact source component (from fit) & 0.01$\sigma$ & 0.23$\sigma$ & 0.25$\sigma$ & 0.16$\sigma$ & 0.05$\sigma$ \\ 
Systematic bias in mean flux density of clustered component (from fit) & 0.31$\sigma$ & 0.13$\sigma$ & 0.07$\sigma$ & 0.16$\sigma$ & 0.20$\sigma$ \\ 
\hline
\end{tabular} 
\label{table:stacksims} 
\end{table*}

\begin{figure}{}
\centerline{\rotatebox{0}{
\includegraphics[width=0.1\textwidth]{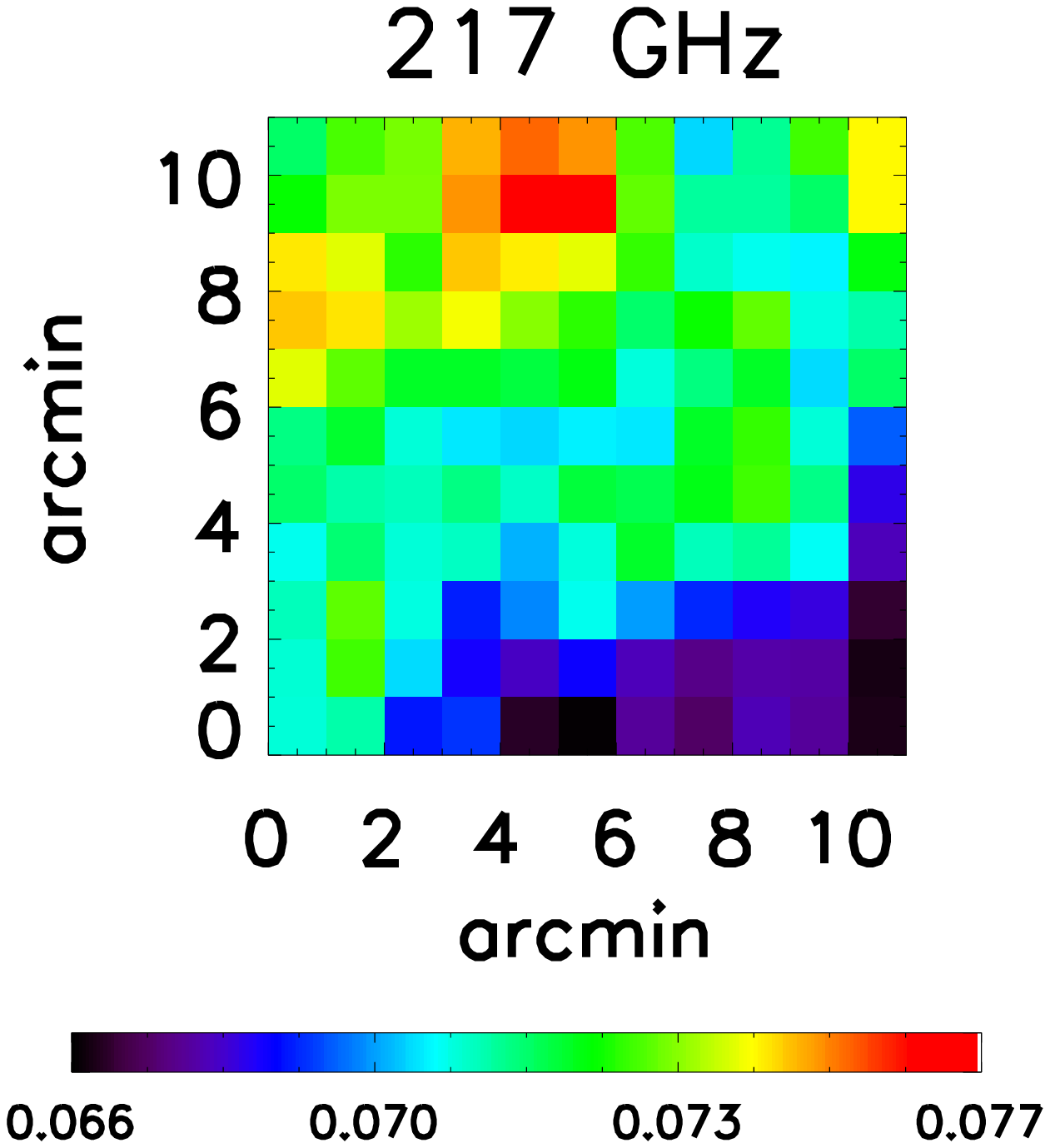}
\includegraphics[width=0.1\textwidth]{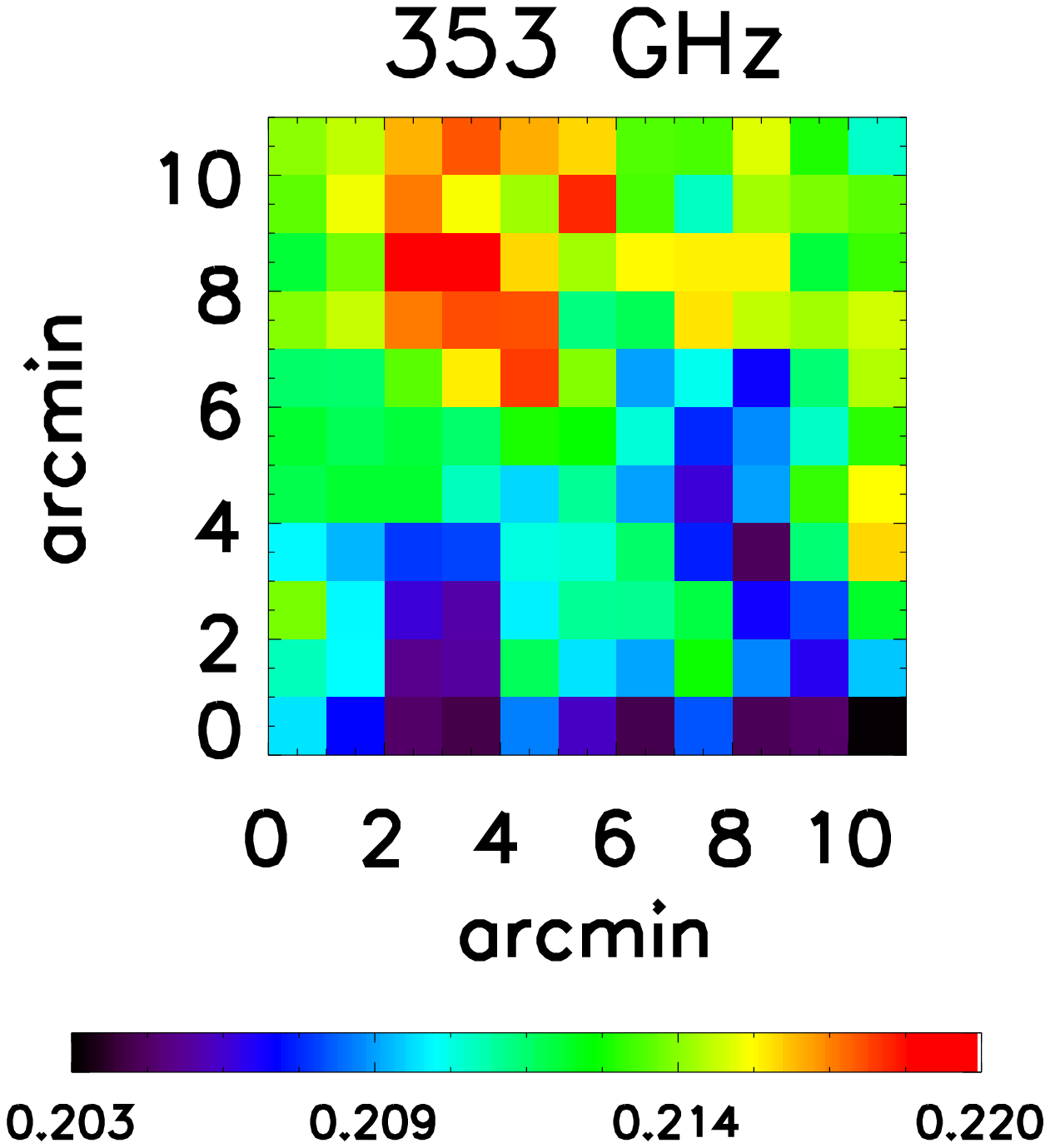}
\includegraphics[width=0.1\textwidth]{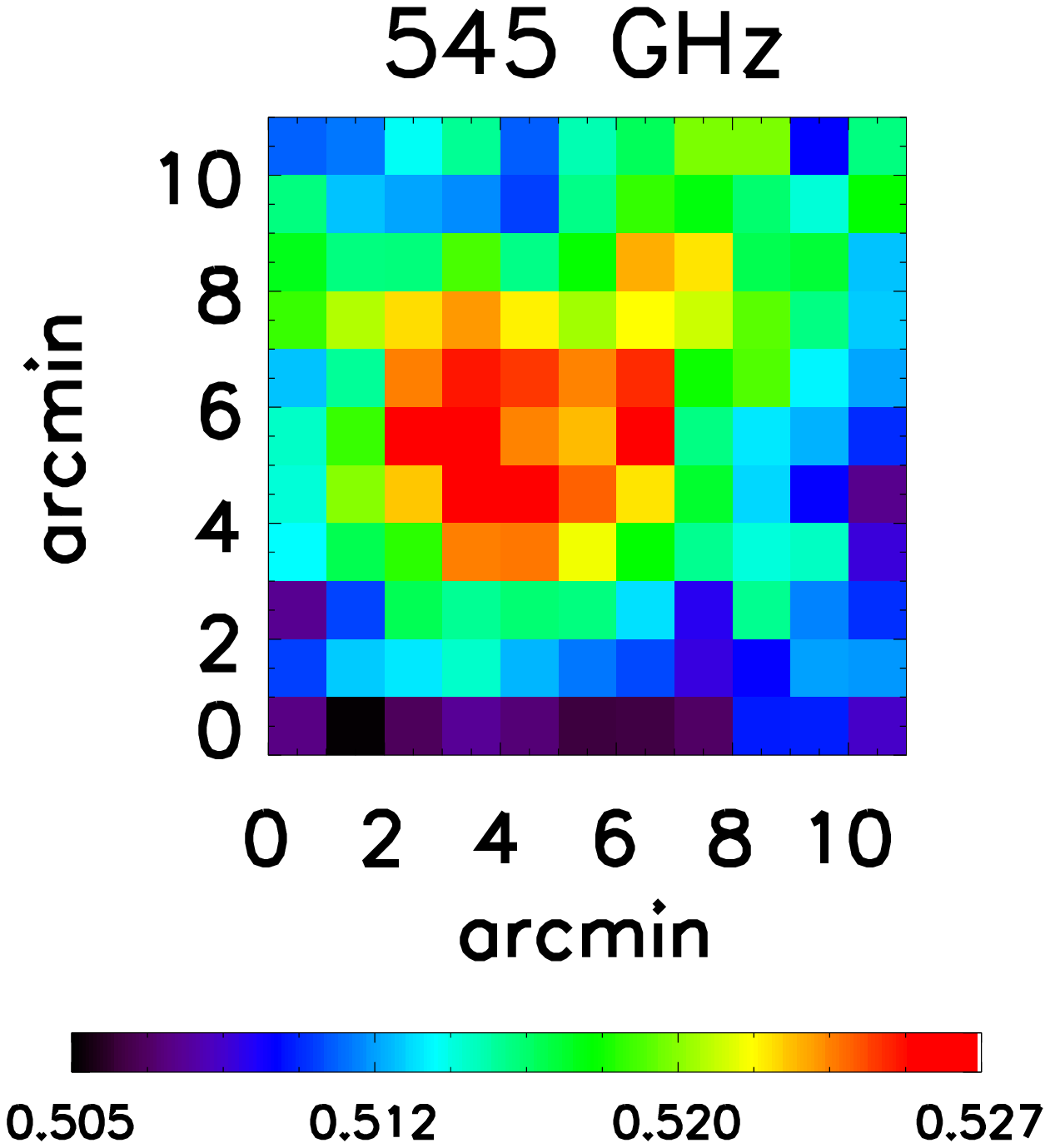}
\includegraphics[width=0.1\textwidth]{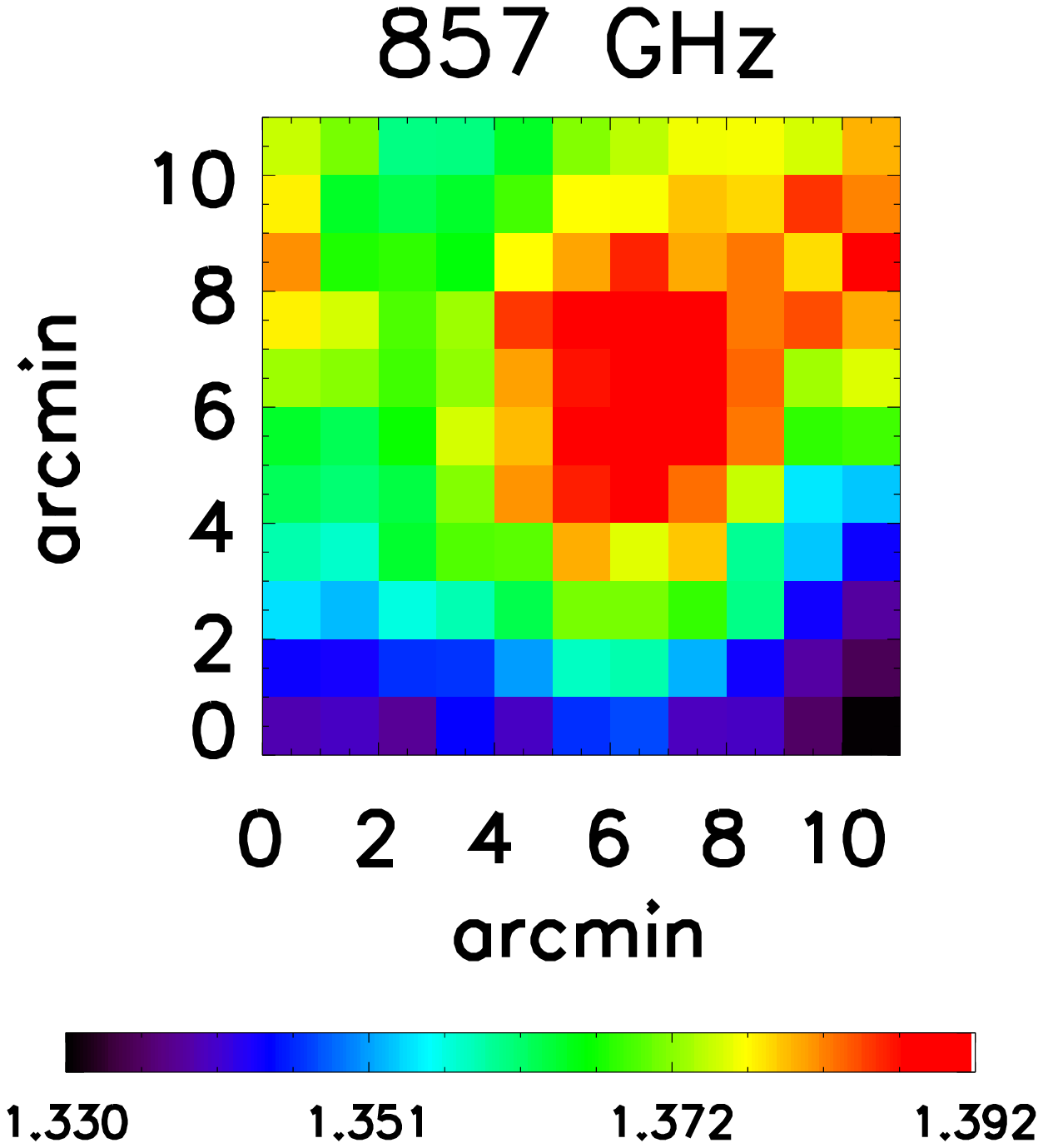}
\includegraphics[width=0.1\textwidth]{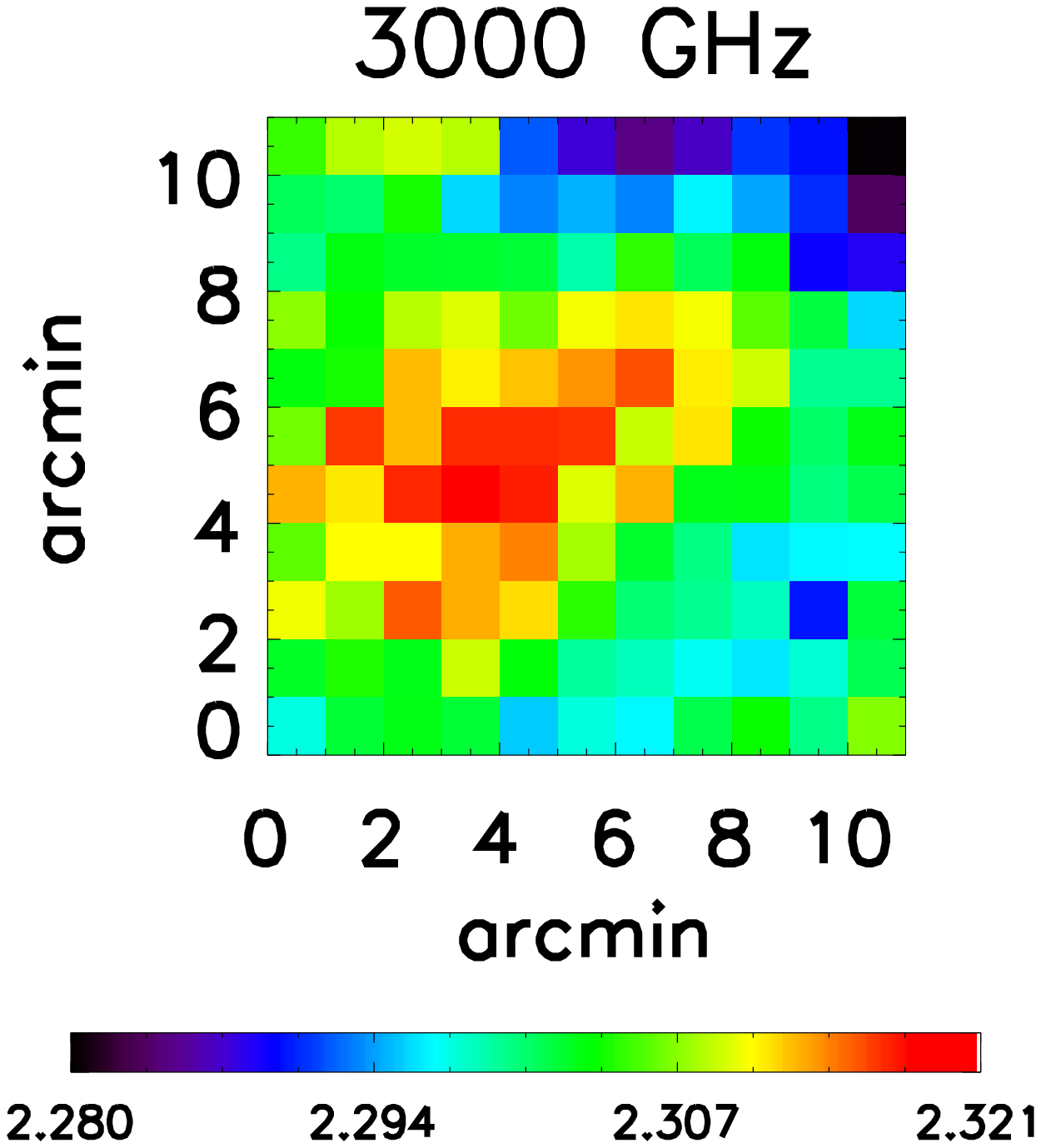}
}}   
\vspace{0.1cm}
\centerline{\rotatebox{0}{
\includegraphics[width=0.1\textwidth]{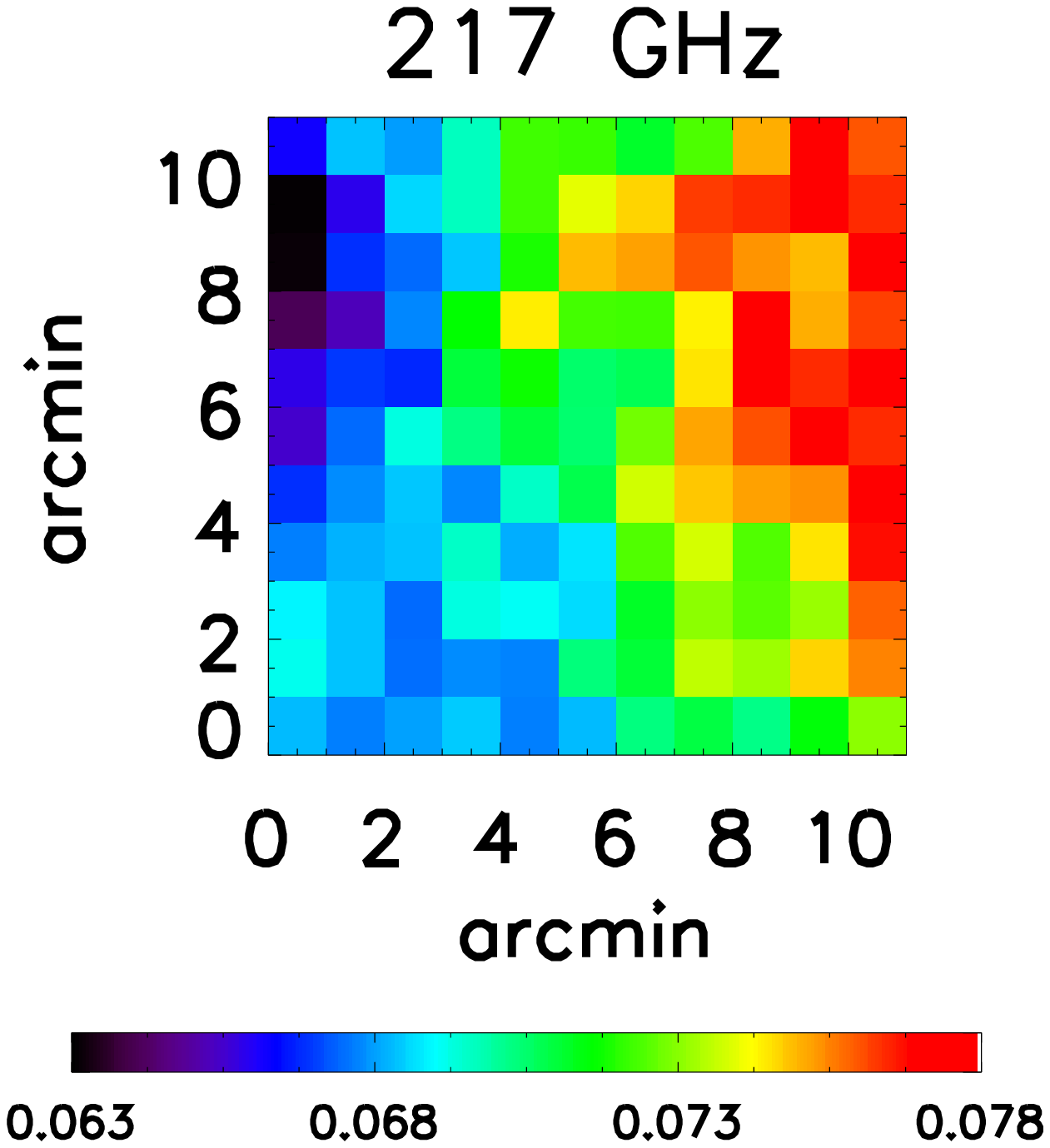}
\includegraphics[width=0.1\textwidth]{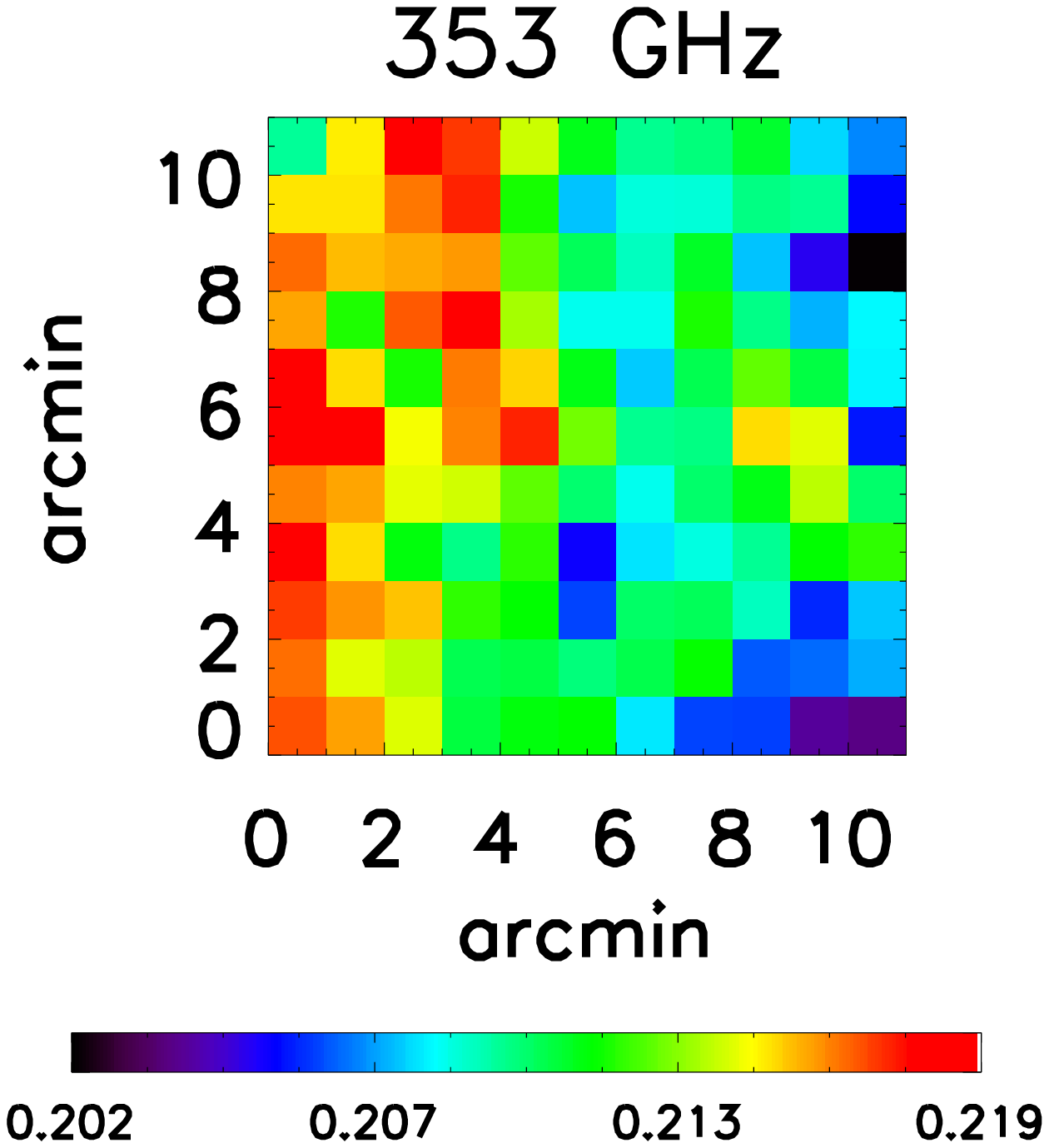}
\includegraphics[width=0.1\textwidth]{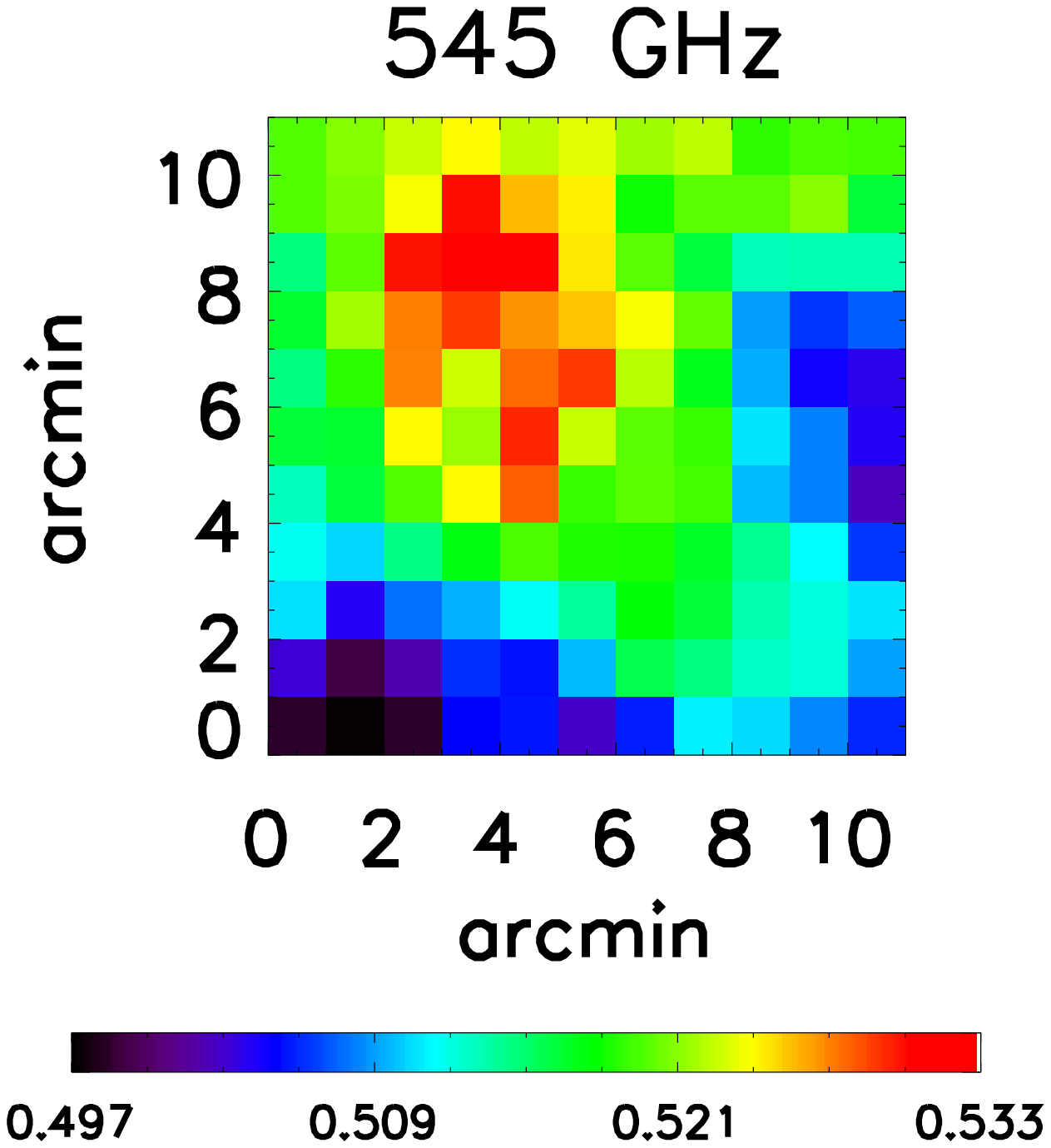}
\includegraphics[width=0.1\textwidth]{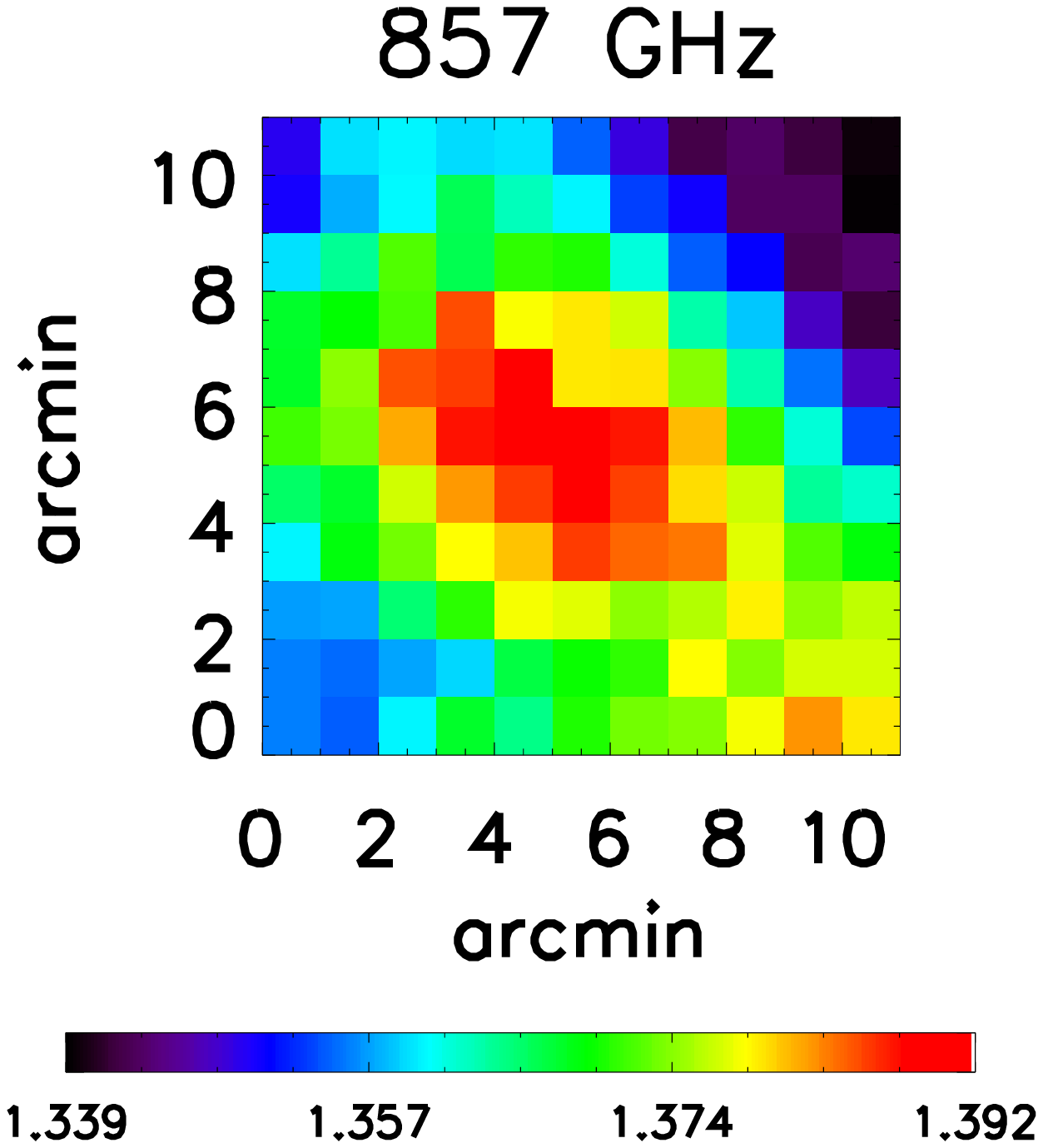}
\includegraphics[width=0.1\textwidth]{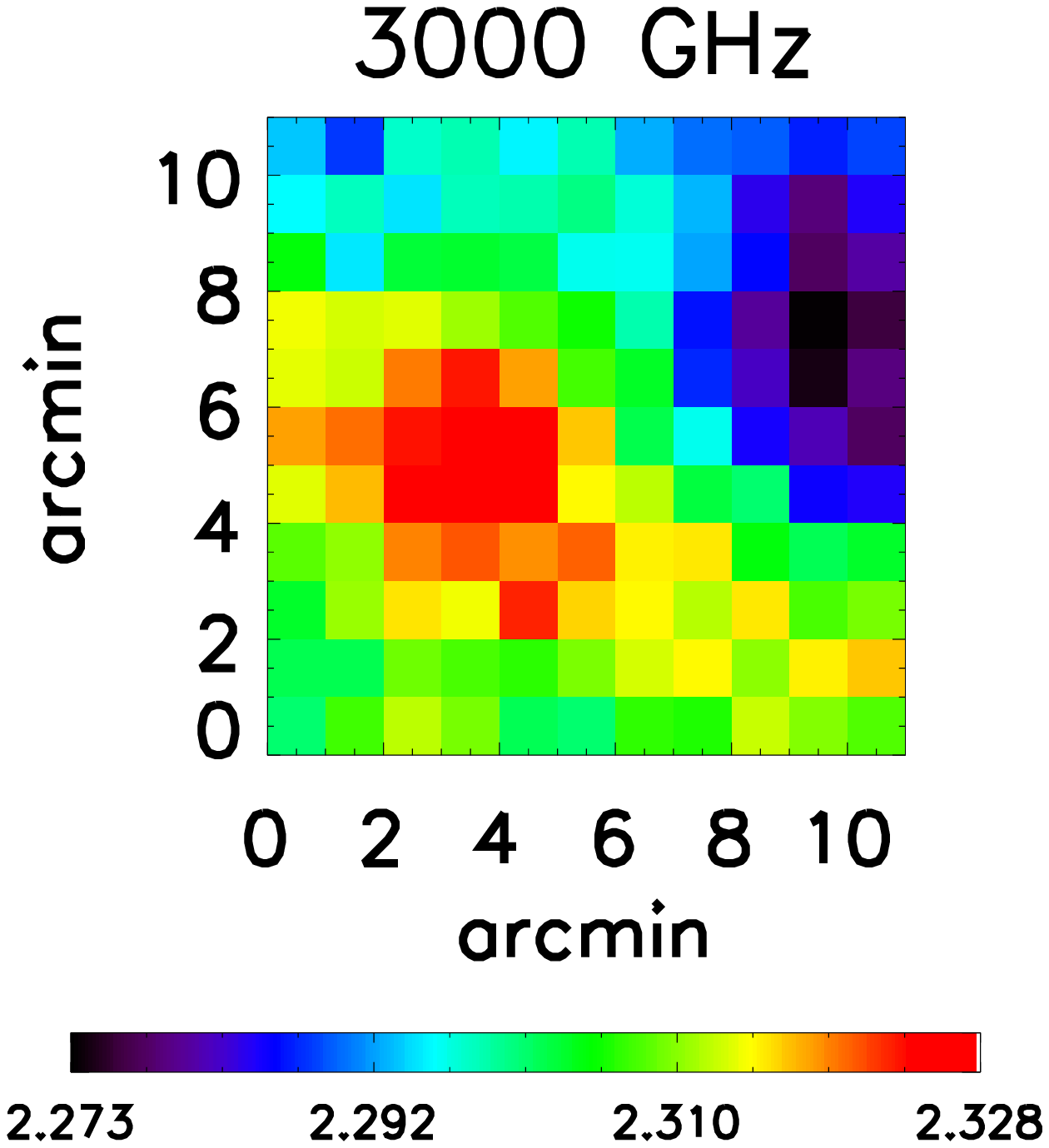}
}}   
\vspace{0.1cm}
\centerline{\rotatebox{0}{
\includegraphics[width=0.1\textwidth]{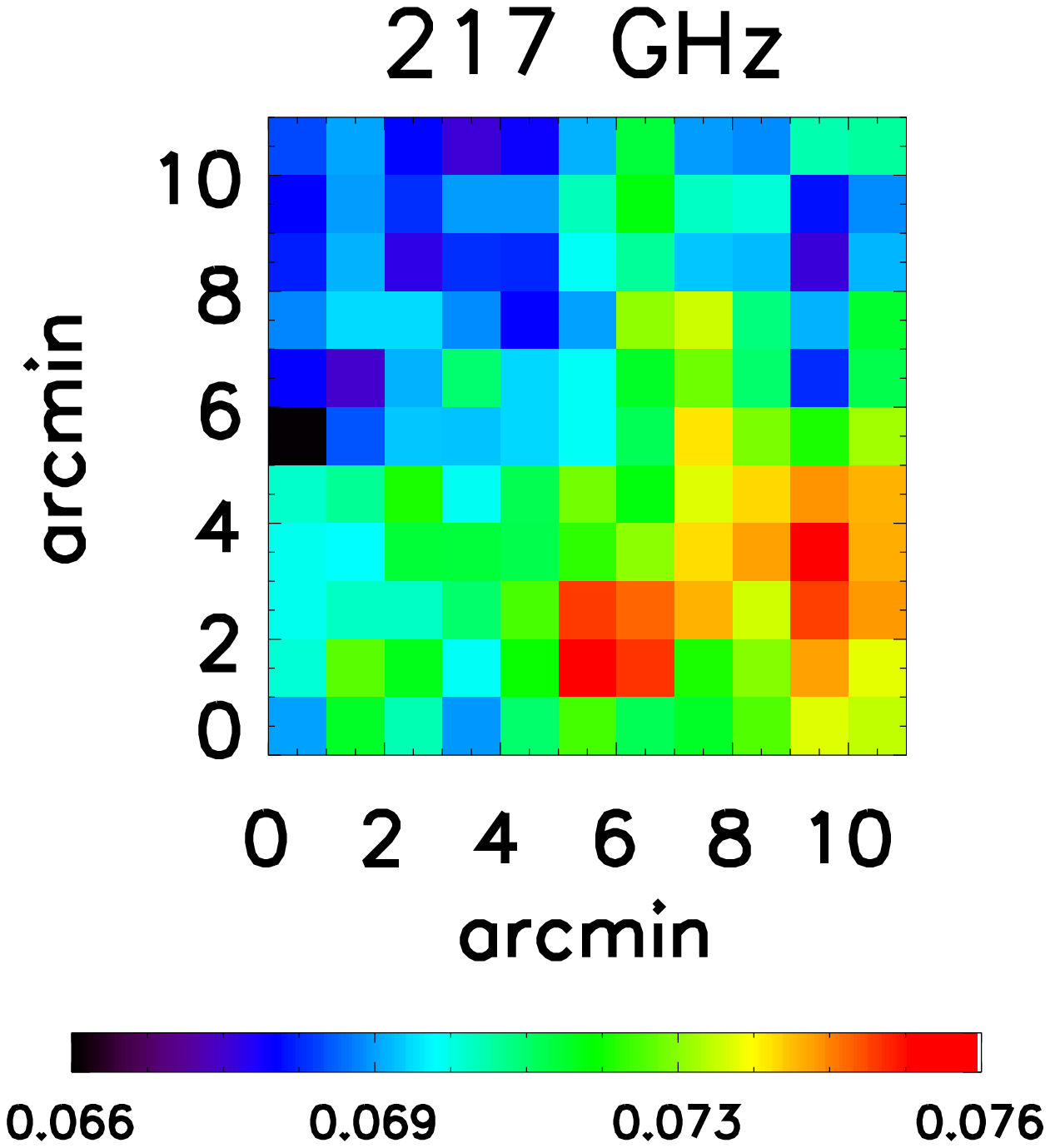}
\includegraphics[width=0.1\textwidth]{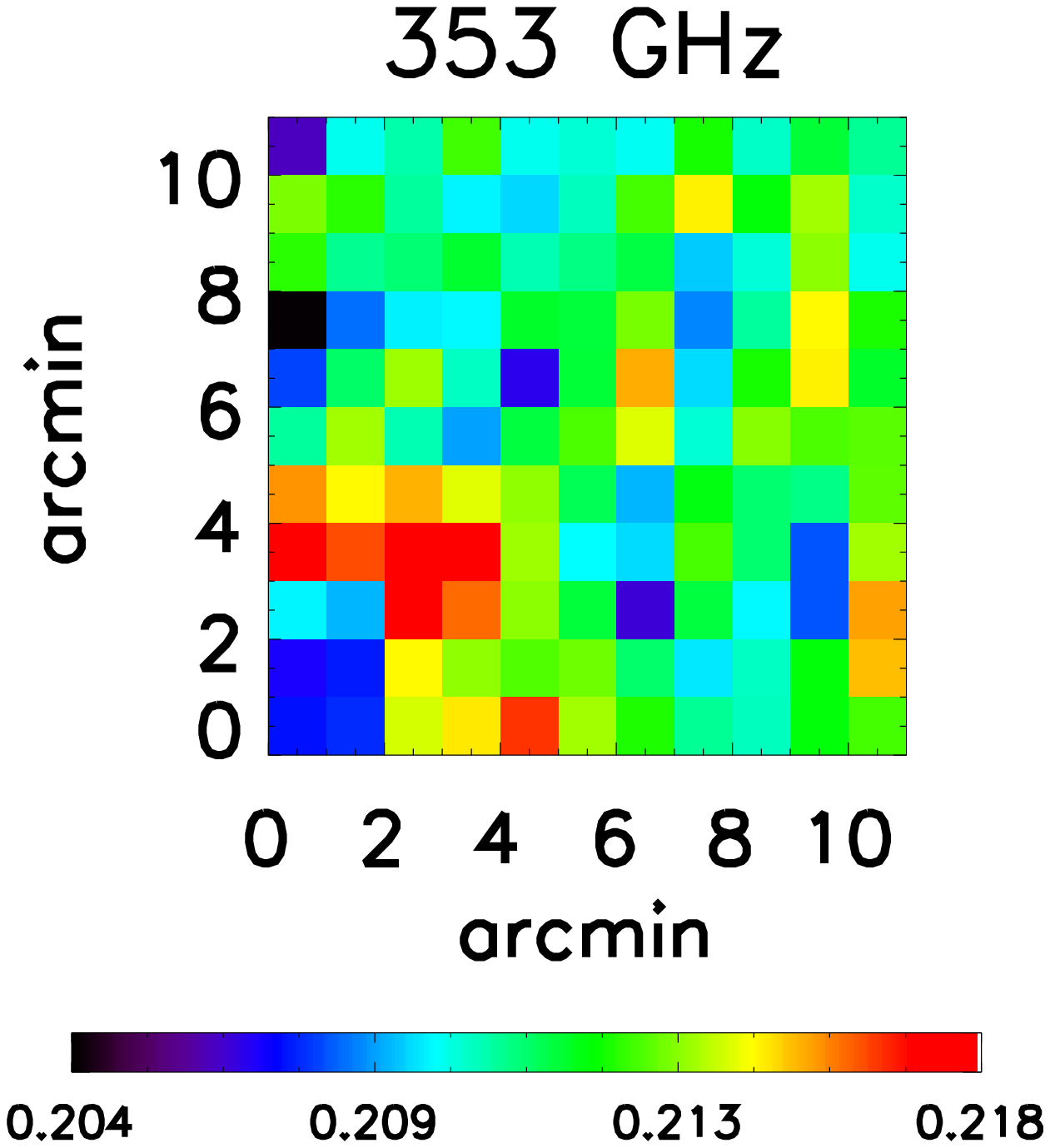}
\includegraphics[width=0.1\textwidth]{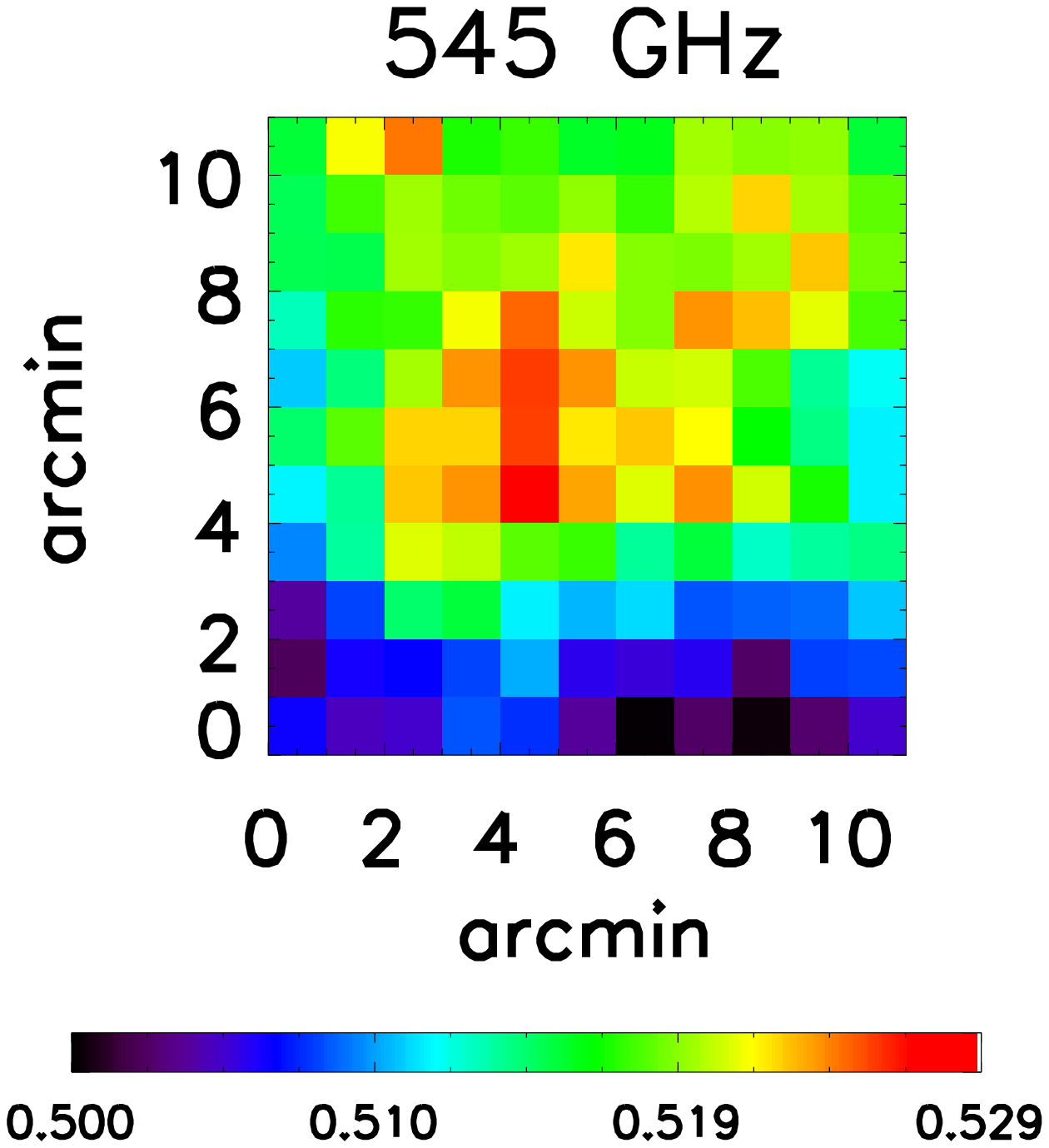}
\includegraphics[width=0.1\textwidth]{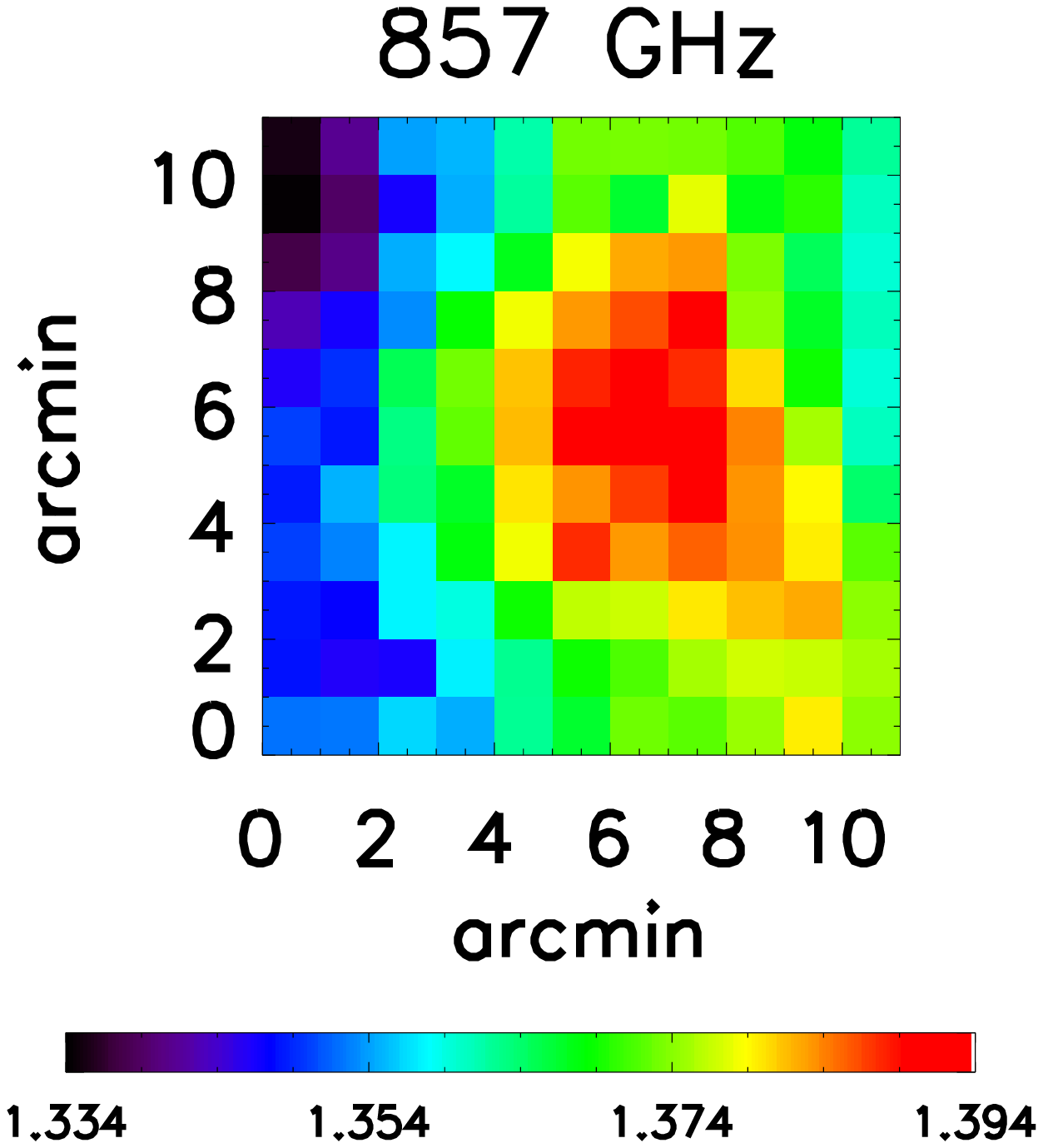}
\includegraphics[width=0.1\textwidth]{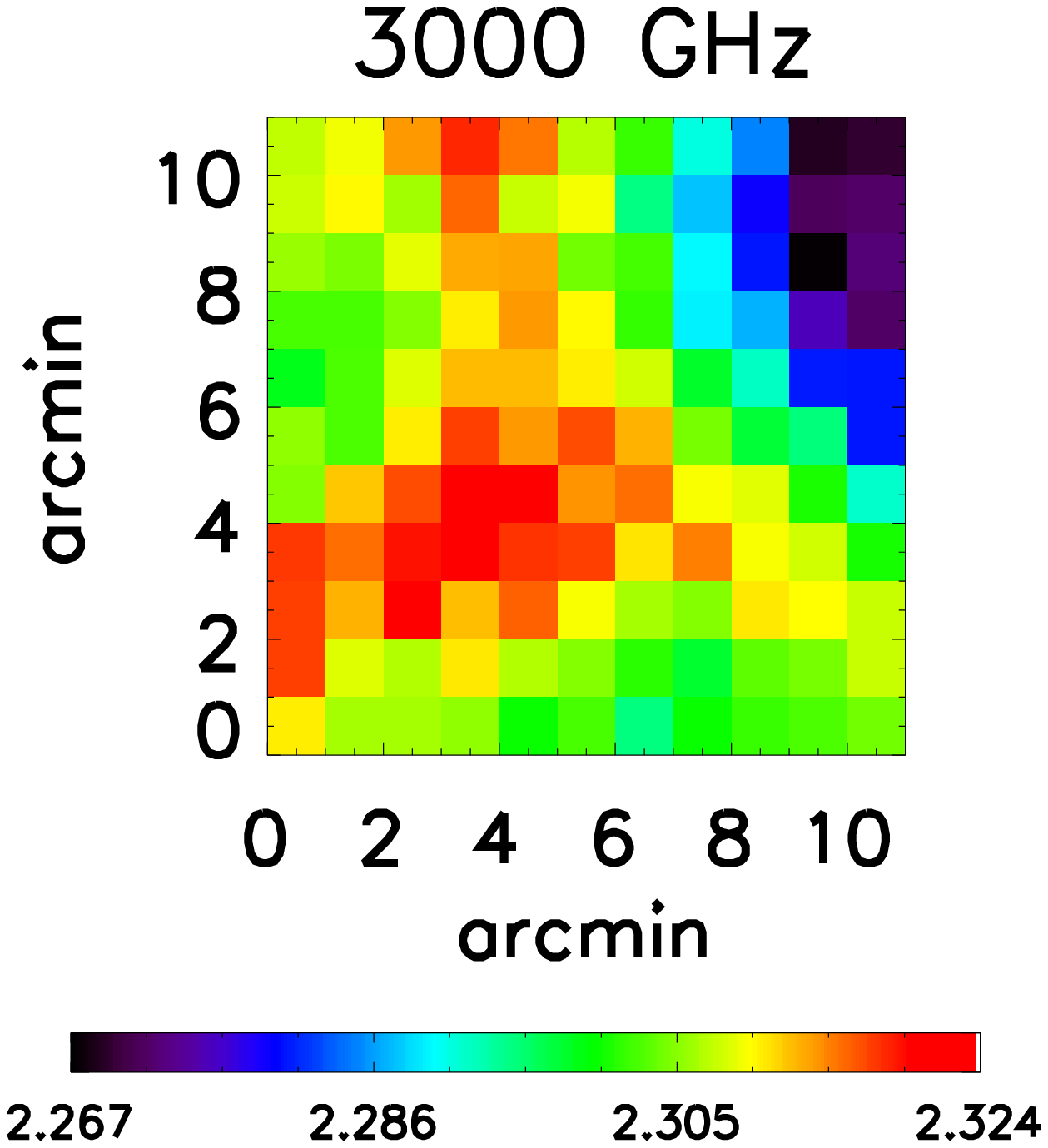}
}}   
\vspace{0.1cm}
\centerline{\rotatebox{0}{
\includegraphics[width=0.1\textwidth]{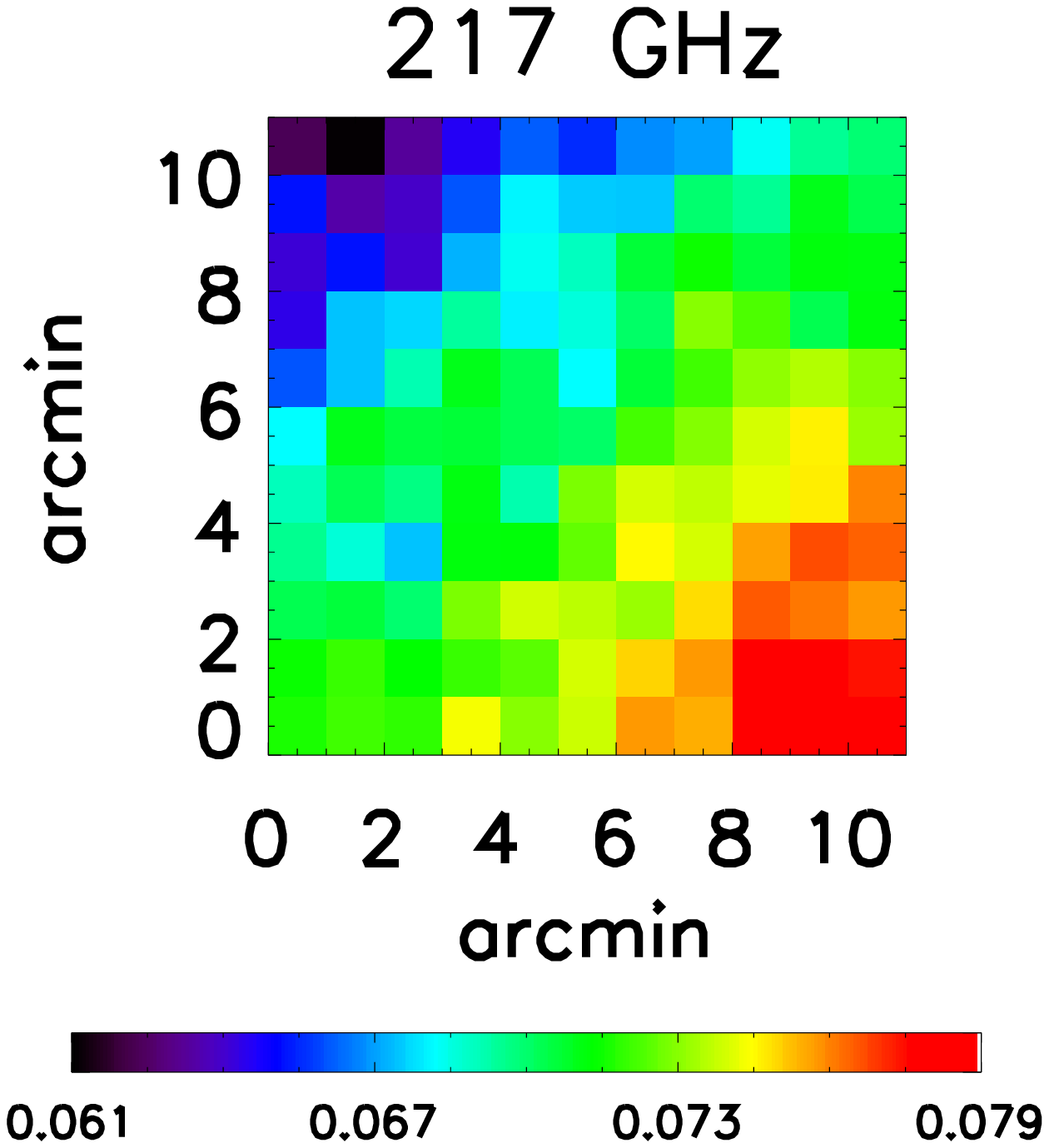}
\includegraphics[width=0.1\textwidth]{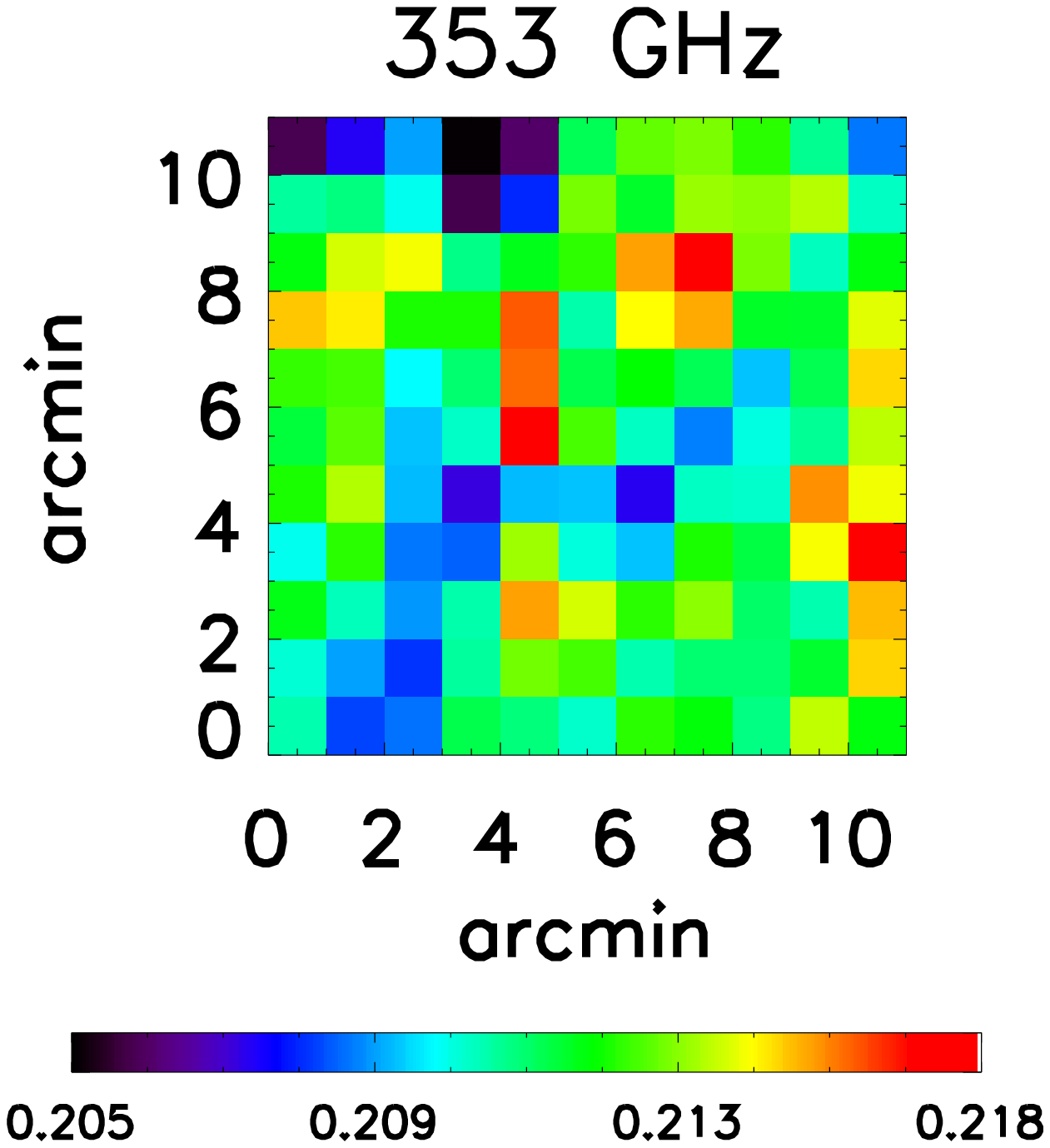}
\includegraphics[width=0.1\textwidth]{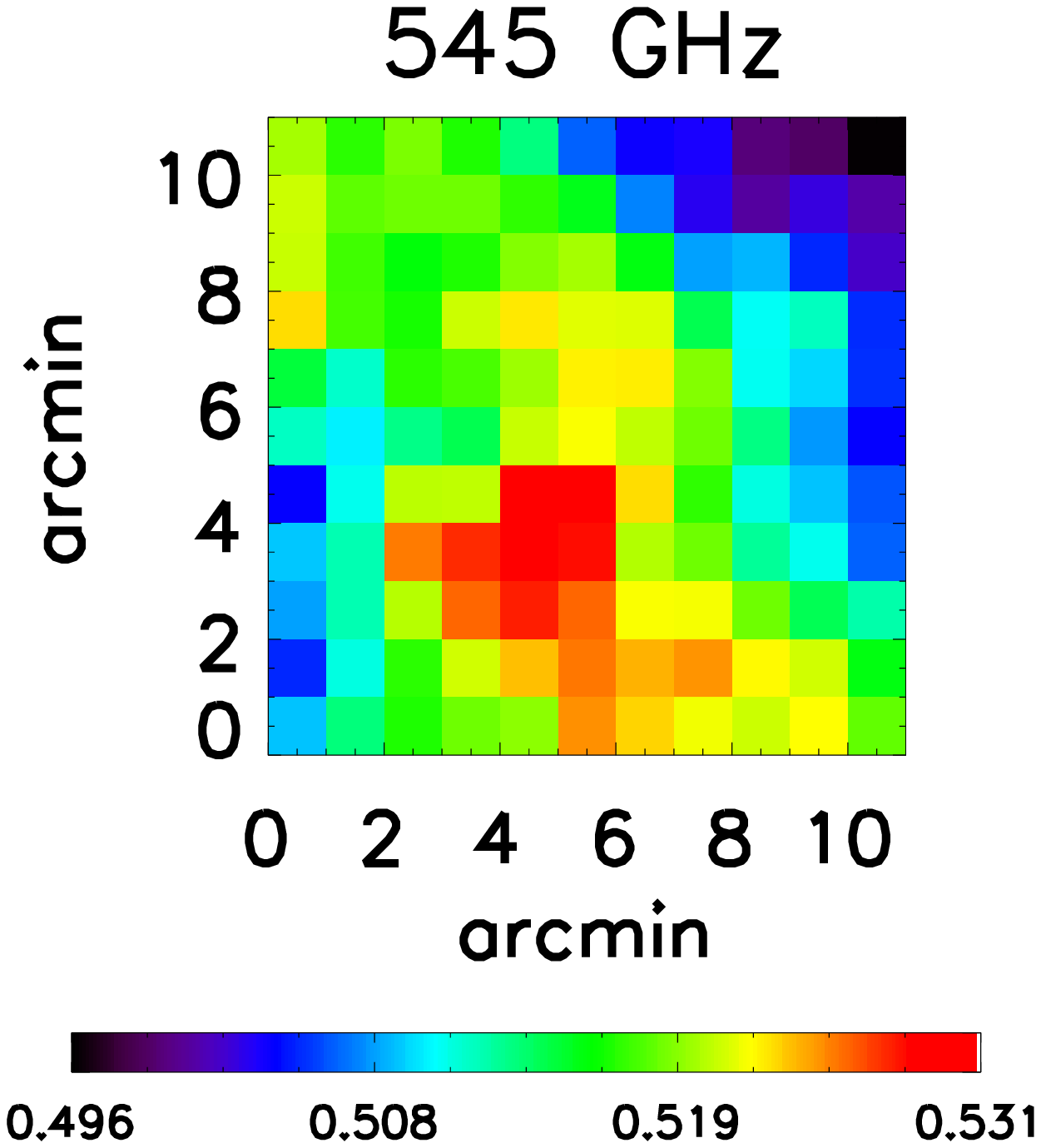}
\includegraphics[width=0.1\textwidth]{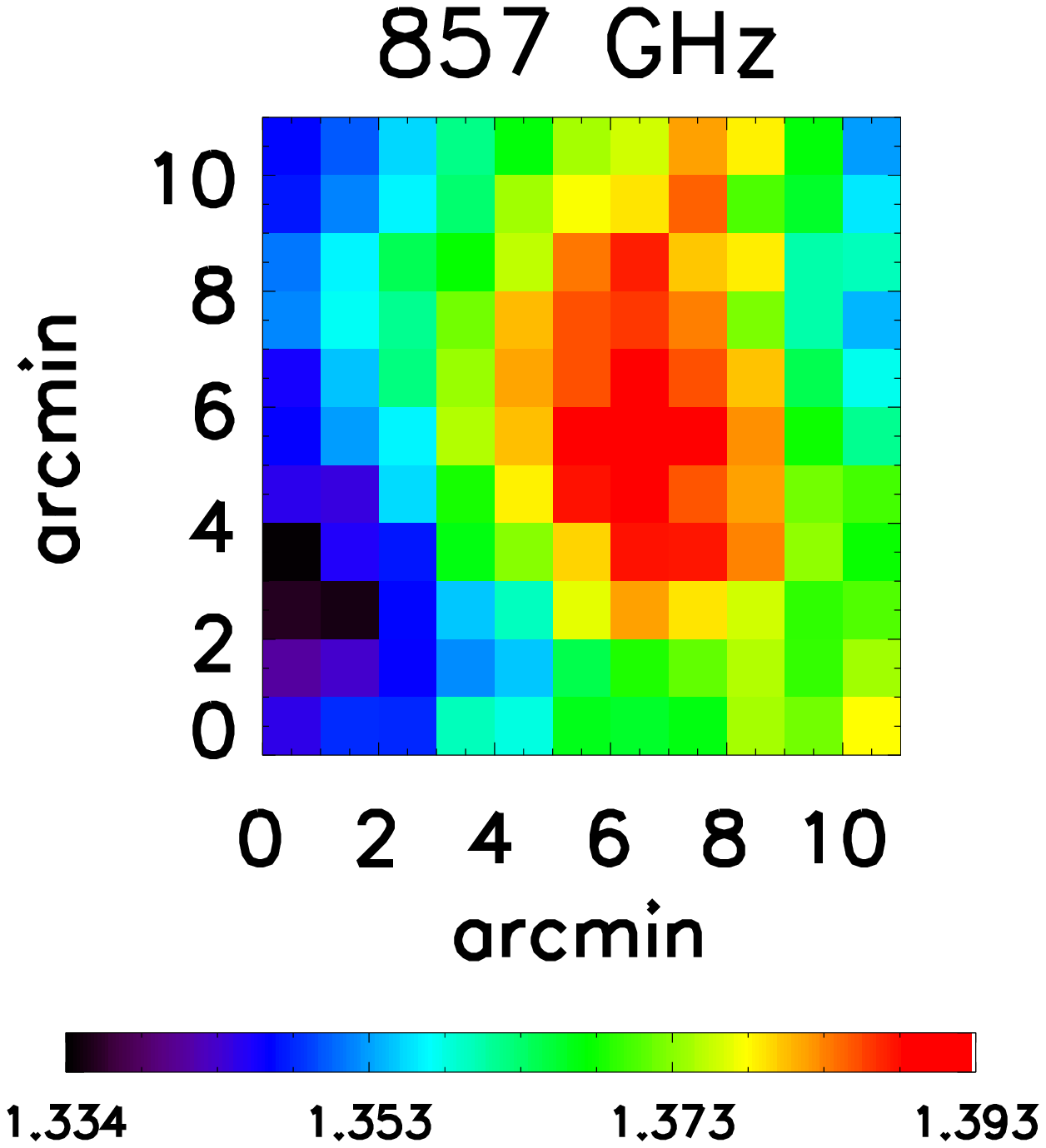}
\includegraphics[width=0.1\textwidth]{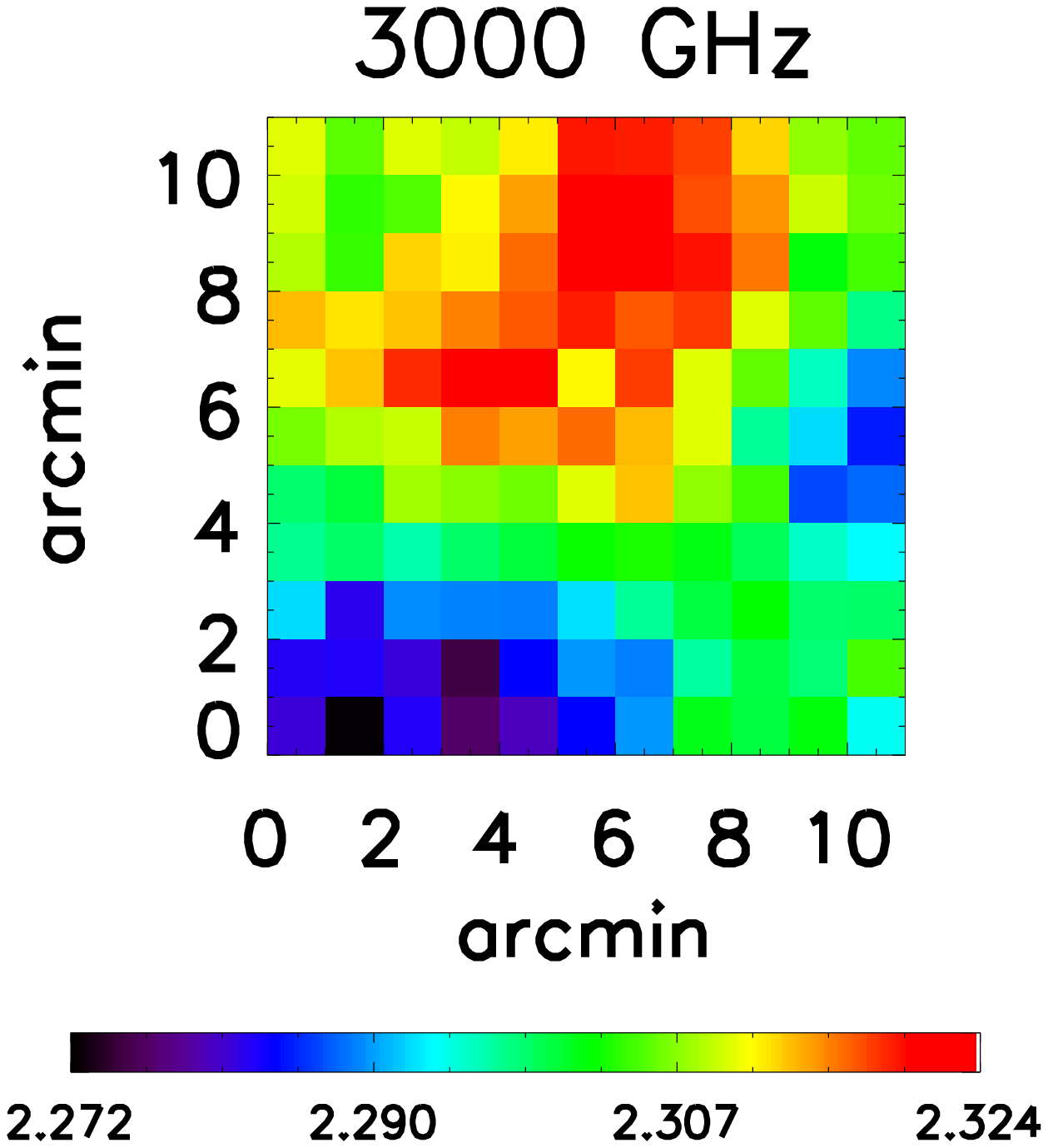}
}}   
\vspace{0.1cm}
\centerline{\rotatebox{0}{
\includegraphics[width=0.1\textwidth]{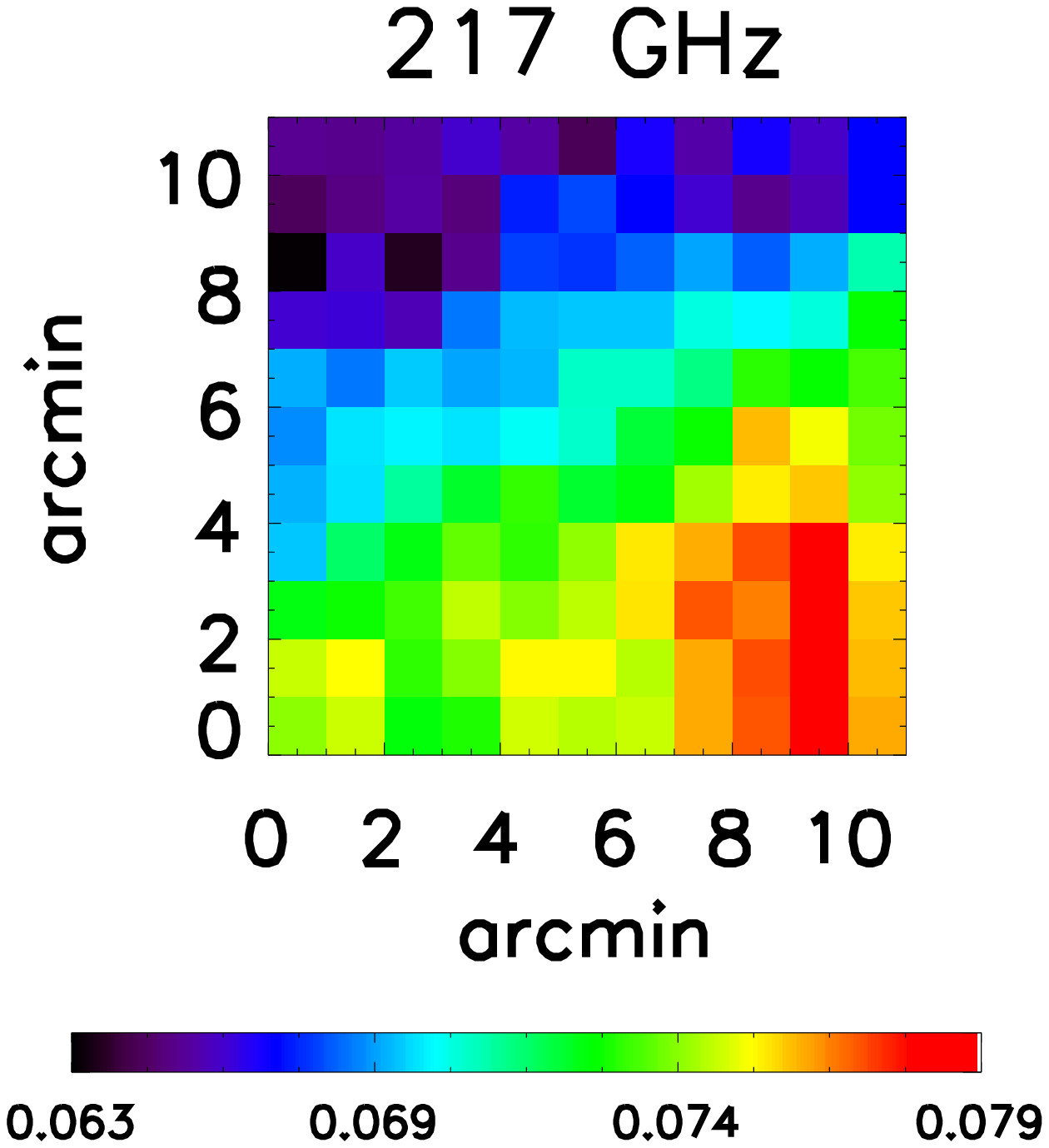}
\includegraphics[width=0.1\textwidth]{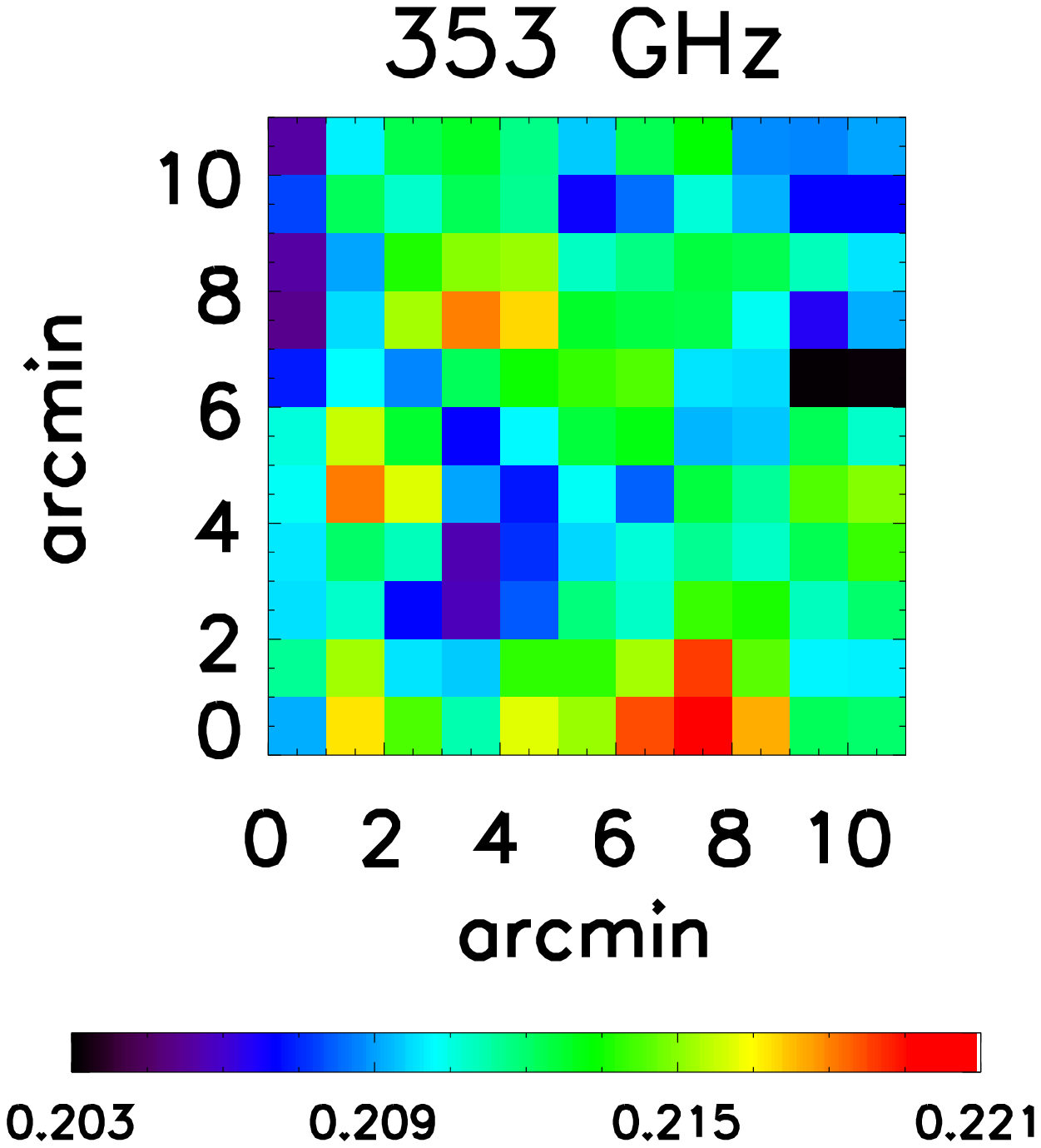}
\includegraphics[width=0.1\textwidth]{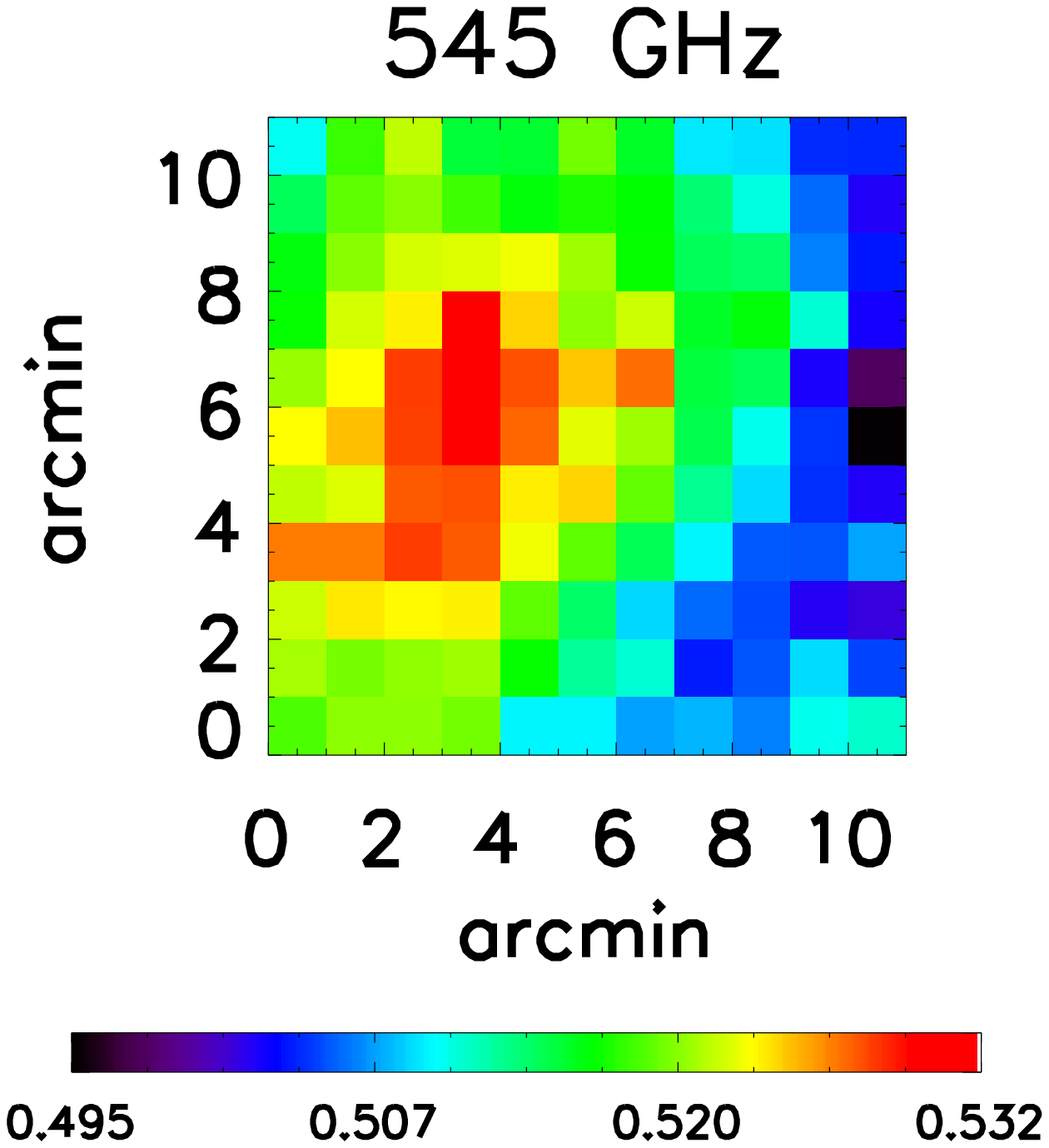}
\includegraphics[width=0.1\textwidth]{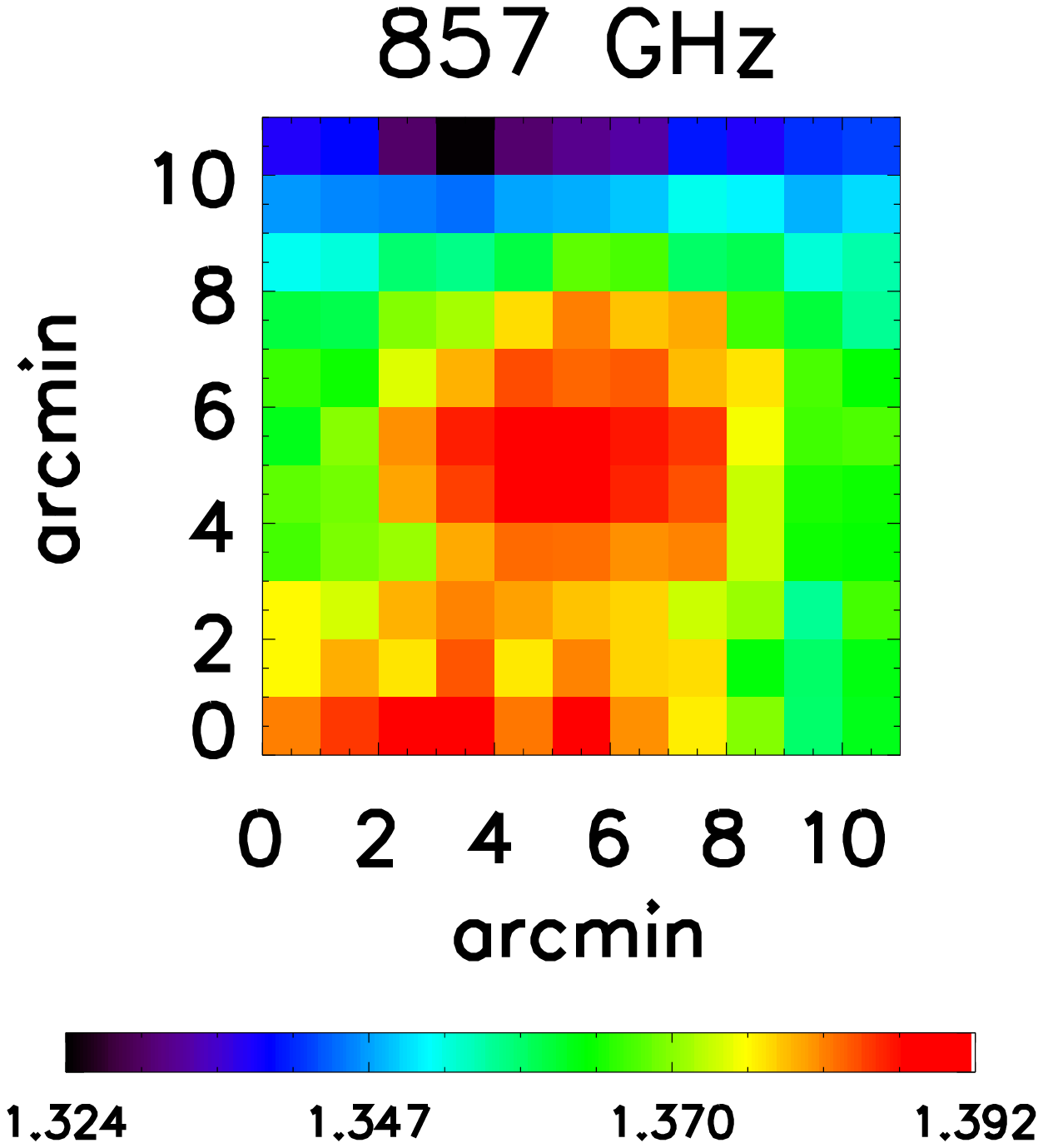}
\includegraphics[width=0.1\textwidth]{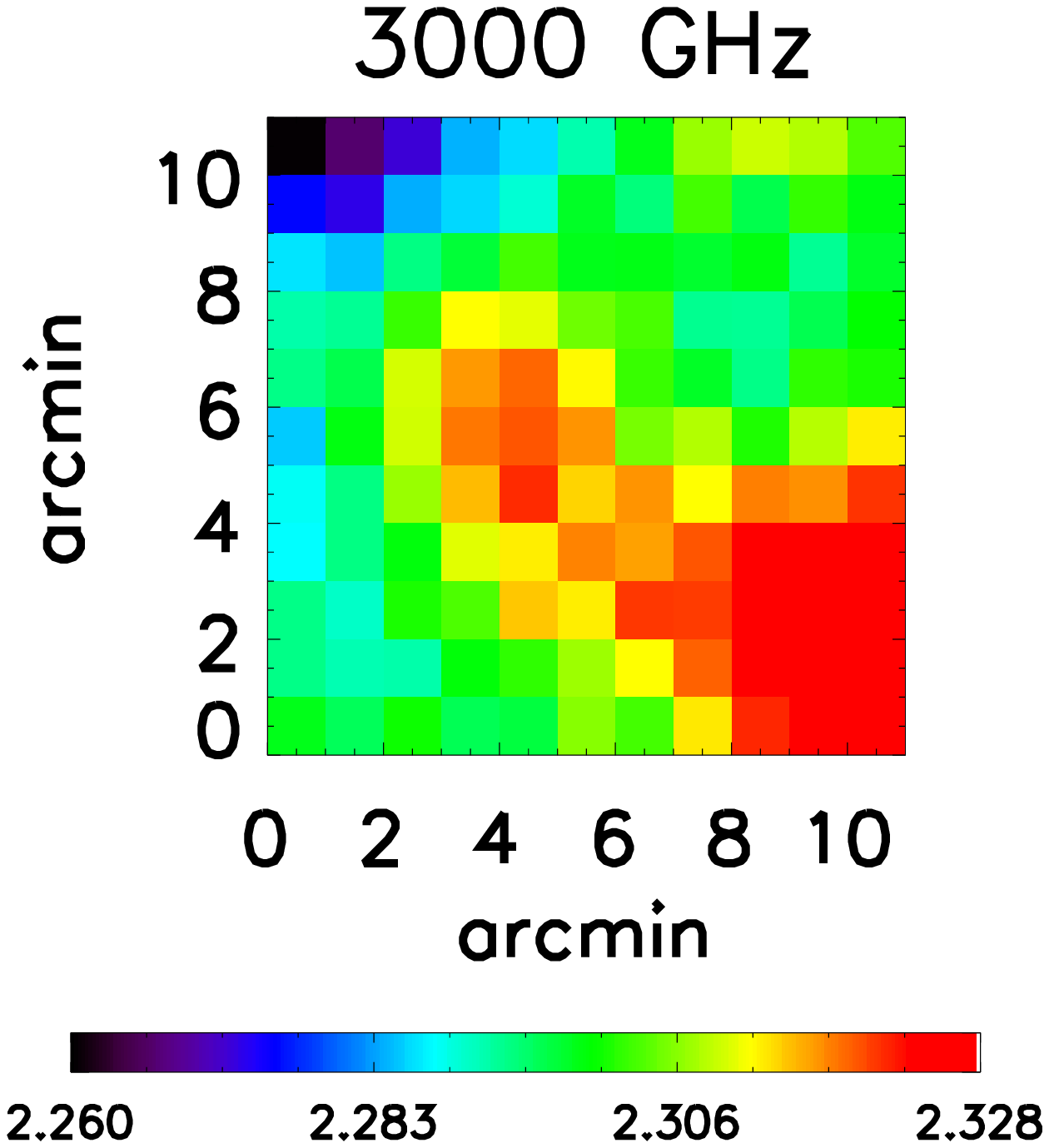}
}}   
\vspace{0.1cm}
\centerline{\rotatebox{0}{
\includegraphics[width=0.1\textwidth]{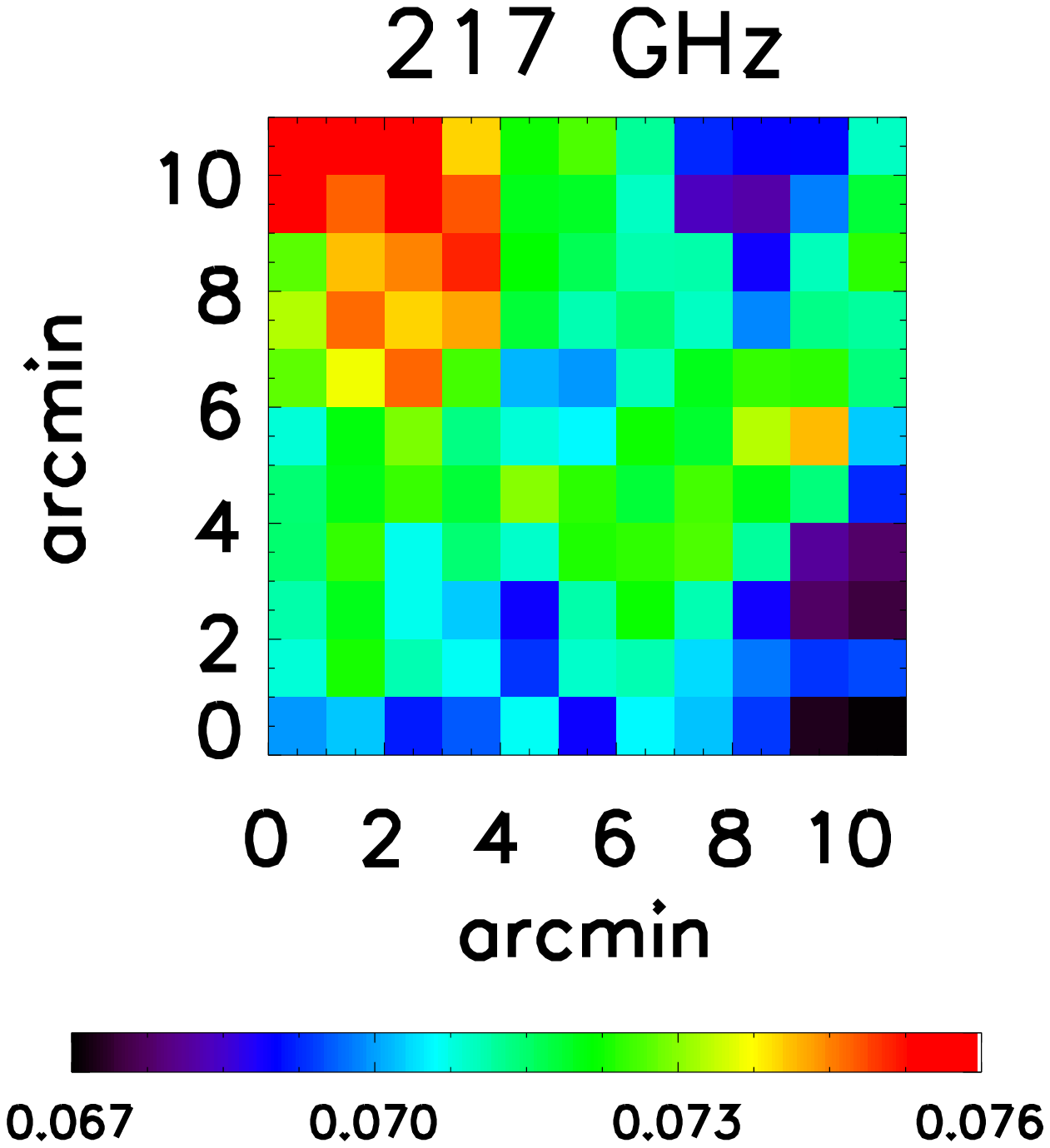}
\includegraphics[width=0.1\textwidth]{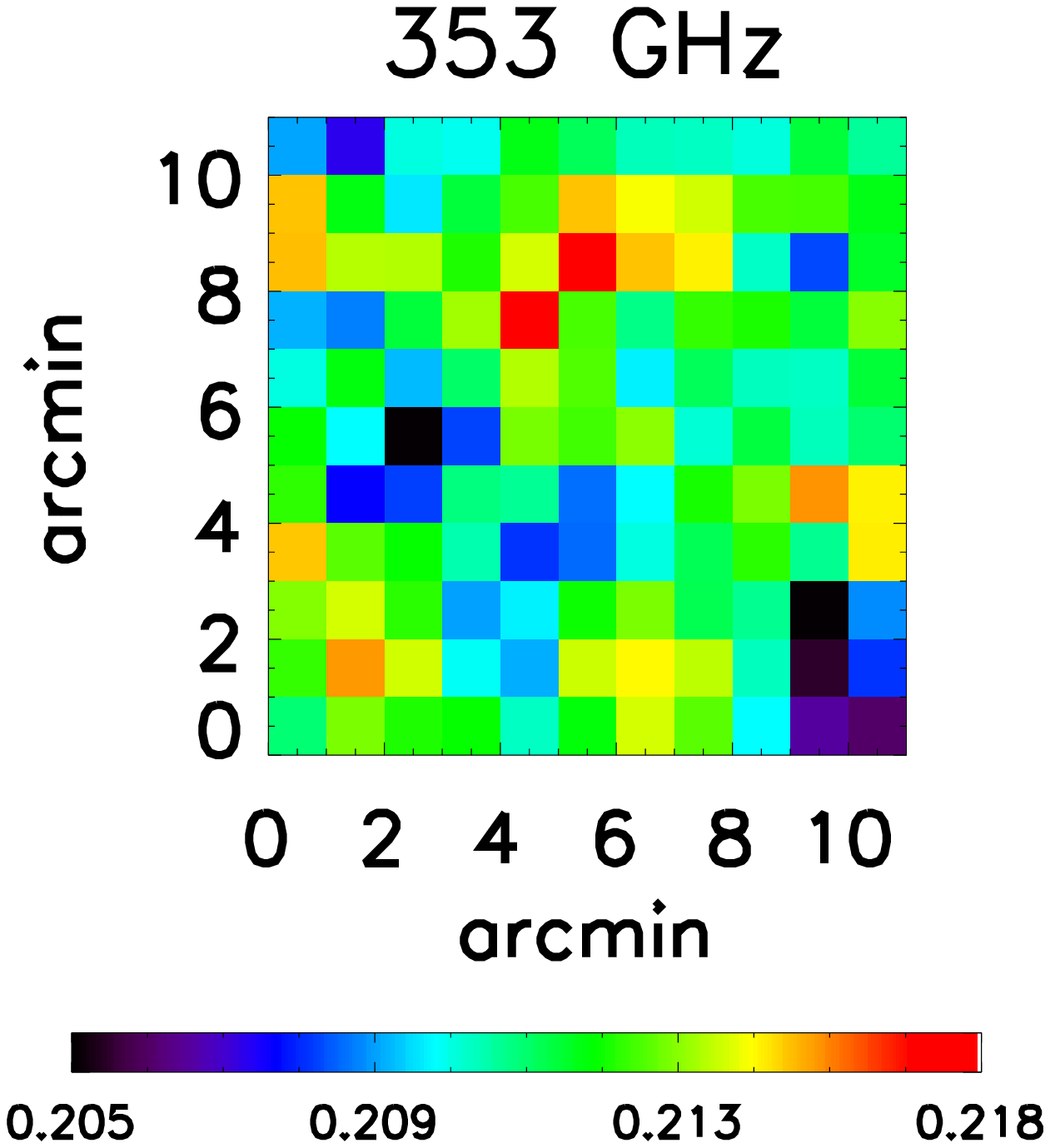}
\includegraphics[width=0.1\textwidth]{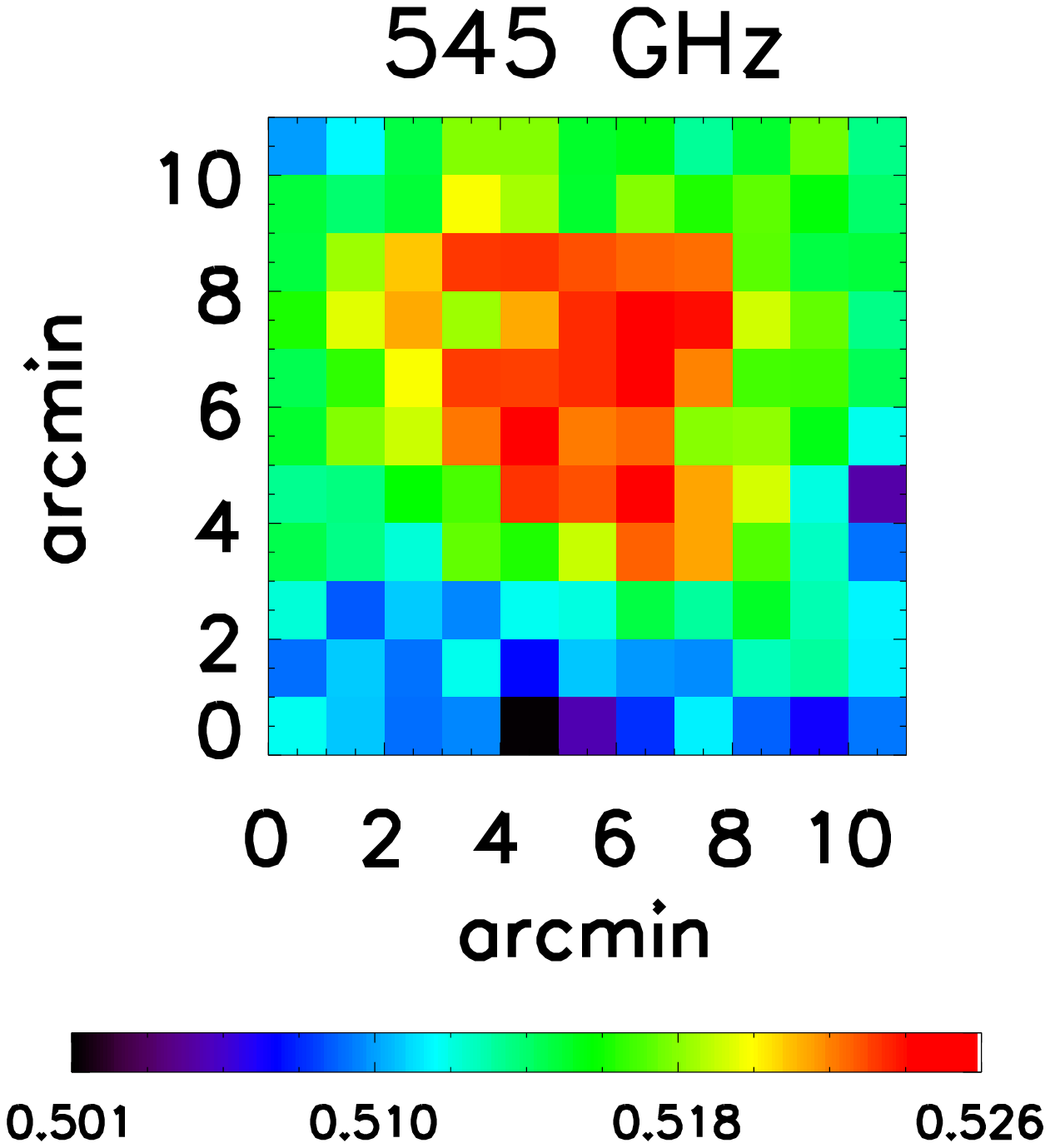}
\includegraphics[width=0.1\textwidth]{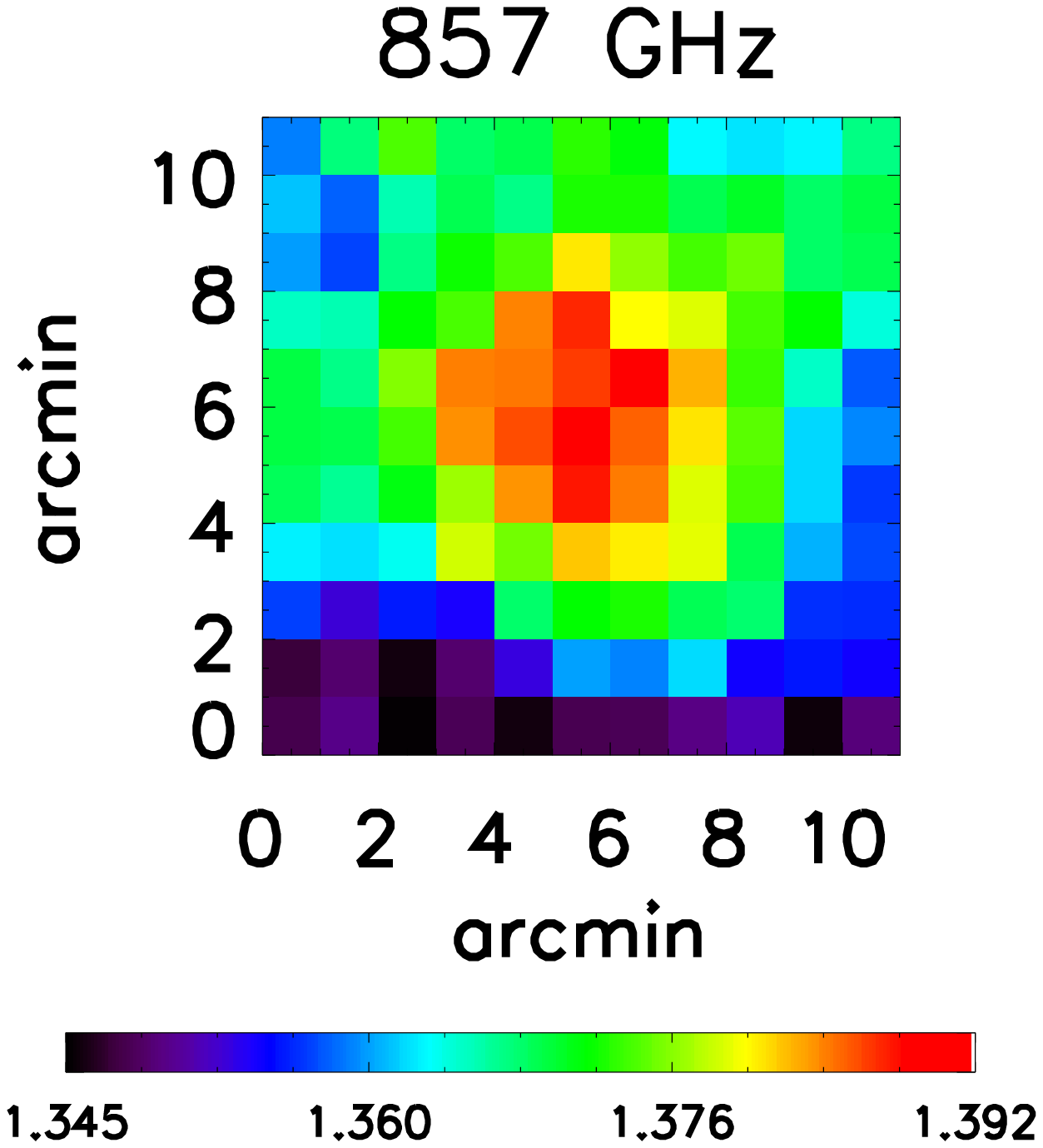}
\includegraphics[width=0.1\textwidth]{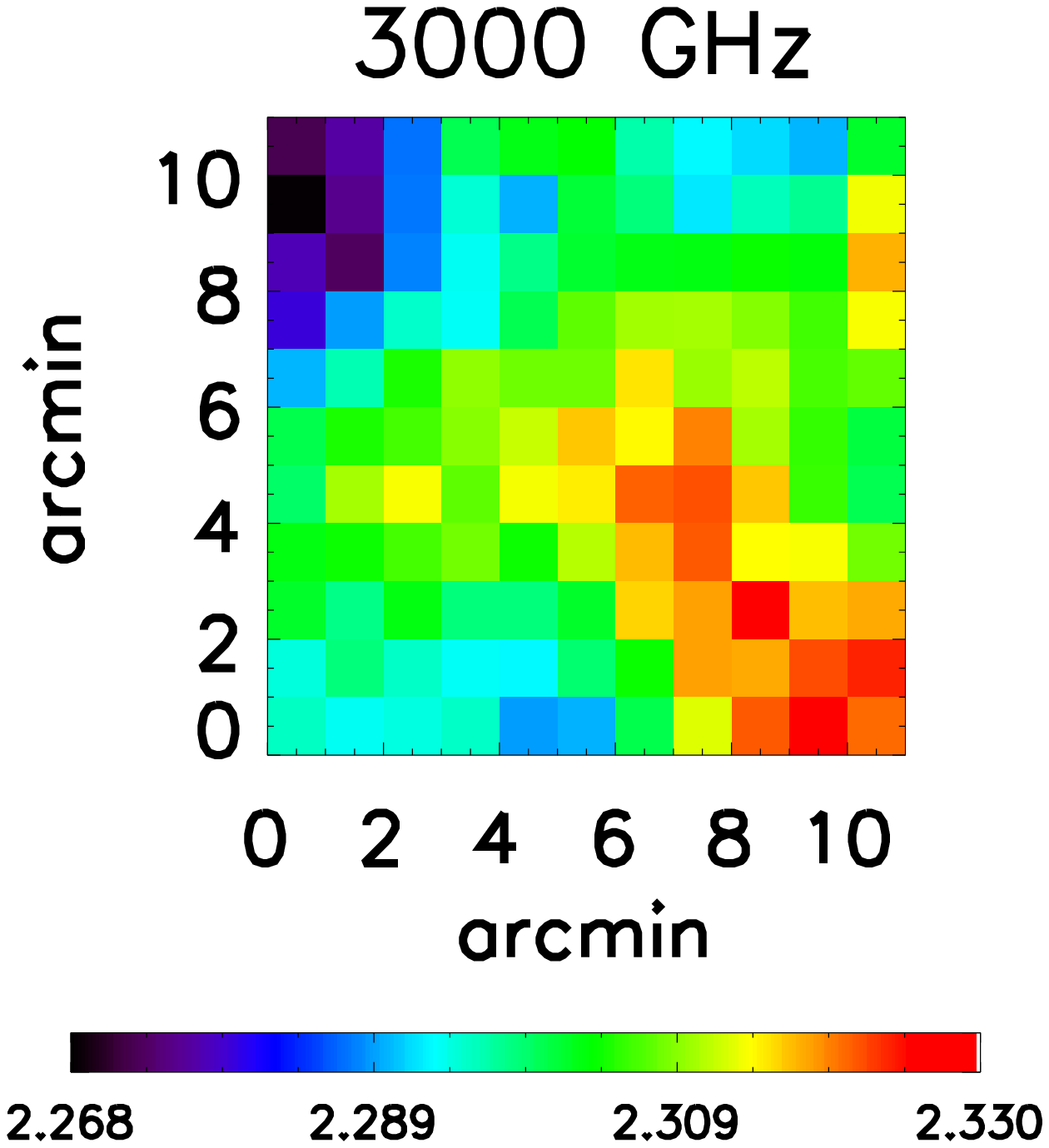}
}}   

\caption{Six realisations of \Planck\ and IRIS residual maps (in units of MJy$sr^{-1}$) after introducing a random rotation of
    90$\deg$ in individual maps of the SPT DSFGs before stacking the maps at
    the positions of the SPT DSFGs, and then removing the central
    compact source component in each realisation of the stacked maps using the formalism in
    Appendix~\ref{app:planck_stack_formalism}. The original size of
    the stacked maps is $1\deg\times1\deg$. Here, we show the
    $10\arcmin\times10\arcmin$ central region in order to see the residual structure more clearly.
\label{fig:random_rotations}}
\end{figure}

\begin{figure}{}
\centerline{\rotatebox{0}{
\includegraphics[width=0.1\textwidth]{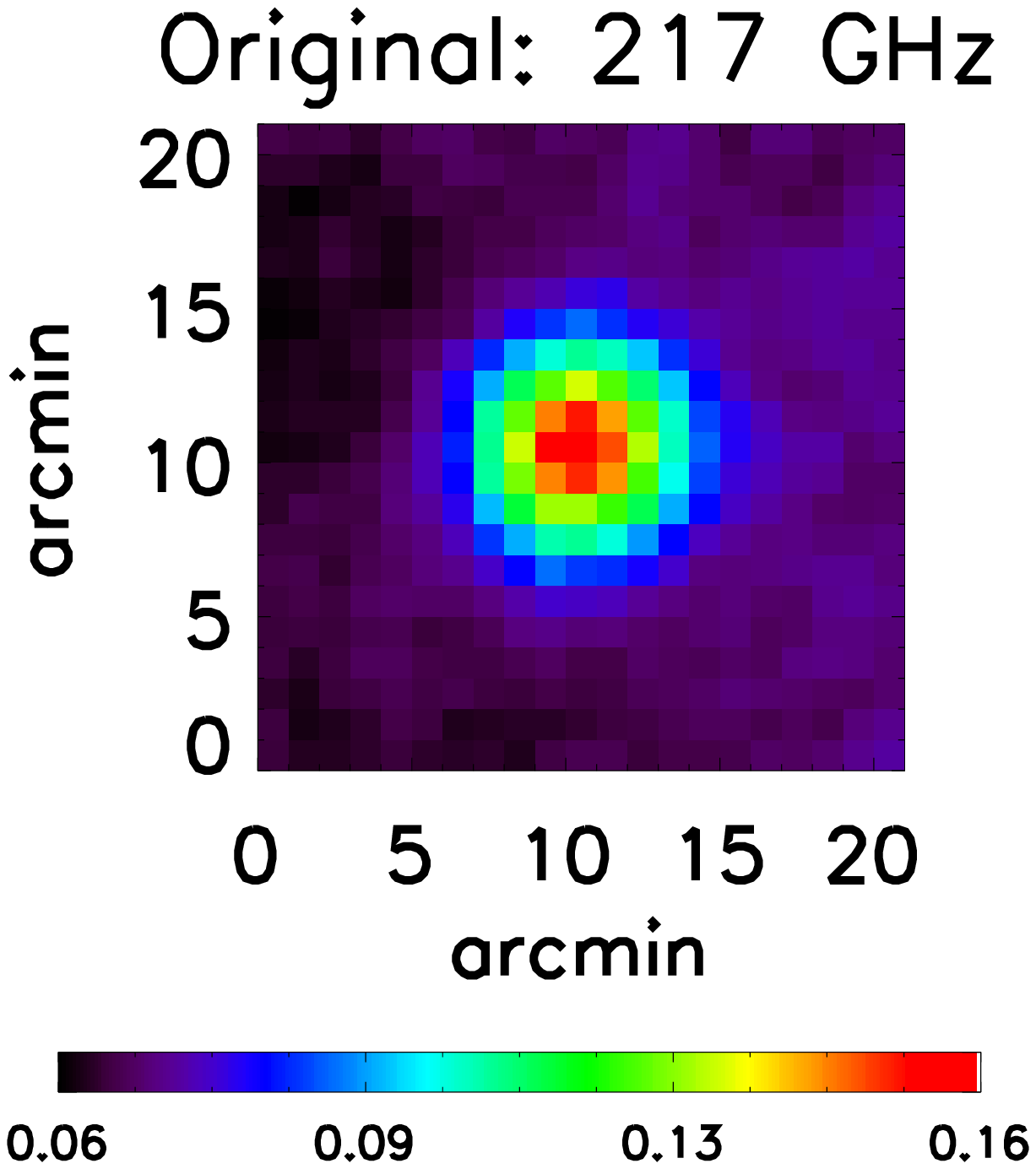}
\includegraphics[width=0.1\textwidth]{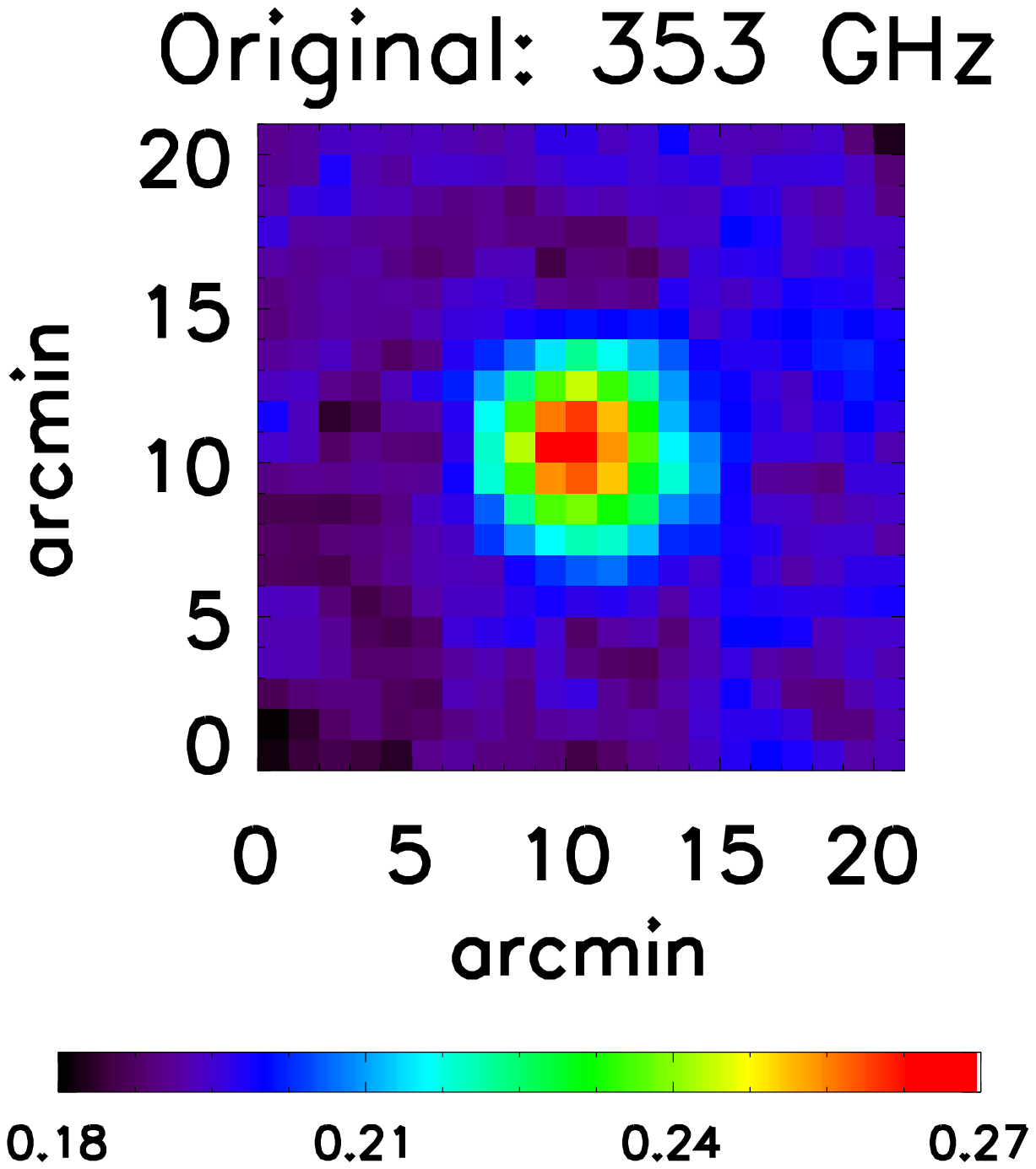}
\includegraphics[width=0.1\textwidth]{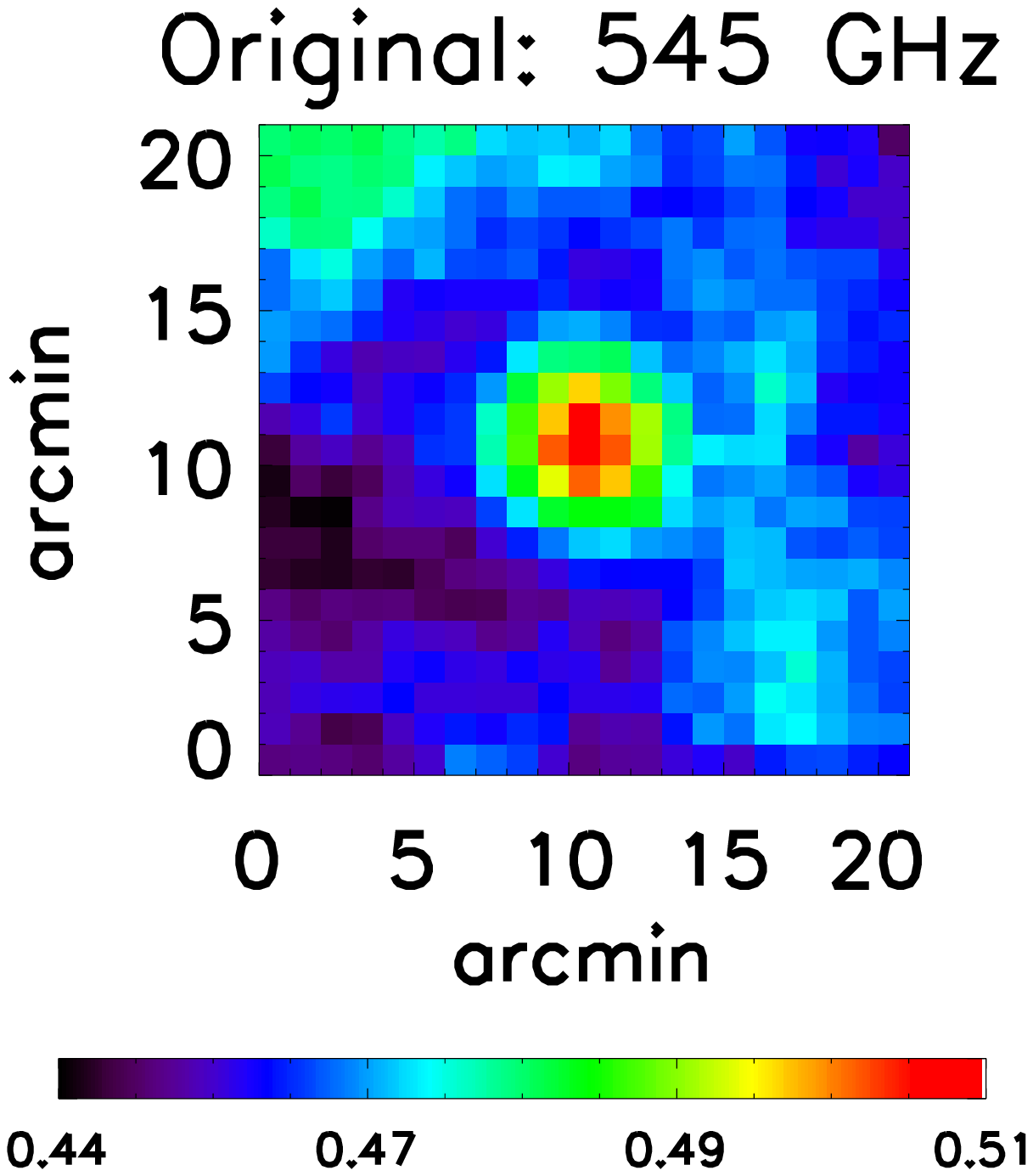}
\includegraphics[width=0.1\textwidth]{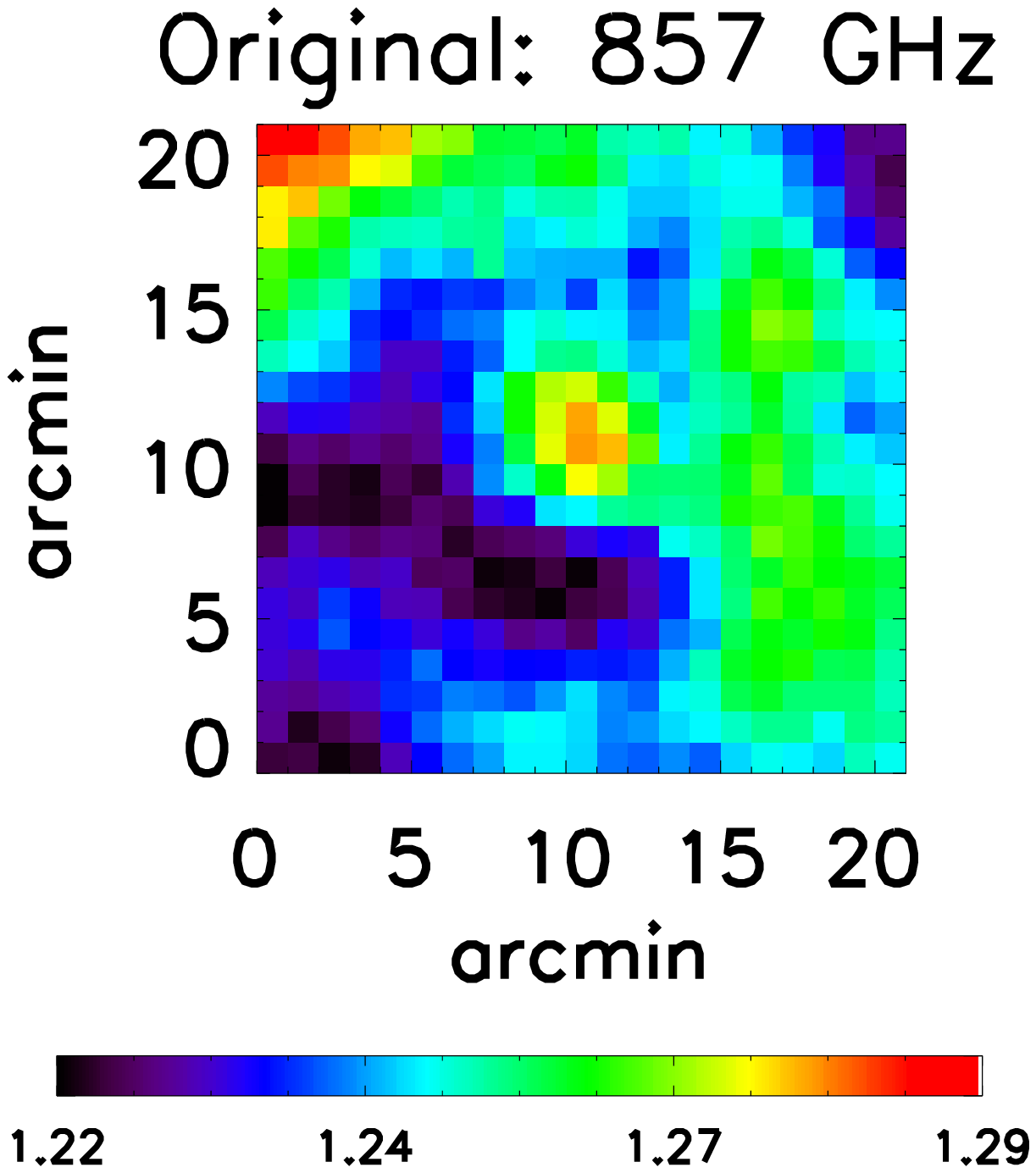}
\includegraphics[width=0.1\textwidth]{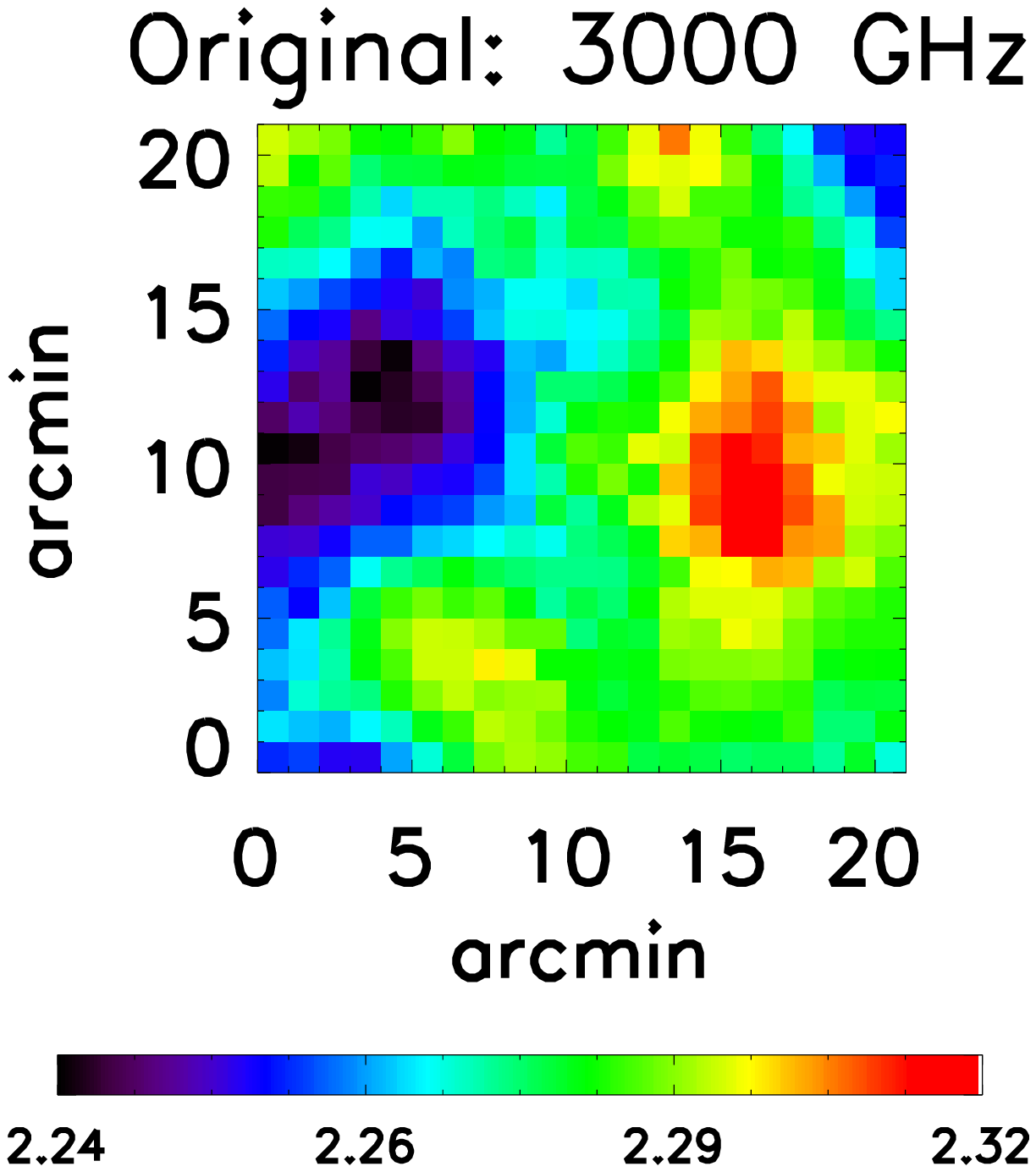}
}}   
\vspace{0.1cm}
\centerline{\rotatebox{0}{
\includegraphics[width=0.1\textwidth]{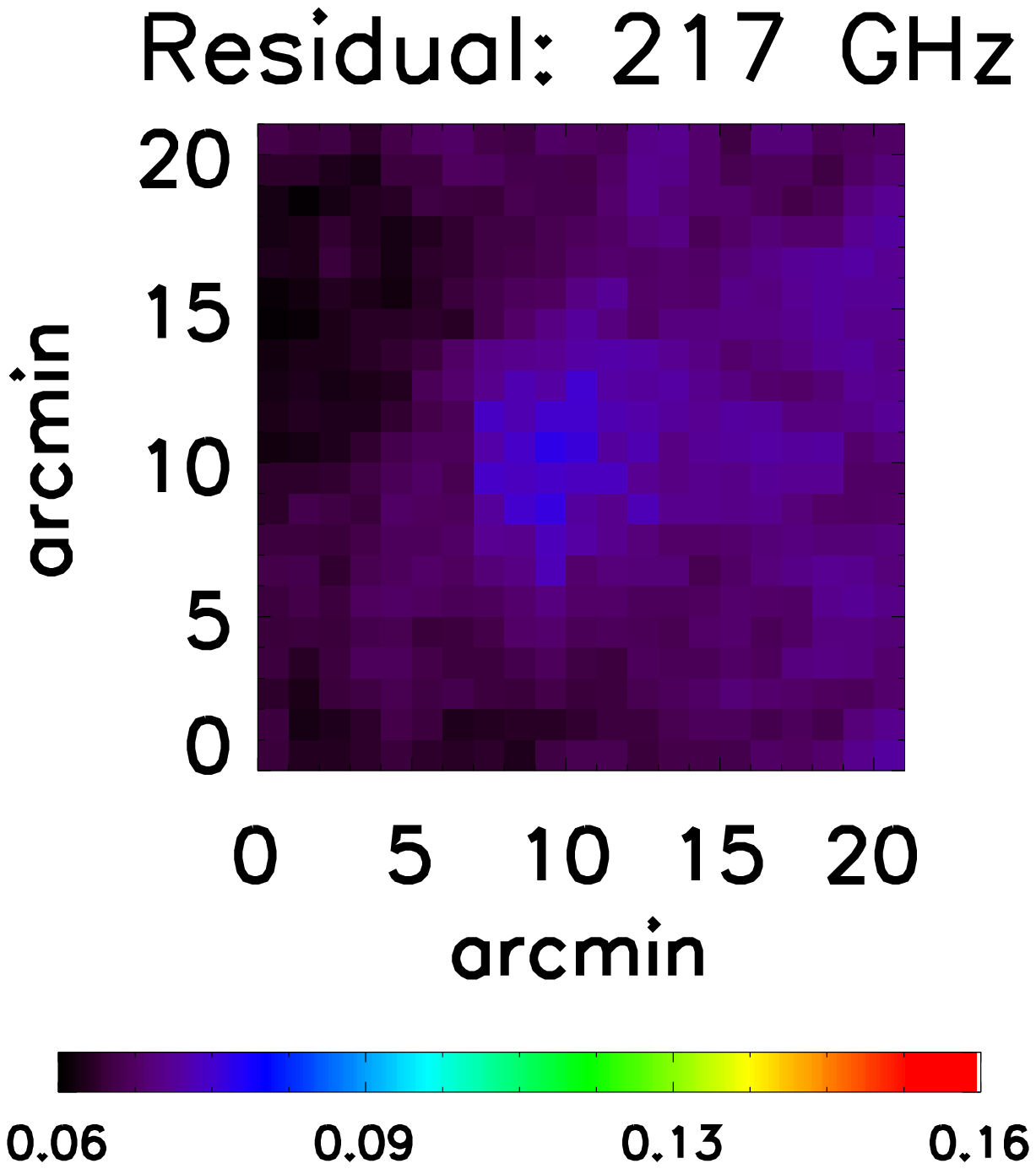}
\includegraphics[width=0.1\textwidth]{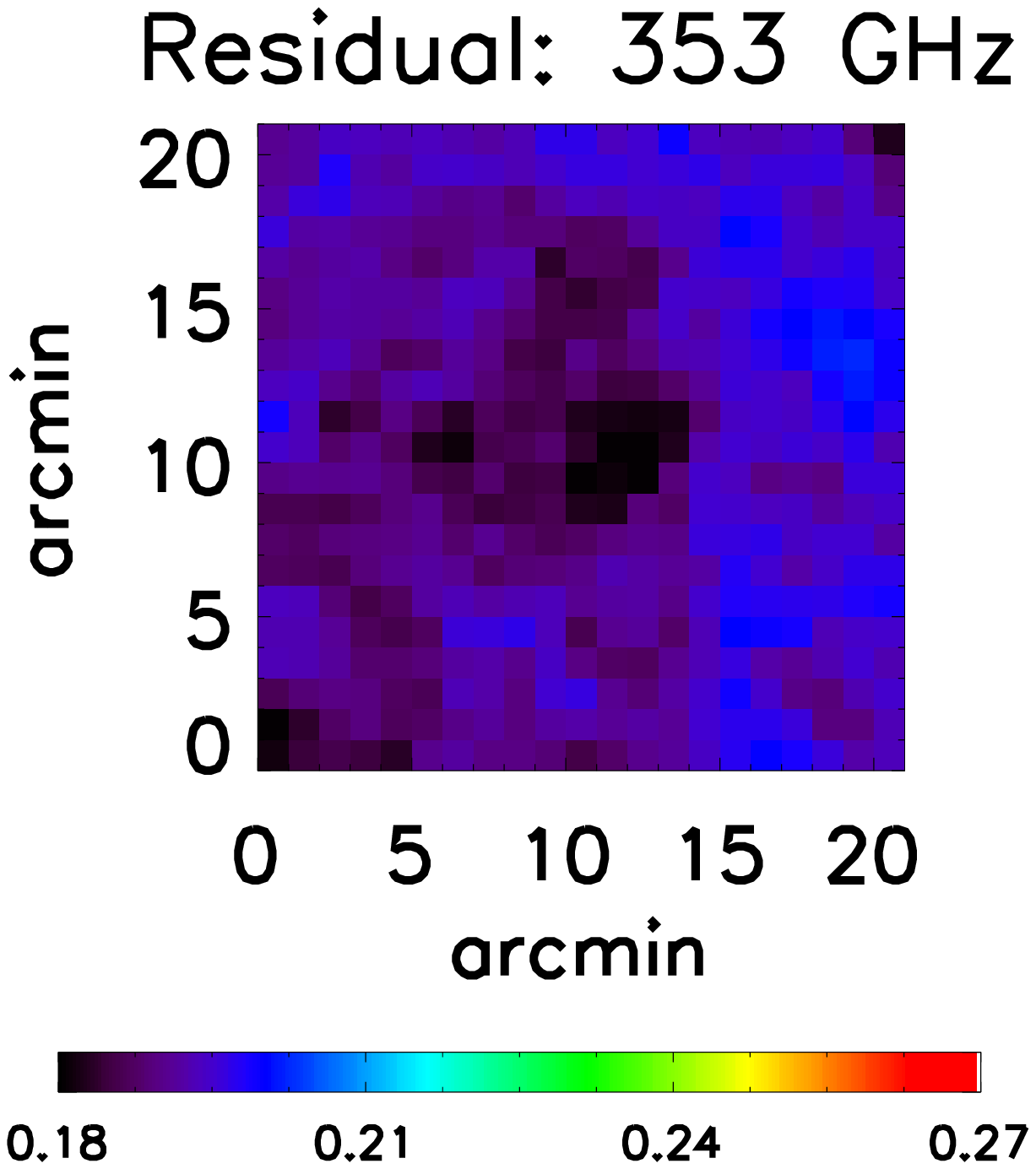}
\includegraphics[width=0.1\textwidth]{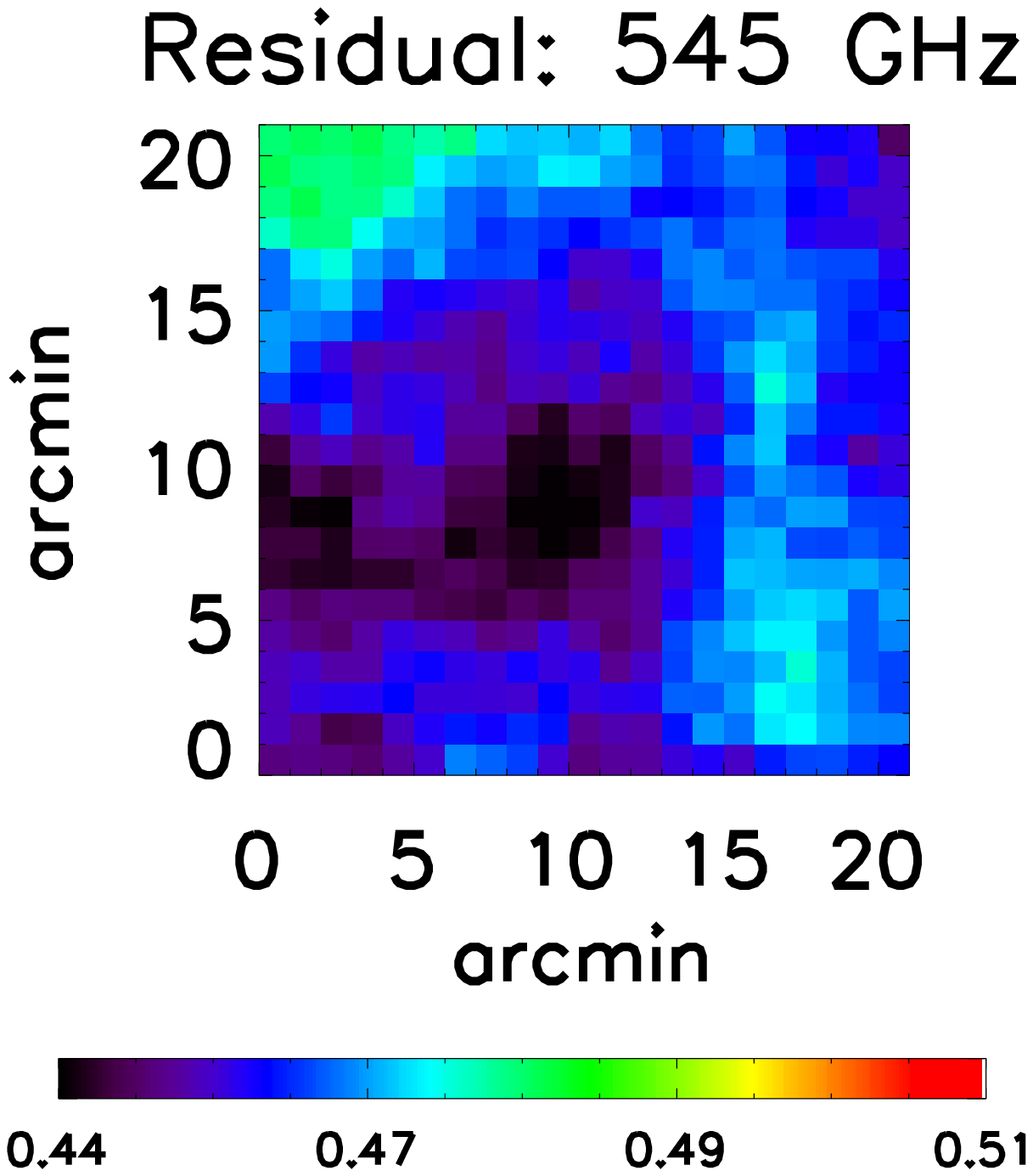}
\includegraphics[width=0.1\textwidth]{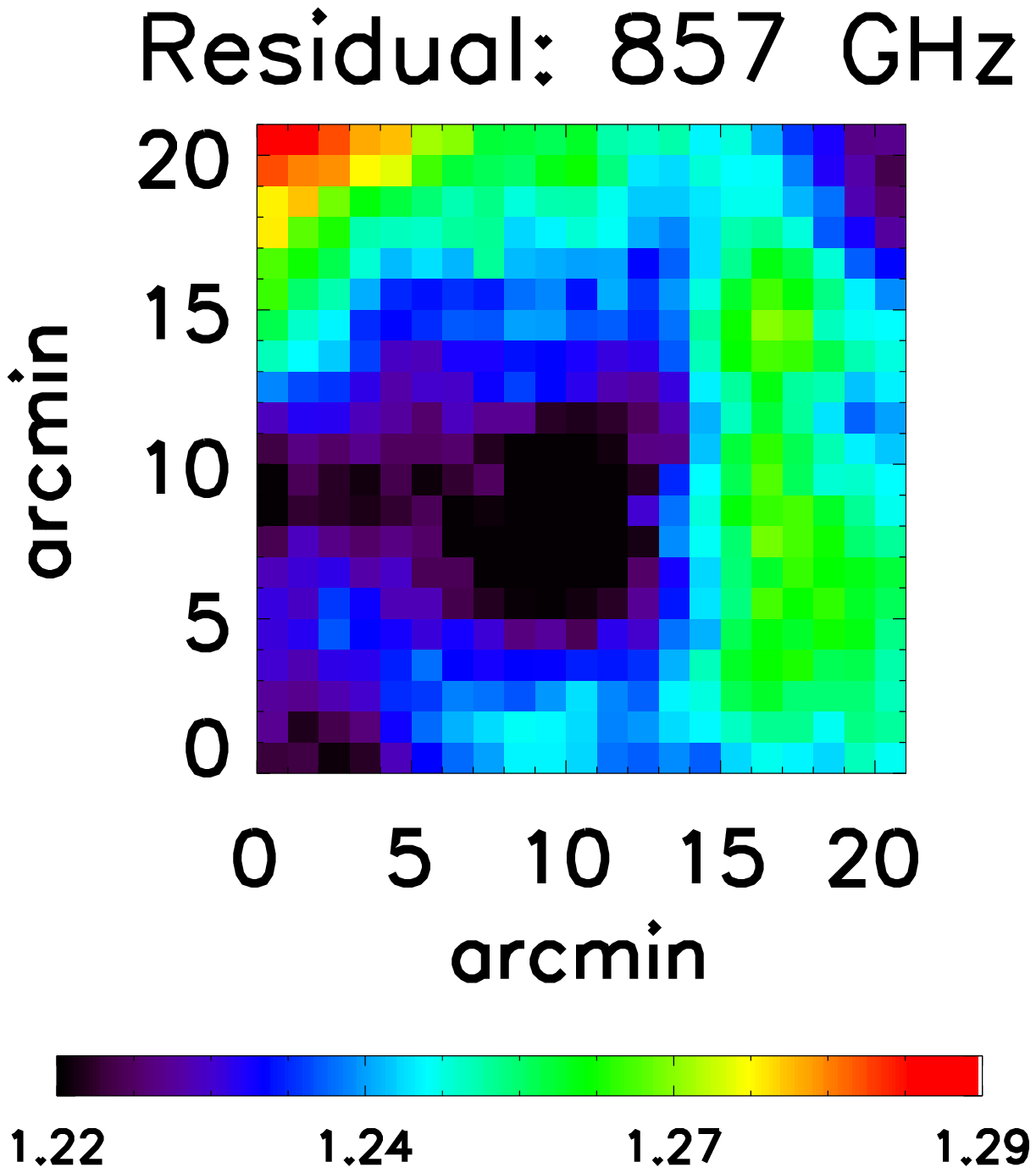}
\includegraphics[width=0.1\textwidth]{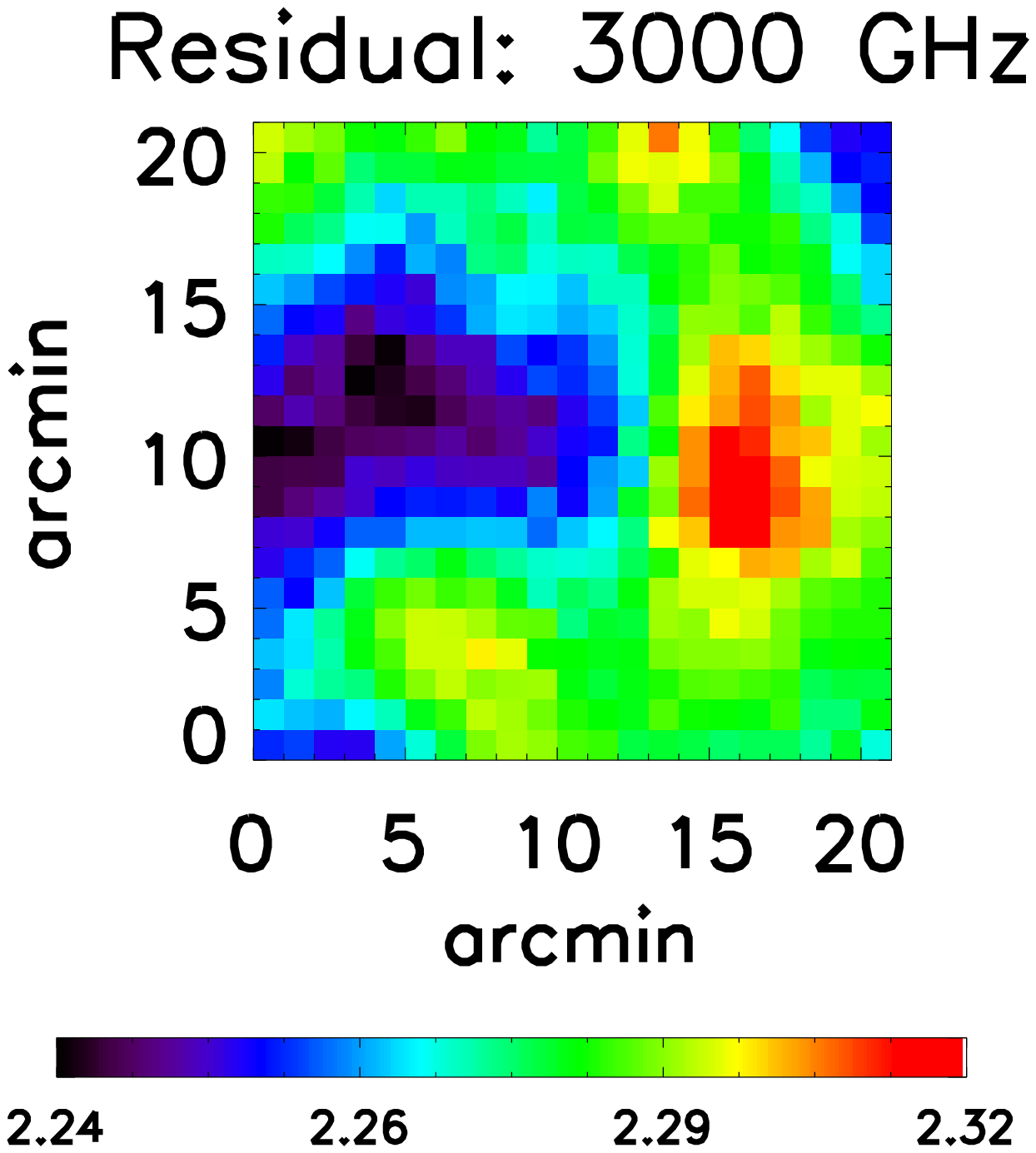}
}}   
\caption{Top panel: \Planck\ and IRIS maps (in
    units of MJy$sr^{-1}$) stacked at the positions of 65 SPT
    synchrotron sources \citep{vieira2010}. Each map in the stack is centred on the
    SPT-derived position of the synchrotron source. The original size of
    the stacked maps is $1\deg\times1\deg$. Here, we show the central
    $20\arcmin\times20\arcmin$ region in order to see the
    residual structure more clearly. Bottom panel: residual
    maps after the compact source (the SPT synchrotron source) at the centre of
  the stacked maps is removed using the formalism in Appendix~\ref{app:planck_stack_formalism}.  
\label{fig:blazars}}
\end{figure}

\subsection{Random rotations of maps}
\label{app:random_rotations}

We make another 1000 realisations of stacking \Planck\ and
  IRIS maps at the positions of the SPT DSFGs by rotating the
  individual maps randomly by 90$\deg$ before
stacking them. In each realisation, we then remove the compact source component from the
stacked map at each frequency using the formalism in Appendix~\ref{app:planck_stack_formalism}. 
Six realisations of the residual maps which are obtained after the
removal of the compact source are chosen at random and displayed
in Figure~\ref{fig:random_rotations}. 
We also verified that the measured mean flux densities in this work did not change significantly when we
introduced the random rotations of the individual maps.

\subsection{Stacking maps of SPT synchrotron sources}
\label{app:blazars}

We stack \Planck\ and IRIS maps at the positions of a
  sample of 65 synchrotron sources detected in the SPT survey \citep{vieira2010}
  and with $S_{220} > 20$ mJy. These sources are not angularly correlated
with foreground structure, unlike the SPT DSFGs. We then remove the compact source component from the
stacked maps at each frequency using the formalism in
Appendix~\ref{app:planck_stack_formalism}. The results are shown in
Figure~\ref{fig:blazars}. We observe no significant excess emission at
217$-$857\,GHz after removal of the central compact source. The residual maps are very different to those for the
SPT DSFGs (Fig~\ref{fig:stack_maps}) where we observe a clear extended emission at 545 and 857\,GHz.

\begin{figure}
\centerline{\rotatebox{0}{
\includegraphics[width=0.525\textwidth]{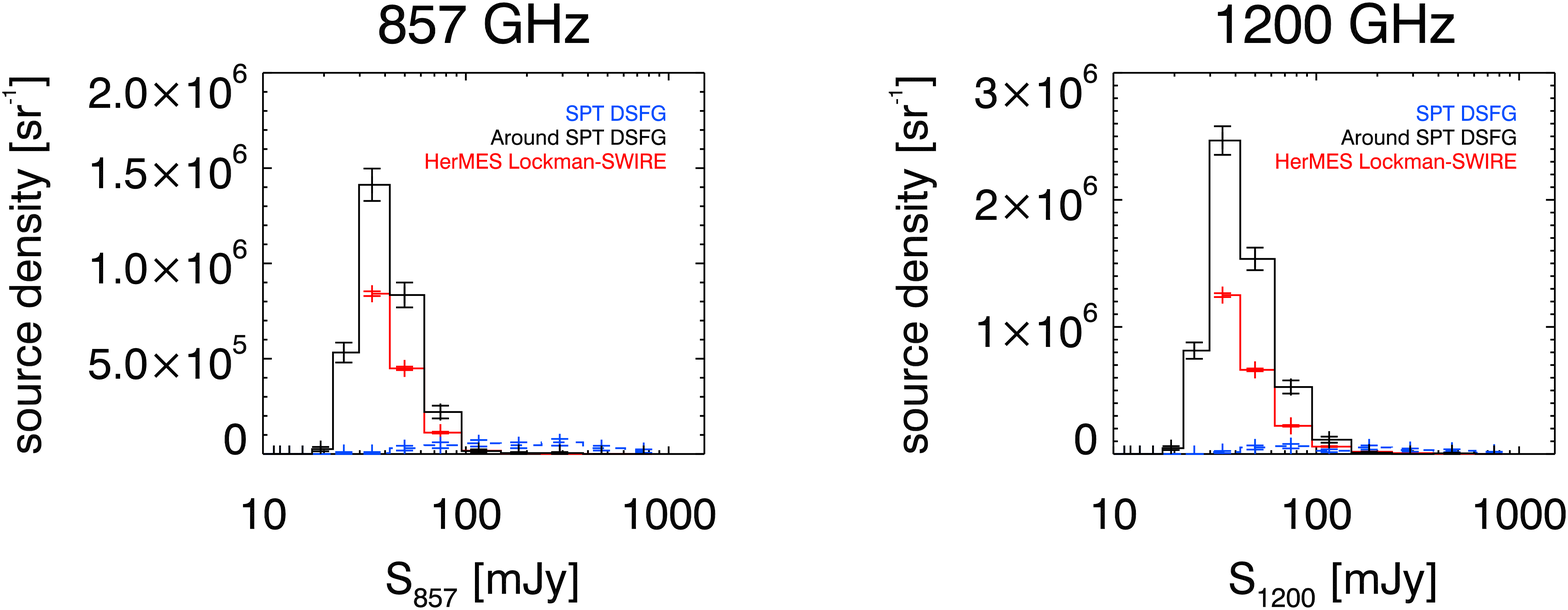}
}}
\caption{Number density of detected
    sources as a function of their flux densities at 857\,GHz (left) and 1200\,GHz (right). 
The sources considered are: (a) those detected at $S_{857}>50$\,mJy
around the DSFG (black); (b) all sources detected at $S_{857}>50$\,mJy 
in the HerMES Lockman-SWIRE field (red); and (c) the SPT DSFG
themselves (blue). This figure is analogous to
  Fig.~\ref{fig:herschel_numberdensity}, which shows the number density
  of the detected sources (of each of the 3 types) by their $S_{857}/S_{1200}$ colour.
\label{fig:herschel_flux_counts}}
\end{figure}

\section{Determining $p(z)$ of the compact source and \textit{Planck} excess from
  the stack}
\label{app:prob_z}

In order to obtain the probability distribution for the redshift $p(z)$ for both the compact source component and
the \Planck\ excess signal, the expected flux density, $T_i$, for each
frequency channel, $i$, is calculated for the template SEDs at a range
of redshifts $z\in [0,6]$. A $\chi^2$ value is computed for each $z$:

\begin{equation}
\label{eq:pz_chi2}
\chi^2(z) = \sum_{i=0}^{N_{\rm f}} {\frac{  (F_i - b(z)T_i(z))^2} {\sigma_i^2}},
\end{equation}

\noindent where $F_i$ is the observed flux density through channel $i$, $\sigma_i$ is the error in $F_i$, $T_i(z)$ is the flux density in the same channel for the template SED at redshift $z$, $N_{\rm f}$ is the number of frequency channels, and $b(z)$ is a scaling factor that normalizes the template to the observed flux density and is determined by minimizing Eq.~\ref{eq:pz_chi2} with respect to $b$ at that redshift, giving
 
\begin{equation}
\label{eq:pz_b}
b(z) = \frac {\sum_{i=0}^{N_{\rm f}} F_iT_i(z)/\sigma_i^2} {\sum_{i=0}^{N_{\rm f}} T_i(z)^2/\sigma_i^2}.
\end{equation}

The probability distribution for the redshift, $p(z)$, will have the form:

\begin{equation}
\label{eq:pz_final}
p(z) \propto  e^{-{\chi^2(z)}}.
\end{equation}


\section{Measuring number densities in \textit{Herschel}}
\label{app:herschel_source_densities}

We estimate the number densities for the three different types of sources
as follows.

\begin{itemize}
\item The number density (per ${\rm sr}^{-1}$), $n_{\rm neighbours}$, of 
sources with $S_{857}>50$\,mJy within $3\parcm5$ of
the DSFGs, defined by:
\begin{equation}
n_{\rm neighbours} = \frac {N_{\rm sources}} {N_{\rm DSFG} \times \omega_{\rm aper}}
\end{equation}
, where $N_{\rm sources}$ is the number of detected sources around the
DSFG, $N_{\rm DSFG}$ is the number of apertures, and $\omega_{\rm aper}$ is the
solid angle subtended by the aperture. The DSFG is not counted in $N_{\rm sources}$.
\item The number density, $n_{\rm null}$, of all 
sources with $S_{857} >50$\,mJy across the entire HerMES Lockman-SWIRE
  field. We will use this as a null test.
\item The number density of SPT DSFGs, $n_{\rm DSFG}$, with
  $S_{220}>20$\,mJy using the same SPIRE maps of the SPT DSFGs. 
\end{itemize}

Figure~\ref{fig:herschel_flux_counts} shows the number density of the
detected sources (for each of the above three classes) per bin of flux
density at 1200 and 857\,GHz. Throughout this paper, we use $S_{220}>20$\,mJy for the SPT flux selection. 

\bsp

\label{lastpage}

\end{document}